\documentclass[sigconf]{acmart}

\usepackage{multirow,tabularx} % multi rows 
\usepackage{adjustbox} % adjusting the table
\usepackage{booktabs}
\usepackage{color, colortbl} % cell colour
\usepackage{array, booktabs, makecell}
\usepackage{comment} 
\usepackage{color,soul}
\usepackage{graphicx}
\usepackage{caption}
\usepackage{subcaption}
\usepackage{nth}
\usepackage{url}
\usepackage{textcomp}
\usepackage[normalem]{ulem}
\usepackage{pgfplots}
\usepackage{tikz}
\usetikzlibrary{patterns}
\usepackage[figuresright]{rotating}
\usepackage[most]{tcolorbox}   % box for things
\usepackage{textcomp}  % add irow
\usepackage{mathtools}
\usepackage{amsmath}
\usepackage{tabularx}
\usepackage{pgfplots}
\usepackage{pgfplotstable}
\usepackage{sfmath}
\usepackage{array, booktabs, makecell}
\usepackage{color}
\usepackage{csquotes}
\usepackage{url}
\usepackage{comment}
\usepackage{tikz}
\usepackage{listings}
\usepackage{breqn}
\usepackage{booktabs} % used for \toprule in tables
\usepackage{multirow}
\usetikzlibrary{fit} %% Used for putting dotted box in image
\usetikzlibrary{positioning}
\usetikzlibrary{arrows}
\usetikzlibrary{shapes.multipart}
\usepackage{float}
\usepackage{enumitem}
\usepackage[small,compact]{titlesec}
\usepackage{mathtools}
\usepackage{pgfplots}
\usepgfplotslibrary{fillbetween}
\pgfplotsset{compat=1.11,width=10cm}
\usepgfplotslibrary{statistics}

\usetikzlibrary{decorations.pathmorphing}
\tikzset{snake it/.style={decorate, decoration=snake}}
\usepackage{xcolor}
\definecolor{MyOrange}{RGB}{237,125,49}
\definecolor{MyBlue}{RGB}{91,155,213}
\usepackage[flushleft]{threeparttable}
\usepackage{tikz}

\newcommand{\RQA}{RQ$_1$: What test smell detection tools are available to the community, and what are the common smell types they support?}%To what extent have test smell detection tools been made available by the research community?}

\newcommand{\RQB}{RQ$_2$: What are the main characteristics of test smell detection tools?}
\setcopyright{acmcopyright}
\copyrightyear{2021}
\acmYear{2021}
\acmDOI{}

\acmConference[EASE '21]{EASE '21: The International Conference on Evaluation and Assessment in Software Engineering}{June 16--18, 2021}{Trondheim, Norway}
\acmBooktitle{EASE '21: The International Conference on Evaluation and Assessment in Software Engineering,
  June 21--23, 2021, Trondheim, Norway}
\acmPrice{}
\acmISBN{}

\begin{document}
\title{Test Smell Detection Tools: A Systematic Mapping Study}


\newcommand\Mark[1]{\textsuperscript#1}

\author{Wajdi Aljedaani}
\email{wajdialjedaani@my.unt.edu}
\affiliation{%
  \institution{University of North Texas}
 \city{Denton}
 \state{Texas}
 \country{USA}}
 
\author{Anthony Peruma}
\email{axp6201@rit.edu}
\affiliation{%
 \institution{Rochester Institute of Technology}
 \city{Rochester}
 \state{New York}
 \country{USA}}
 
\author{Ahmed Aljohani}
\email{aha3089@rit.edu}
\affiliation{%
 \institution{Rochester Institute of Technology}
 \city{Rochester}
 \state{New York}
 \country{USA}}
 
\author{Mazen Alotaibi}
\email{mfa2886@rit.edu}
\affiliation{%
 \institution{Rochester Institute of Technology}
 \city{Rochester}
 \state{New York}
 \country{USA}}

\author{Mohamed Wiem Mkaouer}
\email{mwmvse@rit.edu} 
\affiliation{%
 \institution{Rochester Institute of Technology}
 \city{Rochester}
 \state{New York}
 \country{USA}}

\author{Ali Ouni}
\email{ali.ouni@etsmtl.ca} 
\affiliation{%
  \institution{ETS Montreal, University of Quebec}
  \city{Montreal}
  \state{Quebec}
  \country{Canada}}
  
\author{Christian D. Newman}
\email{cnewman@se.rit.edu} 
\affiliation{%
 \institution{Rochester Institute of Technology}
 \city{Rochester}
 \state{New York}
 \country{USA}}
 
\author{Abdullatif Ghallab}
\email{Abdullatif.Ghallab@unt.edu}
\affiliation{%
  \institution{University of North Texas}
 \city{Denton}
 \state{Texas}
 \country{USA}}

\author{Stephanie Ludi}
\email{Stephanie.Ludi@unt.edu}
\affiliation{%
  \institution{University of North Texas}
 \city{Denton}
 \state{Texas}
 \country{USA}}

\renewcommand{\shortauthors}{W. Aljedaani, A.Peruma, A. Aljohani, M. Alotaibi, M.W. Mkaouer, A. Ouni, C.D. Newman, A. Ghallab, and S. Ludi}

\begin{abstract}
Test smells are defined as sub-optimal design choices developers make when implementing test cases. Hence, similar to code smells, the research community has produced numerous test smell detection tools to investigate the impact of test smells on the quality and maintenance of test suites. However, little is known about the characteristics, type of smells, target language, and availability of these published tools. In this paper, we provide a detailed catalog of all known, peer-reviewed, test smell detection tools.%While prior research has cataloged these tools, the information captured is limited to the tool's name, a brief description, and website.
%In this paper, we perform a comprehensive search of peer-reviewed scientific publications and create a detailed catalog of all known  %more recent collection of 
%test smell detection tools. 

We start with performing a comprehensive search of peer-reviewed scientific publications to construct a catalog of 22 tools. Then, we perform a comparative analysis to identify the smell types detected by each tool and other salient features that include programming language, testing framework support, detection strategy, and adoption, among others. From our findings, we discover tools that detect test smells in Java, Scala, Smalltalk, and C++ test suites, with Java support favored by most tools. These tools are available as command-line and IDE plugins, among others. Our analysis also shows that most tools overlap in detecting specific smell types, such as General Fixture. Further, we encounter four types of techniques these tools utilize to detect smells. %Further, not all these tools are not limited to the command-line; we encounter tools that function as IDE plugins. Our analysis also shows that most tools overlap in detecting specific smell types, such as General Fixture and Eager Test. Further, we encounter four types of techniques these tools utilize to detect smells-- rules, metrics, information retrieval, and dynamic tainting. 
We envision our study as a one-stop source for researchers and practitioners in determining the tool appropriate for their needs. Our findings also empower the community with information to guide future tool development.%Additionally, our findings also empower the community with the necessary information to guide future tool development.

\end{abstract}

%%%%%%%%%%%%%%%%%%%%%%%%%%%%%%%%%%%%
% Remove for camera ready:
\settopmatter{printacmref=false}
\setcopyright{none}
\renewcommand\footnotetextcopyrightpermission[1]{}
\pagestyle{plain}
%%%%%%%%%%%%%%%%%%%%%%%%%%%%%%%%%%%%

\begin{comment}
\begin{CCSXML}
<ccs2012>
   <concept>
       <concept_id>10002944.10011122.10002949</concept_id>
       <concept_desc>General and reference~General literature</concept_desc>
       <concept_significance>500</concept_significance>
       </concept>
   <concept>
       <concept_id>10002944.10011122.10002945</concept_id>
       <concept_desc>General and reference~Surveys and overviews</concept_desc>
       <concept_significance>500</concept_significance>
       </concept>
   <concept>
       <concept_id>10011007.10011006.10011073</concept_id>
       <concept_desc>Software and its engineering~Software maintenance tools</concept_desc>
       <concept_significance>500</concept_significance>
       </concept>
   <concept>
       <concept_id>10011007.10011074.10011099.10011102.10011103</concept_id>
       <concept_desc>Software and its engineering~Software testing and debugging</concept_desc>
       <concept_significance>500</concept_significance>
       </concept>
 </ccs2012>
\end{CCSXML}

\ccsdesc[500]{General and reference~General literature}
\ccsdesc[500]{General and reference~Surveys and overviews}
\ccsdesc[500]{Software and its engineering~Software maintenance tools}
\ccsdesc[500]{Software and its engineering~Software testing and debugging}

\keywords{Test smells; smells; test code; systematic literature review; comparative study; detection tools.}
\end{comment}

\maketitle

\section{Introduction}
\label{sec:Introduction}

Software testing is an essential part of the software development life cycle. %\cite{beck2003test}. 
As part of the software development process, developers create and update their system's test suite to ensure that the system under test adheres to the requirements and provides the expected output \cite{pressman2014software}. However, test code, similar to production code, is subject to bad programming practices (i.e., smells), which hamper the quality and maintainability of the test suite \cite{meszaros2007xunit}. Formally defined in 2001 \cite{van2001refactoring}, the catalog of test smells has been steadily growing throughout the years. While most test smells focus on traditional Java systems, researchers have also studied the impact of these smells on other programming languages, and platforms \cite{reichhart2007rule,breugelmans2008testq,peruma2019CASCON}. With the growth of the test smell catalog, the research community, in turn, has utilized these smells to study the impact test smells have on the maintainability of test suites. These studies show that test smells negatively impact the comprehension of a test suite and increase change- and defect-proneness of the test suite, thereby increasing its flakiness \cite{Luo2014FSE,Spadini2018ICSME}. In addition to defining test smells, researchers have also provided the community with various tools to detect such test smells. Furthermore, research has shown that early detection of bad smells reduces maintenance costs \cite{Lacerda2020JSS}, highlighting the importance of such detection tools.

%\mohamed{this is the most important paragraph in the introduction and it is weak since it is showing our work to be just incremental to the previous SLRs. I will rewrite this and I want you to give your opinion.}These numerous studies on test smells underscore the importance they play in the quality and maintenance of test suites. This, in turn, has spawned work by Garousi and K{\"u}{\c{c}}{\"u}k in cataloging such studies in the form of systematic mapping studies \cite{garousi2018smells,garousi2019smells}. While these studies, along with Santana et al. \cite{Santana2020SBES}, do investigate the availability of test smell detection tools. However, the findings are limited to only a listing of the tools. Our work complements these studies by not only expanding on the set of tools, published in peer-reviewed venues, but also providing more in-depth details about the tools. As part of our tool inventory, we compare and contrast multiple attributes of these tools, such as supported smell types, technologies, and detection mechanisms, among others. Combined with these prior studies, our work provides developers and researchers with a state-of-the-art representation and detection of test smells.

With the growth of test smells studies, recent literature reviews \cite{garousi2018smells,garousi2019smells} have been proposed to study various dimensions related to these anti-patterns. These literature reviews have explored the various definitions of test smells, empirical analysis of their survival, spread, refactoring, and their relationship with change and bug proneness of source code. However, little is known about the toolsets used to detect test smells. The availability of tools is vital for software engineering researchers and practitioners. In research, tools facilitate the reproducibility of studies while developers benefit from improved productivity through tool adoption \cite{Siegmund2015ICSE}. Without a thorough understanding of available tools and how these tools compare to one another, it will be difficult to conduct future research that uses the right toolset for a given research problem. Therefore, our work complements these reviews by not only extracting all the test smell detection tools published in peer-reviewed venues, but also providing more in-depth details about them. To facilitate their adoption, we compare and contrast multiple attributes of these tools, such as supported smell types, target environment, detection mechanisms, etc. Hence, our work provides a catalog for developers and researchers to support the adoption of these tools. %for test smells with a state-of-the-art representation and detection of test smells. 

%In addition to defining test smells, researchers have also provided the community with various tools to detect such test smells. However, not all tools support all test smell types. Additionally, there also exists overlaps in smell support between some of these tools. In other words, there does not exist a single tool that supports the detection of all known test smell types. Hence, in this paper, we provide an inventory of all known test smell detection tools that have appeared in peer-reviewed publications. Additionally, we compare and contrast multiple attributes of these tools.

\subsection{Goal \& Research Questions}
The goal of this study is to provide developers and researchers with a one-stop source that offers a comprehensive insight into test smell detection tools. The information in this study will \textit{allow researchers to select the right tool for their research task and provide data-driven advice on how test smell tools can be advanced through future work}. Through this study, the community will be better equipped to determine the correct tool they need to utilize to satisfy their requirement, along with the shortcomings of these tools. This work also provides the research community with insight into areas that require improved automation.  Hence, we aim at addressing the following research questions (RQs):

%\anthony{RQ 1 and 3 need to be merged since. RQ1 contains a listing of the smells, which is repeated in RQ . I propose that RQ1 should be composed of two sub-RQs. RQ1.1 is the results of the SLR search, and RQ1.2 are the smell types detected by each app \newline \newline
%RQ 2 and 4 also can be merged since these two RQs look at the features of the tools. RQ 2.1 will look at the general/platform features, while RQ2.2 will look a the detection architecture the tools utilize. \newline \newline
%The new RQ breakdown improves the flow of the paper. RQ 1 will be a high-level intro to the tools, and RQ2 will go deeper into the features/architectures of the tools.}
%\christian{Maybe change RQ1 to something like: "What test smell detection tools are available to the community and what level of support do they provide for detecting test smells?" - at the end of the paragraph we say we'll discuss the spectrum of test smells but it's nowhere in the question} 
\smallskip
\textbf{\RQA}
This RQ investigates the volume of test smell detection tools released by the research community. We answer this RQ by performing an extensive and comprehensive search on six popular digital libraries among the software engineering community. We investigate the frequency of tool release, and the spectrum of test smells detected by the tools.

%\christian{RQ2 is somewhat abstract. But maybe this is the best way to word it. I'll think about it after I read the evaluation}\mohamed{after reading the RQs again, I totally agree with Christian, both RQs are obscure, especially the second one}
\smallskip
\textbf{\RQB}
In this RQ, we examine the design-level features that are common to test smell detection tools, such as platform support and smell detection mechanisms. This RQ provides us with details into how the research community constructs such tools and provides insight into the development of future tools.

\subsection{Contributions}
Through this study, we provide the community of researchers and practitioners with a view and insights on the history of the availability of test smell detection tools. More specifically, our contributions are outlined below:
\begin{itemize}[wide, labelwidth=1pt, labelindent=0pt]
    \item A catalog of 22, peer-reviewed, test smell detection tool publications, and publications that utilize these tools. These publications provide an initial platform for future research in this area.
    \item A series of experiments highlighting the growth of such tools, along with a comparison of key tool attributes.
    \item A discussion of how our findings provide insight into future research areas in this field along with details that need to be considered when selecting a test smell detection tool.%\mohamed{I do not like this, it is too repetitive:} Our discussion provides insight to the researcher and practitioner community around the scope for improvements in this field and decisions that they need to keep in mind when selecting a test smell detection tool.
     \item A replication package of our survey for extension purposes \cite{ProjectWebiste}.
\end{itemize}

\section{Research Methodology}
\label{sec:ResearchMethod}

Being a Systematic Mapping Study (SMS), our research explores published scientific literature to gather information about a specific topic in software engineering, and to provide a high-level understanding and/or answering exploratory research questions \cite{Petersen2008EASE,Barn2017ISEC}. To this extent, our SMS aims at proving a high-level understanding of the existence of test smell detection tools, their characteristics, and their adoption in academic studies. Prior studies in cataloging detection-based tools, were associated with technical debt \cite{avgeriou2020overview}, bad smells \cite{fernandes2016review}, bug localization \cite{akbar2020large}, and architectural smells \cite{azadi2019architectural}. Further, while Garousi and K{\"u}{\c{c}}{\"u}k \cite{garousi2018smells} provide a list of test smell detection tools as part of their SMS on test smells, our study aims to expand on this listing. Hence, in this section, we describe the procedure we adopt to search and select the relevant publications for analysis. In brief, our methodology consists of three phases-- (1) planning, (2) execution, and (3) synthesis. In the following subsections, we elaborate on these phases. 

\subsection{Planning}
In this phase, we detail our publications search strategy. In conformance with systematic mapping studies, we utilize a specific set of domain-specific (i.e., test smell related) keywords to search, in popular digital libraries, for publications that meet our requirements.

\subsubsection*{\textbf{Digital Libraries.}}
To locate publications for our study, we search six digital libraries. These libraries, listed in Table \ref{tab:digital_libraries}, either contain or index publications from computer science and software engineering venues and are utilized by similar studies (e.g., \cite{fernandes2016review}). %\cite{Barn2017ISEC,fernandes2016review}.

\begin{table}
\centering
\caption{The digital libraries queried in our study. }
\vspace{-0.3cm}
\small
\label{tab:digital_libraries}
\begin{tabular}{|l|l|}\hline
\rowcolor{gray!60}\multicolumn{1}{|c|}{\textbf{Digital Library}} & \multicolumn{1}{c|}{\textbf{URL}} \\
ACM Digital Library                            & \url{https://dl.acm.org/}               \\
\rowcolor{gray!30} IEEE Xplore                                     & \url{https://ieeexplore.ieee.org/}      \\
Science Direct                                 & \url{https://www.sciencedirect.com/}    \\
\rowcolor{gray!30} Scopus                                         & \url{https://www.scopus.com/}           \\
Springer Link                                  & \url{https://link.springer.com/}        \\
\rowcolor{gray!30} Web of Science                                 & \url{https://webofknowledge.com/}       \\ \hline
\end{tabular}
\vspace{-0.5cm}
\end{table}

\subsubsection*{\textbf{Inclusion/Exclusion Criteria.}}

%Through the inclusion and exclusion criterion, we retrieve vial peer-reviewed scientific publications. This criterion also helps in guiding the search process of backward and forward snowballing and also helps in creating the search string for electronic libraries. Table \ref{table:selection_criteria} lists the inclusion and exclusion we utilize in this study. With regards to the time range, we did not set a starting date. However, our end date was set to the end of December 2020. Hence, the date range for our inclusion criteria is all publications till 31-December 2020.

Inclusion and exclusion criteria are crucial in pruning our search space, reducing bias, and retrieving relevant peer-reviewed scientific publications. Selected publications of these criteria become our starting point for manual filtering, to see whether they fit in our study, i.e., propose or adopt a test smells tool. The initial pool of publications also serves for: (1) backward snowballing, i.e., analyzing publications cited by the selected pool; and (2) forward snowballing, i.e., analyzing publications citing our pool publications. %along This criterion also helps in guiding the search process of backward and forward snowballing and also helps in creating the search string for electronic libraries. 
Table \ref{table:selection_criteria} lists the inclusion and exclusion criteria considered in this study. With regards to the time range, we did not set a starting date. However, our end date was set to the end of December 2020. Hence, the date range criterion allows the selection of any tool as long as it appeared before December 31, 2020.

%%%%%%%%%%%%%%%%
\begin{table}[h!]
\centering
\caption{Our inclusion and exclusion search criteria.}  % title of Table
\vspace{-0.3cm}
\resizebox{\columnwidth}{!}{%
\begin{tabular} {|l|l|}\hline

\rowcolor{gray!60}
 \multicolumn{1}{|c|}{\textbf{Inclusion}} & \multicolumn{1}{|c|}{\textbf{Exclusion}} \\ %[0.5ex]
% inserts table
%heading
\hline % inserts single horizontal line
Published in Computer Science	& Websites, leaflets, and grey literature \\ % inserting body of the table
%%%%%%%%%%
\rowcolor{gray!30}
Written in English           & Published in 2021  \\ %Papers shorter than two pages
%%%%%%%%%%
Available in digital format  	& Full-text not available online\\
%%%%%%%%%%
\rowcolor{gray!30}
Propose or use test smell detect tool   & Tools not associated with peer-reviewed papers\\%Student (PhD and Master) thesis papers

\hline %inserts single line
\end{tabular}
}
\label{table:selection_criteria} % is used to refer this table in the text
\vspace{-0.2cm}
\end{table}
%%%%%%%%%%%%%%%%%%%%%%%%%

\subsubsection*{\textbf{Search Keywords.}}
To determine the optimal set of search keywords, we conducted a pilot search \cite{Parkkila2015CSR} on two well-known digital libraries, i.e., IEEE and ACM. This process intends to identify relevant words or synonyms utilized in test smell publications. We performed multiple instances of the pilot search, where each instance involved refinement of the keyword terms in the search query. We conduct our query only on the title and abstract of the publication. We decided to apply the search on a publication's metadata instead of on the full-text to avoid false positives. The finalized search string is presented below.

\begin{tcolorbox}[colback=black!5!white,colframe=black!50!black,top=2mm, bottom=2mm, left=2mm, right=2mm]
  \texttt{\emph{Title:("tool*" OR "detect*" OR "test smell" OR "test smells") AND Abstract:("test smell" OR "test smells" OR "test code" OR "unit test smell")}}
\end{tcolorbox}

\subsection{Execution}

\begin{figure*}
	\centering
    \includegraphics[width=\textwidth]{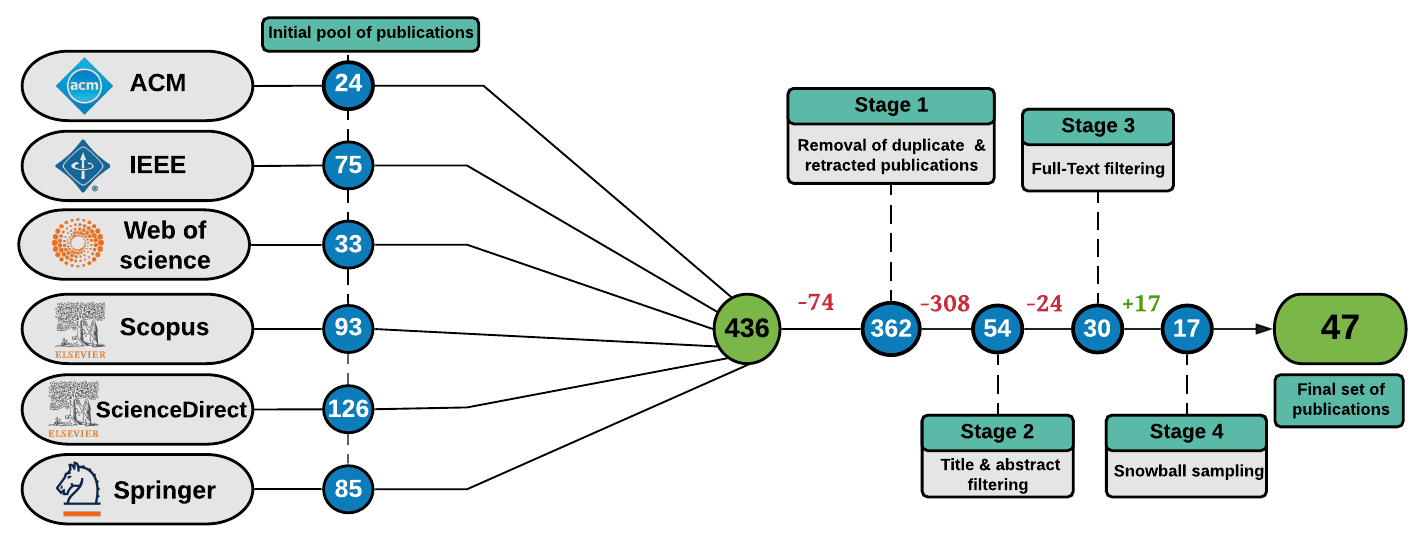}   
    \vspace{-0.3cm}
    \caption{Overview of the volume of publications resulting from our filtering process.}
    \label{fig:SearchProcess}
    \vspace{-0.4cm}
\end{figure*}

In this phase, we detail how we process and filter the publications we obtain from our digital library search. Our initial search of the six digital libraries results in 436 publications, with ScienceDirect resulting in the highest number of publications (126). Next, we employ a four-stage quality control process to filter out publications that were not part of our inclusion criteria. Figure \ref{fig:SearchProcess} depicts the volume of publications filtered at each stage. This quality control process involves three authors manually reviewing the publications to determine if a publication can pass from one stage to another. The first stage starts with removing duplicate and retracted publications; 74 publications were removed, and 362 candidate publications made it to the next stage. In the second stage, we apply our inclusion and exclusion criteria to the title and abstract. For instance, we discard all publications that were not peer-reviewed or did not propose or adopt a tool. This thorough procedure resulted in only including 54 publications. The third stage involves a full-text analysis of each selected publication. Using our inclusion and exclusion criteria, we retain only 30 publications. In the last stage, we perform the forward and backward snowball sampling \cite{10.1145/2601248.2601268}, resulting in 17 publications. Therefore, we ended up with a final set of 47 publications.

%Using the citations associated with each tool development publication, we perform a snowballing analysis to identify further publications that utilize these tools (which were not captured in our digital library search). Furthermore, we review each publication in our pool with a focus on each of our RQs to extract the required information to answer our RQs. As part of this review, we evaluate each study based on concrete evidence present in the publication (and supporting artifacts) without any vague assertions. Finally, all RQ-related data collected during the publication review task was reviewed to ensure bias mitigation. 

\subsection{Synthesis}

This phase synthesizes the extracted data to answer our RQs. First, we classify the primary set of publications into one of two types, i.e., tool development or tool adoption. Tool development publications are studies that propose a test smell detection tool, either as part of proposing a new catalog of smells or detecting existing smell types. Tool adoption publications are studies that utilize an existing test smell detection tool as part of their study design. Additionally, we also classify studies by publication year and venue; this helps with partly answering RQ$_1$. For the remainder of RQ$_1$ and RQ$_2$, we manually review the full-text of each tool development publication. As part of this review, we evaluate each study based on concrete evidence present in the publication (and its supporting artifacts, if any) without any vague assertions. For each tool, we extract the types of test smells the tool detects and other tool features, such as supported programming language, testing framework, correctness, etc. We elaborate further on the tool features in Section \ref{SubSection:RQB} when presenting our findings for RQ$_2$.

Finally, all RQ-related data collected during the publication review task was peer-reviewed to ensure bias mitigation, with conflicts resolved through discussions. We utilized a spreadsheet to hold the manually extracted data to facilitate collaboration during the author-review process. The authors, participating in the filtering stages and manual review, are experienced with this research. They have published work in this area, including defining test smell types, tool development, and adoption \cite{peruma2020FSE,peruma2019CASCON}. %individuals that performed the publication review and data extraction are Ph.D. students that hold an MSc in software engineering and industry experience.  

\section{Research Findings}
\label{sec:ResultsAndAnalysis}

In this section, we present the findings for our proposed RQs based on the synthesis of the finalized set of 47 test smell tool detection-related publications, which are composed of \textbf{22 publications that propose new tools and 25 publications that adopt these tools.} 

\subsection{\textbf{\RQA}}
\label{SubSection:RQA}

This RQ comprises of four parts. In the first part, we provide a breakdown of the publications by publication date and venue. In the second part, we present the tools identified in our systematic search, while in the third part, we provide insight into the types of test smells detected by the identified tools. Finally, the fourth part looks at the programming languages supported by the test smells.

\subsubsection{\textbf{Publication Years \& Venues}}
\noindent

\begin{figure}[h]
\centering 

\begin{tikzpicture}
    \pgfplotsset{every tick label/.append style={font=\footnotesize}}
    \pgfplotsset{
        show sum on top/.style={
            /pgfplots/scatter/@post marker code/.append code={%
                \node[
                    at={(normalized axis cs:%
                            \pgfkeysvalueof{/data point/x},%
                            \pgfkeysvalueof{/data point/y})%
                    },
                    anchor=south,
                ]
                {\pgfmathprintnumber{\pgfkeysvalueof{/data point/y}}};
            },
        },
    }

  \begin{axis}[
    ybar stacked, ymin=0, ymax=14,  
    width=0.5\textwidth,
    height=.5\textwidth,
    legend style={at={(0.5,1)},
    anchor=north,legend columns=-1},
    symbolic x coords={2006,2007,2008,2010,2012,2013,2014,2015,2016,2018,2019,2020},
    xtick=data,xlabel=\bf Year,ylabel=\bf Publication Count,
    ytick={0,2,4,6,8,10,12,14},
    nodes near coords, 
  ]
  %Tool Development
  \addplot [pattern color=blue, pattern=north east lines] coordinates {
({2006},1)
({2007},1)
({2008},1)
({2010},2)
({2012},1)
({2013},2)
({2014},2)
({2015},2)
({2016},0)
({2018},2)
({2019},4)
({2020},4)};
  %Tool Adoption
  \addplot [pattern=horizontal lines,pattern color=red] coordinates {
({2006},1)
({2007},2)
({2008},1)
({2010},0)
({2012},0)
({2013},0)
({2014},0)
({2015},1)
({2016},3)
({2018},1)
({2019},7)
({2020},9)};
  \legend{Tool Development,Tool Adoption}
  \end{axis}
  \end{tikzpicture}
   	\vspace{-0.3cm}
    \caption{Yearly breakdown of tool publications.}
 	\label{fig:ToolsDistribution}
 	\vspace{-0.4cm}
\end{figure}
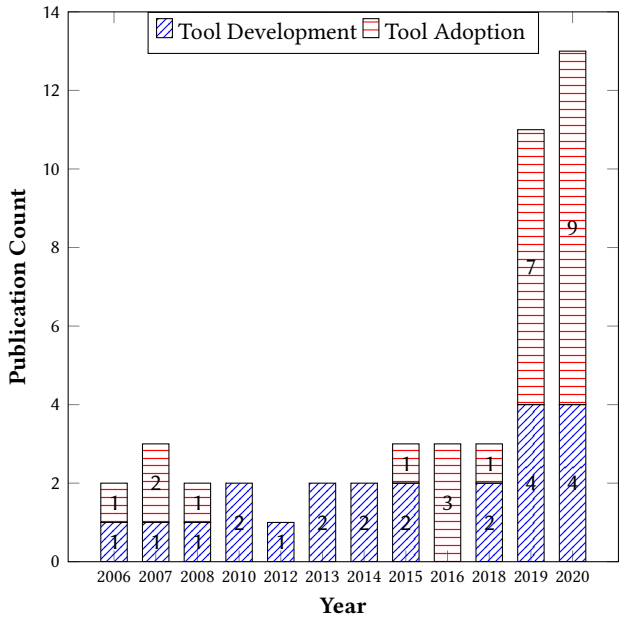

\begin{figure*}
	\centering
    \includegraphics[width=\textwidth]{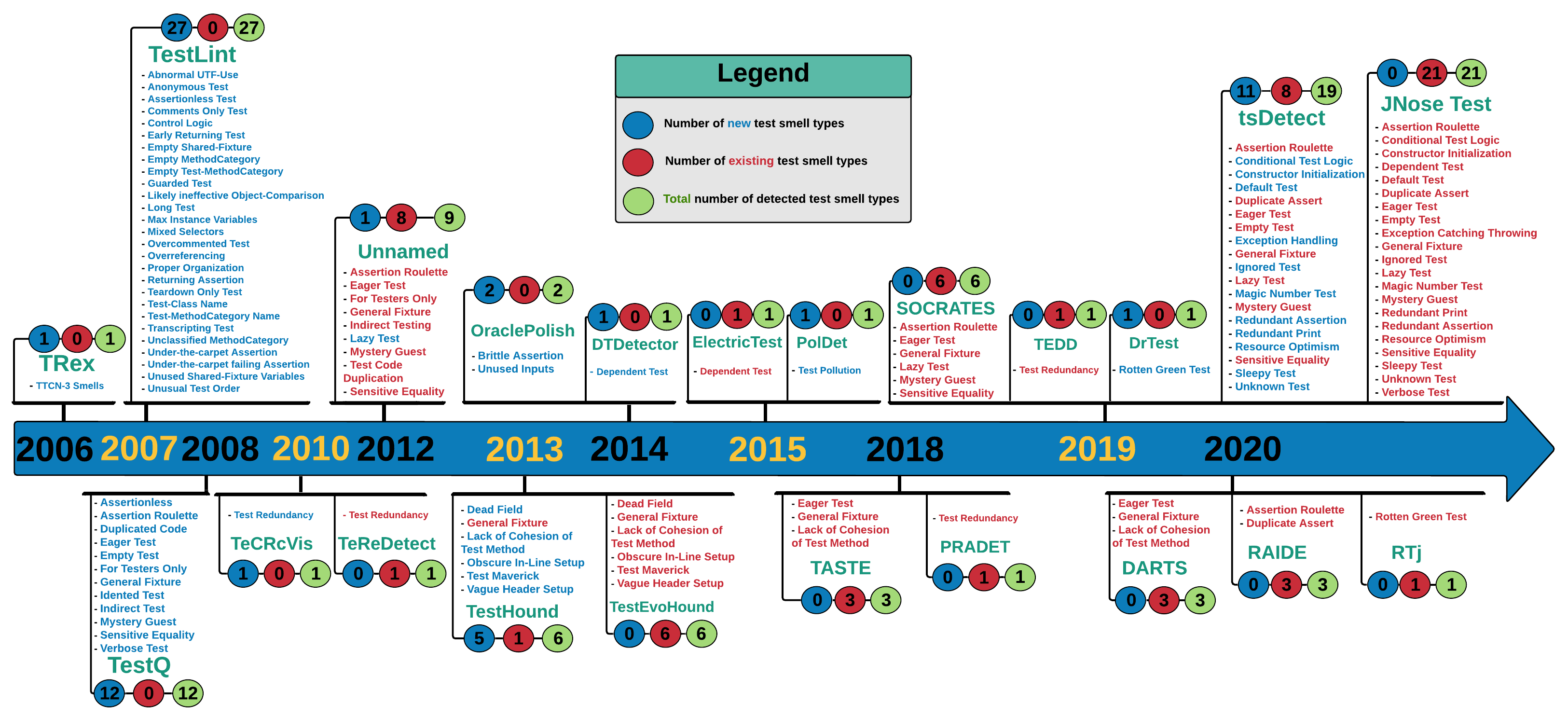}   
    \vspace{-0.3cm}
    \caption{Timeline of the release of test smell detection tools by the research community.} 
    \label{fig:Literature_TimeLine}
    \vspace{-0.4cm}
\end{figure*}

Figure \ref{fig:ToolsDistribution} depicts the yearly breakdown of tool publications. The first test smell tool, TRex \cite{baker2006trex}, appeared in 2006. Since then, there was a steady trend of one or two tools appearing every one or two years, until 2018. The years 2019 and 2020 witnessed a notable increase in tool-based publications compared to the prior years, with approximately 51\% of tool development and adoption publications occurring in these two years. %\mohamed{this seems wrong to me:}\hl{Furthermore, to the latter part of the period (i.e., 2016 onward), researchers have focused on using existing tools in studies than releasing new detection tools}. 
There can be many factors influencing this recent hype. We have observed the following: The dynamic nature of detection mechanisms of traditional state-of-the-art tools made them require compilable projects with constraints over how test files should be written and located. Therefore, traditional tools are implemented to run as standalone applications or plugins in Integrated Development Environments (IDEs). Besides being constrained to their environments, they are not intended to run on large-scale software systems. However, the tools that appeared in recent years were developed as APIs, facilitating their deployment to mine software repositories. While their detection strategies carry the false-positiveness of static analysis, they allowed the analysis of a wide variety of software systems. Therefore, the number of empirical publications, adopting these tools, has significantly increased, reaching up to 16 in two years, higher than the number of all previous tool adoption publications combined. These studies have explored various characteristics of test smells, including co-occurrence, survivability, severity, refactoring, impact on flaky tests, proneness to changes and bugs, etc. Next, in Figure \ref{fig:Literature_TimeLine}, we provide a pictorial representation, in the form of a timeline, depicting the release of the 22 test smell detection tools. Analyzing the smell types detected by each tool, we specify, in green, the total number of smell types detected by each tool. Additionally, we also indicate, in red, the number of net new test smell types and the number of existing smell types in blue. Reading Figure \ref{fig:Literature_TimeLine} from left to right (i.e., the oldest tool to newest), a smell type first introduced by a tool is in blue text, while its subsequent appearance in another tool is in red. For example, the General Fixture smell first appears in TestQ (hence it is shown in blue text), this smell next appears in the unnamed tool (hence it is shown in red text), and so on.    

% \begin{figure}
%  	\centering
%  	\includegraphics[trim=1.3cm 3.5cm 2.5cm 0.5cm,width=1\linewidth]{Charts/publication_by_year.pdf}
%  	\vspace{-0.3cm}
%  	\caption{Yearly breakdown of tool publications.} 
%  	\label{fig:ToolsDistribution}
%  	\vspace{-0.4cm}
% \end{figure}

In terms of venues, looking at the complete set of primary publications, 40 publications are associated with a conference/workshop/symposium, while seven publications appear in journals. Looking at just tool development publications, 20 of these publications are associated with a conference/workshop/symposium. Finally, the most popular venue for a tool development publication is the Joint European Software Engineering Conference and Symposium on the Foundations of Software Engineering, with four publications. The complete breakdown is available in our replication package. 

\subsubsection{\textbf{Test Smell Detection Tools}}
\noindent
% \subsection{List of Contributors in Test Smells Tools}

%\subsection{List of Test Smell Detection Tools}
%\textbf{RQ1:} \textsl{What are the test smell detection tools proposed or used in literature papers?}

In this part of the RQ, based on available documentation (i.e., full-text of the publication and any of its supporting artifacts), we provide an overview of each tool, from the oldest to the most recent.

Released in 2006 is \textbf{TRex} by Baker et al. \cite{baker2006trex}. This tool analyzes for TTCN-3 test suites for issues specific to this testing framework. The tool also provides developers with the ability to correct identified issues. \textbf{TestLint}, released by Reichhart et al. \cite{reichhart2007rule} in 2007, is a rules-based tool that detects 27 quality violations in the unit test code in Smalltalk systems. %Due to the interconnection of specific types of test smells, a rule can be reused to detect several test smells. 
In 2008, Breugelmans and Van Rompaey released \textbf{TestQ} \cite{breugelmans2008testq}. The tool provides a visual interface for developers to explore test suites and detects 12 test smells in C++ test suites. The tool facilitates customizations such as smell prioritization. 

In 2009 Koochakzadeh and Garousi released \textbf{TeReDetect
} \cite{koochakzadeh2010tester} (Test Redundancy Detection), a test redundancy detection tool for JUnit tests that work in conjunction with a code coverage tool. The authors then released \textbf{TeCReVis} (Test Coverage and Test Redundancy Visualization) in 2010 \cite{koochakzadeh2010tecrevis}, an Eclipse plugin that provides developers with a visualization of a project's test coverage and test redundancy. \textbf{Bavota et al.} \cite{bavota2012empirical} released an unnamed test smell detection tool in 2012. The tool detects nine test smell types in Java test suites. The tool prioritizes recall over precision resulting in a long list of potential issues and thereby require manual reviews. 

2013 saw the release of two detection tools. Greiler et al. introduce \textbf{TestHound} \cite{greiler2013automated}. This static analysis tool focuses on test smells related to test fixtures in Java test suites and recommends refactorings to address the detected issues. In a user study, the authors show that developers are appreciative of the tool with regards to understanding test fixture code. Greiler et al. improve on their prior tool by releasing \textbf{TestEvoHound} \cite{greiler2013strategies}. This improved tool analyzes Git or SVN repositories to analyze the evolution of a system's test fixture code. As part of the analysis process, the tool does a checkout and build of each revision of the project and then passes the revision to TestHound to detect test fixture smells.  

Zhang et al. released \textbf{DTDetector}, a JUnit supported test dependency detection tool in 2014 \cite{zhang2014empirically}. Also released in 2014 by Huo et al. \cite{huo2014improving}, is \textbf{OraclePolish}, which utilizes a dynamic tainting-based technique for the detection of two test smell types in JUnit test suites. The tool's empirical evaluation demonstrates that it can detect both brittle assertions and unused inputs in real tests at a reasonable cost. In 2015, Bell et al. released \textbf{ElectricTest} \cite{bell2015efficient}, another dependency detection tool for JUnit test suites. The authors demonstrate that their tool outperforms DTDetector in test parallelization. Also released in 2015 was \textbf{\textsc{PolDet}} by Gyori et al. \cite{gyori2015reliable}, a test pollution smell detection tool. The tool analyses heap-graphs and file-system states during test execution for instances of state pollution (e.g., tests reading/writing shared resources).

%\textbf{HistoryMiner}, developed by Tufano et al. \cite{tufano2016empirical} in 2014 mines a project's repository to detect all Java files to identify the existence of test and code smells in the lifetime of the system. The tool depends on  Bavota et al. \cite{bavota2012empirical} tool to detect five test smell types, and DECOR \cite{Moha2010TSE} to detect code smells. 

Palomba et al. \cite{palomba2018automatic} released \textbf{\textsc{Taste}} (Textual AnalySis for Test smEll detection) in 2018. This tool utilizes information retrieval techniques to detect three test smell types in Java test suites. Results from an empirical study show that the tool is 44\% more effective in detecting test smells when compared to structural-based detection tools. Also released in 2018 is \textbf{\textsc{PraDeT}}, by Gambi et al. \cite{gambi2018practical}. This tool detects manifest test dependencies and can analyze large projects containing a vast quality of tests.

There were four test smell detection tools released in 2019. \textbf{\textsc{tsDetect}}, released by Peruma et al. \cite{peruma2020FSE}, detects a total of 19 test smell types. The smell types comprise of 11 newly introduced types and 8 existing types. The tool utilizes an abstract syntax tree to analyzes JUnit test suites and reports an average F-score of 96.5\% for each smell type. Further, one smell type (i.e., \textit{Default Test}) is exclusive to Android applications, while the remaining types apply to all Java systems. \textbf{\textsc{SoCRATES}} (SCala RAdar for TEst Smells) by De Bleser et al. \cite{de2019socrates} detects the presence of six smell types in Scala systems using static analysis. Virginio et al. released \textbf{JNose Test}, a tool with the ability to detect 21 test smell types in Java systems. Additionally, the tool also provides ten metrics around code coverage. Biagiola et al. released \textbf{TEDD} (Test Dependency Detector), a  tool to detect test dependencies in end-to-end web test suites \cite{biagiola2019web}. The tool presents a list of manifest dependencies as output from its execution. Delplanque et al. \cite{Delplanque2019ICSE} released \textbf{DrTest}, a tool that detects \textit{Rotten Green Test} smell in the Pharo ecosystem.    
  
2020 saw the release of two IDE plugins and one command-line tool. Lambiase et al. \cite{lambiase2020just} released \textbf{DARTS} (Detection And Refactoring of Test Smells), a plugin that utilizes information retrieval to detect three smell types. The tool also offers refactoring support. \textbf{RAIDE}, an Eclipse plugin was released by Santana et al. \cite{Santana2020SBES}. This plugin detects and provides semi-automated refactoring support for two test smell types in JUnit test suites. Martinez et al. \cite{Martinez2020ICSE} released \textbf{RTj}, a command-line tool that supports the detection and refactoring of \textit{Rotten Green Test} smells.

Finally, when compared against the catalog of Garousi and K{\"u}{\c{c}}{\"u}k \cite{garousi2018smells}, our dataset contains ten of the 12 listed tools. The tools excluded from our study are not proposed in peer-reviewed literature. The common set of tools are indicated in RQ2.

\subsubsection{\textbf{Detected Test Smell Types}}
\noindent

%\subsection{Types of Test Smells Detected}
%\textbf{RQ3:} \textsl{What are types of test smells the existence tools aim to detect?} \anthony{This RQ overlaps with RQ 1. The timeline figure in RQ1 also shows th smell types detected by the tools}
%\wajdi{The figure 3 (RQ1), showed the tools with the new test smell types introduced in each new tools , however, RQ3 -- we aim to obverse the similarity of test smell types across the 11 tools (Table 6). For example GF used in 10 tools.}

\begin{table*}
\centering
\caption{Definition of the test smells detected by the tools in our dataset.}
\vspace{-0.3cm}
% \resizebox{\columnwidth}{!}{%
\begin{adjustbox}{width=1.1\textwidth,center}
% \begin{threeparttable}
\begin{tabular} {|l|l|l|l|l|}\hline
\rowcolor{gray!60}
\textbf{No.} & \multicolumn{1}{c|}{\textbf{Test Smell Name}} & \multicolumn{1}{c|}{\textbf{Abbreviation}} & \multicolumn{1}{c|}{\textbf{Definition}} & \multicolumn{1}{c|}{\textbf{Ref.}} \\ \hline   

%%%%%%%%%%%%%%%
\textbf{01} & \textbf{Abnormal UTF-Use} & \textbf{AUU} & Overriding the default behavior of the testing framework by test-suite.   & \cite{reichhart2007rule} \\ \hline
%%%%%%%%%%%%%%%
\rowcolor{gray!30}
\textbf{02} & \textbf{Anonymous Test} & \textbf{AT} &   A test method with a meaningless and unclear method name.      & \cite{reichhart2007rule} \\ \hline
%%%%%%%%%%%%%%%
\textbf{03} & \textbf{Assertion Roulette} & \textbf{AR} & A test method with multiple assertions without explanation messages. &  \cite{bavota2012empirical} \\ \hline
%%%%%%%%%%%%%%%
\rowcolor{gray!30}
\textbf{04} & \textbf{Assertionless} & \textbf{AL} &   A test that is acting to assert data and functionality but does not.    & \cite{breugelmans2008testq} \\ \hline
%%%%%%%%%%%%%%%
\textbf{05} & \textbf{Assertionless Test} & \textbf{ALT} &  A test that does not contain at least one valid assertion.     & \cite{reichhart2007rule} \\ \hline
%%%%%%%%%%%%%%%
\rowcolor{gray!30}
\textbf{06} & \textbf{Brittle Assertion} & \textbf{BA} &  A test method that has assertions that check data input.     & \cite{huo2014improving} \\ \hline
%%%%%%%%%%%%%%%
\textbf{07} & \textbf{Comments Only Test} & \textbf{COT} &   A test that has been put into comments.  & \cite{reichhart2007rule} \\ \hline
%%%%%%%%%%%%%%%
\rowcolor{gray!30}
\textbf{08} & \textbf{Conditional Test Logic} & \textbf{CTL} & A test method that contains a conditional statement as a prerequisite to executing the test statement.   & \cite{peruma2020FSE} \\ \hline
%%%%%%%%%%%%%%%
\textbf{09} & \textbf{Constructor Initialization} & \textbf{CI} &   A test class that contains a constructor.  & \cite{peruma2020FSE} \\ \hline
%%%%%%%%%%%%%%%
\rowcolor{gray!30}
\textbf{10} & \textbf{Control Logic} & \textbf{ConL} & A test method that controls test data flow by methods such as debug or halt. & \cite{reichhart2007rule} \\ \hline
%%%%%%%%%%%%%%%
\textbf{11} & \textbf{Dead Field} & \textbf{DF} &  When a class has a field that is never used by any test methods.   & \cite{greiler2013automated} \\ \hline
%%%%%%%%%%%%%%%
\rowcolor{gray!30}
\textbf{12} & \textbf{Default Test} & \textbf{DT} &   Default or an example test suite created by Android Studio.    & \cite{peruma2020FSE} \\ \hline
%%%%%%%%%%%%%%%
\textbf{13} & \textbf{Dependent Test} & \textbf{DepT} &  A test that only executes on the successful execution of other tests.      & \cite{virginio2019influence} \\ \hline
%%%%%%%%%%%%%%%
\rowcolor{gray!30}
\textbf{14} & \textbf{Duplicate Assert} & \textbf{DA} &   Occurs when a test method has the exact assertion multiple times within the same test method.    & \cite{peruma2020FSE} \\ \hline
%%%%%%%%%%%%%%%
\textbf{15} & \textbf{Duplicated Code} & \textbf{DC} &  A test method that has redundancy in the code. & \cite{breugelmans2008testq} \\ \hline
%%%%%%%%%%%%%%%
\rowcolor{gray!30}
\textbf{16} & \textbf{Eager Test} & \textbf{ET} &  A test method that calls several methods of the object to be tested.   & \cite{bavota2012empirical} \\ \hline
%%%%%%%%%%%%%%%
\textbf{17} & \textbf{Early Returning Test} & \textbf{ERT} & A test method that returns a value too early which may drop assertions. & \cite{reichhart2007rule} \\ \hline
%%%%%%%%%%%%%%%
\rowcolor{gray!30}
\textbf{18} & \textbf{Empty Method Category} & \textbf{EMC} &  A test method with an empty method category.  & \cite{reichhart2007rule} \\ \hline
%%%%%%%%%%%%%%%
\textbf{19} & \textbf{Empty Shared-Fixture} & \textbf{ESF} &   A test that defines a fixture with an empty body.  & \cite{reichhart2007rule} \\ \hline
%%%%%%%%%%%%%%%
\rowcolor{gray!30}
\textbf{20} & \textbf{Empty Test} & \textbf{EmT} &  A test method that is empty or does not have executable statements.  & \cite{peruma2020FSE} \\ \hline
%%%%%%%%%%%%%%%
\textbf{21} & \textbf{Empty Test-Method Category} & \textbf{ETMC} &  A test method with an empty test method category.   & \cite{reichhart2007rule} \\ \hline
%%%%%%%%%%%%%%%
\rowcolor{gray!30}
\textbf{22} & \textbf{Exception Handling} & \textbf{EH} &  Occurs when custom exception handling is utilized instead of using JUnit's exception handling feature.  & \cite{peruma2020FSE} \\ \hline
%%%%%%%%%%%%%%%
\textbf{23} & \textbf{For Testers Only} & \textbf{FTO} &  A production class that contains methods that are only used for test methods.  & \cite{bavota2012empirical} \\ \hline
%%%%%%%%%%%%%%%
\rowcolor{gray!30}
\textbf{24} & \textbf{General Fixture} & \textbf{GF} &  This smell emerges when \texttt{setUp()} fixture creates many objects, and test methods only use a subset.   & \cite{bavota2012empirical} \\ \hline
%%%%%%%%%%%%%%%
\textbf{25} & \textbf{Guarded Test} & \textbf{GT} &  A test that has conditional branches like \texttt{ifTrue:aCode} or \texttt{ifFalse:aCode}.  & \cite{reichhart2007rule} \\ \hline
%%%%%%%%%%%%%%%
\rowcolor{gray!30}
\textbf{26} & \textbf{Ignored Test} & \textbf{IgT} &   A test method that uses an ignore annotation which prevents the test method from running.      & \cite{peruma2020FSE} \\ \hline
%%%%%%%%%%%%%%%
\textbf{27} & \textbf{Indented Test} & \textbf{InT} &  A test method that contains a large number of decision points, loops, and conditional statements.    & \cite{breugelmans2008testq} \\ \hline
%%%%%%%%%%%%%%%
\rowcolor{gray!30}
\textbf{28} & \textbf{Indirect Testing} & \textbf{IT} &   A test that interacts with a corresponding class by using another class.   & \cite{bavota2012empirical} \\ \hline
%%%%%%%%%%%%%%%
\textbf{29} & \textbf{Lack of Cohesion of Methods} & \textbf{LCM} &  When test methods are grouped in one test class, but they are not cohesive. & \cite{greiler2013automated} \\ \hline
%%%%%%%%%%%%%%%
\rowcolor{gray!30}
\textbf{30} & \textbf{Lazy Test} & \textbf{LT} &  Occurs when multiple test methods check the same method of production object.  & \cite{bavota2012empirical} \\ \hline
%%%%%%%%%%%%%%%
\textbf{31} & \textbf{Likely Ineffective Object-Comparison} & \textbf{LIOC} & A test that performs a comparison between objects will never fail.    & \cite{reichhart2007rule} \\ \hline
%%%%%%%%%%%%%%%
\rowcolor{gray!30}
\textbf{32} & \textbf{Long Test} & \textbf{LoT} & A test with many statements.       & \cite{reichhart2007rule} \\ \hline
%%%%%%%%%%%%%%%
\textbf{33} & \textbf{Magic Number Test} & \textbf{MNT} &   A test method that contains undocumented numerical values.    & \cite{peruma2020FSE} \\ \hline
%%%%%%%%%%%%%%%
\rowcolor{gray!30}
\textbf{34} & \textbf{Max Instance Variables} & \textbf{MIV} &  A test method that has a large fixture.  & \cite{reichhart2007rule} \\ \hline
%%%%%%%%%%%%%%%
\textbf{35} & \textbf{Mixed Selectors} & \textbf{MS} & Violates test conventions by mixing up testing and non-testing methods.  & \cite{reichhart2007rule}   \\ \hline
%%%%%%%%%%%%%%%
\rowcolor{gray!30}
\textbf{36} & \textbf{Mystery Guest} & \textbf{MG} &  A test that uses external resources, such as a database, that contains test data. & \cite{bavota2012empirical} \\ \hline  
%%%%%%%%%%%%%%%
\textbf{37} & \textbf{Obscure In-line Setup} & \textbf{OISS} &  A test that has too much setup functionality in the test method.  & \cite{greiler2013automated}  \\ \hline
%%%%%%%%%%%%%%%
\rowcolor{gray!30}
\textbf{38} & \textbf{Overcommented Test} & \textbf{OCT} &  A test with numerous comments.      & \cite{reichhart2007rule} \\ \hline
%%%%%%%%%%%%%%%
\textbf{39} & \textbf{Overreferencing} & \textbf{OF} &  A test that causes duplication by creating unnecessary dependencies.       & \cite{reichhart2007rule} \\ \hline
%%%%%%%%%%%%%%%
\rowcolor{gray!30}
\textbf{40} & \textbf{Proper Organization} & \textbf{PO} &  Bad organization of methods     & \cite{reichhart2007rule} \\ \hline
%%%%%%%%%%%%%%%
\textbf{41} & \textbf{Redundant Assertion} & \textbf{RA} &  A test method that has an assertion statement that is permanently true or false.  & \cite{peruma2020FSE} \\ \hline
%%%%%%%%%%%%%%%
\rowcolor{gray!30}
\textbf{42} & \textbf{Redundant Print} & \textbf{RP} &   A test method that has print statement.    & \cite{peruma2020FSE} \\ \hline
%%%%%%%%%%%%%%%
\textbf{43} & \textbf{Resource Optimism} & \textbf{RO} &  A test that make an assumption about the existence of external resources.    & \cite{bavota2012empirical} \\ \hline
%%%%%%%%%%%%%%%
\rowcolor{gray!30}
\textbf{44} & \textbf{Returning Assertion} & \textbf{RA} &  A test method that has an assertion and returns a value. & \cite{reichhart2007rule} \\ \hline
%%%%%%%%%%%%%%%
\textbf{45} & \textbf{Rotten Green Tests} & \textbf{RT} &  Occurs when intended assertions in a test are never executed.  & \cite{Delplanque2019ICSE} \\ \hline
%%%%%%%%%%%%%%%
\rowcolor{gray!30}
\textbf{46} & \textbf{Sensitive Equality} & \textbf{SE} &  Occurs when an assertion has an equality check by using the \texttt{toString} method.  & \cite{bavota2012empirical} \\ \hline
%%%%%%%%%%%%%%%
\textbf{47} & \textbf{Sleepy Test} & \textbf{ST} &  Occurs when a test method has an explicit wait.  & \cite{peruma2020FSE} \\ \hline
%%%%%%%%%%%%%%%
\rowcolor{gray!30}
\textbf{48} & \textbf{Teardown Only Test} & \textbf{TOT} &  Exists when a test-suite is only specifying teardown.   & \cite{reichhart2007rule} \\ \hline
%%%%%%%%%%%%%%%
\textbf{49} & \textbf{Test Code Duplication} & \textbf{TCD} &  Occurs when code clones contained inside the test.   & \cite{bavota2012empirical} \\ \hline
%%%%%%%%%%%%%%%
\rowcolor{gray!30}
\textbf{50} & \textbf{Test Maverick} & \textbf{TM} &  Exists when a test class has a test method with an implicit setup; however, the test methods are independent.   & \cite{greiler2013automated} \\ \hline
%%%%%%%%%%%%%%%
\textbf{51} & \textbf{Test Pollution}  & \textbf{TP} & Test that introduces dependencies such as reading/writing a shared resource. & \cite{gyori2015reliable} \\ \hline
%%%%%%%%%%%%%%%
\rowcolor{gray!30}
\textbf{52} &  \textbf{Test Redundancy}  & \textbf{TR} &  Occurs when the removal of a test does not impact the effectiveness of the test suite. & \cite{koochakzadeh2010tecrevis} \\ \hline
%%%%%%%%%%%%%%%
\textbf{53} & \textbf{Test Run War} & \textbf{TRW} &   A test that fails when more than one programmer runs them.      & \cite{bavota2012empirical} \\ \hline
%%%%%%%%%%%%%%%
\rowcolor{gray!30}
\textbf{54} & \textbf{Test-Class Name} & \textbf{TCN} &  A test that has a class with a meaningless name.   & \cite{reichhart2007rule} \\ \hline
%%%%%%%%%%%%%%%
\textbf{55} & \textbf{Test-Method Category Name} & \textbf{TMC} &  A test method has a meaningless name.   & \cite{reichhart2007rule} \\ \hline
%%%%%%%%%%%%%%%
\rowcolor{gray!30}
\textbf{56} & \textbf{Transcripting Test} & \textbf{TT} &  A test that is printing and logging to the console. & \cite{reichhart2007rule} \\ \hline
%%%%%%%%%%%%%%%
\textbf{57} & \textbf{TTCN-3 Smells}  & \textbf{TTCN} &  Collection of smells specific to TTCN-3 test suites. & \cite{baker2006trex} \\ \hline
%%%%%%%%%%%%%%%
\rowcolor{gray!30}
\textbf{58} & \textbf{Unclassified Method Category} & \textbf{UMC} &  A test method that is not organized by a method category.  & \cite{reichhart2007rule} \\ \hline
%%%%%%%%%%%%%%%
\textbf{59} & \textbf{Under-the-carpet Assertion} & \textbf{UCA} &  A test that has assertions in the comments.   & \cite{reichhart2007rule} \\ \hline
%%%%%%%%%%%%%%%
\rowcolor{gray!30}
\textbf{60} & \textbf{Under-the-carpet failing Assertion} & \textbf{UCFA} &  A test method that has  failing assertions in the comments.   & \cite{reichhart2007rule} \\ \hline
%%%%%%%%%%%%%%%
\textbf{61} & \textbf{Unknown Test} & \textbf{UT} &  A test method without an assertion statement and non-descriptive name.     & \cite{peruma2020FSE} \\ \hline
%%%%%%%%%%%%%%%
\rowcolor{gray!30}
\textbf{62} & \textbf{Unused Inputs} & \textbf{UI} &   Inputs that are controlled by the test.    & \cite{huo2014improving} \\ \hline
%%%%%%%%%%%%%%%
\textbf{63} & \textbf{Unused Shared-Fixture Variables} & \textbf{USFV} &  Occurs when a piece of the fixture is never used.  & \cite{reichhart2007rule}  \\ \hline
%%%%%%%%%%%%%%%
\rowcolor{gray!30}
\textbf{64} & \textbf{Unusual Test Order} & \textbf{UTO} &   A test that is calling other tests explicitly.  & \cite{reichhart2007rule} \\ \hline
%%%%%%%%%%%%%%%
\textbf{65} & \textbf{Vague Header Setup} & \textbf{VHS} &  A field that is initialized in the class header but not explicitly defined in code.    & \cite{greiler2013automated} \\ \hline
%%%%%%%%%%%%%%%
\rowcolor{gray!30}
\textbf{66} & \textbf{Verbose Test} & \textbf{VT} &   Test code that is complex and not simple or clean. & \cite{breugelmans2008testq} \\ \hline
%%%%%%%%%%%%%%%

\end{tabular}

% \begin{tablenotes}
%             \item $\star\,$ These test smells are related to Samlltalk \cite{reichhart2007rule}.
%             \item \textdaggerdbl$\,$ This test smell occurs in  Smalltalk \& Java with having same purpose and definition.
%         \end{tablenotes}
        
% \end{threeparttable}

\end{adjustbox}
% }

\label{tab:Test_Definitions}
\end{table*}

\begin{table*}
\centering
\caption{Distribution of test smells detected by the test smell detection tools.}
\vspace{-0.3cm}
\label{tab:TestSmellTypes_v3}
\resizebox{\textwidth}{!}{%
\begin{tabular}{|lcccccccccccccccccccccccccccccccccr|}
\rowcolor{gray!60}
\multicolumn{1}{|c}{\textbf{Tool \textbackslash{} Smell Type}} &
  \textbf{AL} &
  \textbf{AR} &
  \textbf{CI} &
  \textbf{CTL} &
  \textbf{DA} &
  \textbf{DC} &
  \textbf{DepT} &
  \textbf{DF} &
  \textbf{DT} &
  \textbf{EH} &
  \textbf{EmT} &
  \textbf{ET} &
  \textbf{FTO} &
  \textbf{GF} &
  \textbf{IgT} &
  \textbf{InT} &
  \textbf{IT} &
  \textbf{LCM} &
  \textbf{LT} &
  \textbf{MG} &
  \textbf{MNT} &
  \textbf{OISS} &
  \textbf{RA} &
  \textbf{RO} &
  \textbf{RP} &
  \textbf{RT} &
  \textbf{SE} &
  \textbf{ST} &
  \textbf{TM} &
  \textbf{TR} &
  \textbf{TRW} &
  \textbf{UT} &
  \textbf{VHS} &
  \textbf{VT} 
  %\multicolumn{1}{c|}{\textit{\textbf{Total}}} 
  \\ \hline
\multicolumn{1}{|l|}{\begin{tabular}[c]{@{}l@{}}\textbf{DARTS}       {\cite{lambiase2020just}}\end{tabular}} &
  \multicolumn{1}{c|}{} &
  \multicolumn{1}{c|}{} &
  \multicolumn{1}{c|}{} &
  \multicolumn{1}{c|}{} &
  \multicolumn{1}{c|}{} &
  \multicolumn{1}{c|}{} &
  \multicolumn{1}{c|}{} &
  \multicolumn{1}{c|}{} &
  \multicolumn{1}{c|}{} &
  \multicolumn{1}{c|}{} &
  \multicolumn{1}{c|}{} &
  \multicolumn{1}{c|}{$\surd$} &
  \multicolumn{1}{c|}{} &
  \multicolumn{1}{c|}{$\surd$} &
  \multicolumn{1}{c|}{} &
  \multicolumn{1}{c|}{} &
  \multicolumn{1}{c|}{} &
  \multicolumn{1}{c|}{$\surd$} &
  \multicolumn{1}{c|}{} &
  \multicolumn{1}{c|}{} &
  \multicolumn{1}{c|}{} &
  \multicolumn{1}{c|}{} &
  \multicolumn{1}{c|}{} &
  \multicolumn{1}{c|}{} &
  \multicolumn{1}{c|}{} &
  \multicolumn{1}{c|}{} &
  \multicolumn{1}{c|}{} &
  \multicolumn{1}{c|}{} &
  \multicolumn{1}{c|}{} &
  \multicolumn{1}{c|}{} &
  \multicolumn{1}{c|}{} &
  \multicolumn{1}{c|}{} &
  \multicolumn{1}{c|}{} &
  \multicolumn{1}{c|}{} 
  %\textbf{3} 
  \\
\rowcolor{gray!30}
\multicolumn{1}{|l|}{\begin{tabular}[c]{@{}l@{}}\textbf{DrTest} \cite{Delplanque2019ICSE} \end{tabular}} &
  \multicolumn{1}{c|}{} &
  \multicolumn{1}{c|}{} &
  \multicolumn{1}{c|}{} &
  \multicolumn{1}{c|}{} &
  \multicolumn{1}{c|}{} &
  \multicolumn{1}{c|}{} &
  \multicolumn{1}{c|}{} &
  \multicolumn{1}{c|}{} &
  \multicolumn{1}{c|}{} &
  \multicolumn{1}{c|}{} &
  \multicolumn{1}{c|}{} &
  \multicolumn{1}{c|}{} &
  \multicolumn{1}{c|}{} &
  \multicolumn{1}{c|}{} &
  \multicolumn{1}{c|}{} &
  \multicolumn{1}{c|}{} &
  \multicolumn{1}{c|}{} &
  \multicolumn{1}{c|}{} &
  \multicolumn{1}{c|}{} &
  \multicolumn{1}{c|}{} &
  \multicolumn{1}{c|}{} &
  \multicolumn{1}{c|}{} &
  \multicolumn{1}{c|}{} &
  \multicolumn{1}{c|}{} &
  \multicolumn{1}{c|}{} &
  \multicolumn{1}{c|}{$\surd$} &
  \multicolumn{1}{c|}{} &
  \multicolumn{1}{c|}{} &
  \multicolumn{1}{c|}{} &
  \multicolumn{1}{c|}{} &
  \multicolumn{1}{c|}{} &
  \multicolumn{1}{c|}{} &
  \multicolumn{1}{c|}{} &
  \multicolumn{1}{c|}{} 
  %\textbf{1} 
  \\
\multicolumn{1}{|l|}{\begin{tabular}[c]{@{}l@{}}\textbf{DTDetector}       {\cite{zhang2014empirically}}\end{tabular}} &
  \multicolumn{1}{c|}{\textbf{}} &
  \multicolumn{1}{c|}{\textbf{}} &
  \multicolumn{1}{c|}{\textbf{}} &
  \multicolumn{1}{c|}{\textbf{}} &
  \multicolumn{1}{c|}{\textbf{}} &
  \multicolumn{1}{c|}{\textbf{}} &
  \multicolumn{1}{c|}{\textbf{$\surd$}} &
  \multicolumn{1}{c|}{\textbf{}} &
  \multicolumn{1}{c|}{\textbf{}} &
  \multicolumn{1}{c|}{\textbf{}} &
  \multicolumn{1}{c|}{\textbf{}} &
  \multicolumn{1}{c|}{\textbf{}} &
  \multicolumn{1}{c|}{\textbf{}} &
  \multicolumn{1}{c|}{\textbf{}} &
  \multicolumn{1}{c|}{\textbf{}} &
  \multicolumn{1}{c|}{\textbf{}} &
  \multicolumn{1}{c|}{\textbf{}} &
  \multicolumn{1}{c|}{\textbf{}} &
  \multicolumn{1}{c|}{\textbf{}} &
  \multicolumn{1}{c|}{\textbf{}} &
  \multicolumn{1}{c|}{\textbf{}} &
  \multicolumn{1}{c|}{\textbf{}} &
  \multicolumn{1}{c|}{\textbf{}} &
  \multicolumn{1}{c|}{\textbf{}} &
  \multicolumn{1}{c|}{\textbf{}} &
  \multicolumn{1}{c|}{\textbf{}} &
  \multicolumn{1}{c|}{\textbf{}} &
  \multicolumn{1}{c|}{\textbf{}} &
  \multicolumn{1}{c|}{\textbf{}} &
  \multicolumn{1}{c|}{\textbf{}} &
  \multicolumn{1}{c|}{\textbf{}} &
  \multicolumn{1}{c|}{\textbf{}} &
  \multicolumn{1}{c|}{\textbf{}} &
  \multicolumn{1}{c|}{\textbf{}} 
  %\textbf{1} 
  \\
\rowcolor{gray!30}
\multicolumn{1}{|l|}{\begin{tabular}[c]{@{}l@{}}\textbf{ElectricTest}       {\cite{bell2015efficient}}\end{tabular}} &
  \multicolumn{1}{c|}{\textbf{}} &
  \multicolumn{1}{c|}{\textbf{}} &
  \multicolumn{1}{c|}{\textbf{}} &
  \multicolumn{1}{c|}{\textbf{}} &
  \multicolumn{1}{c|}{\textbf{}} &
  \multicolumn{1}{c|}{\textbf{}} &
  \multicolumn{1}{c|}{\textbf{$\surd$}} &
  \multicolumn{1}{c|}{\textbf{}} &
  \multicolumn{1}{c|}{\textbf{}} &
  \multicolumn{1}{c|}{\textbf{}} &
  \multicolumn{1}{c|}{\textbf{}} &
  \multicolumn{1}{c|}{\textbf{}} &
  \multicolumn{1}{c|}{\textbf{}} &
  \multicolumn{1}{c|}{\textbf{}} &
  \multicolumn{1}{c|}{\textbf{}} &
  \multicolumn{1}{c|}{\textbf{}} &
  \multicolumn{1}{c|}{\textbf{}} &
  \multicolumn{1}{c|}{\textbf{}} &
  \multicolumn{1}{c|}{\textbf{}} &
  \multicolumn{1}{c|}{\textbf{}} &
  \multicolumn{1}{c|}{\textbf{}} &
  \multicolumn{1}{c|}{\textbf{}} &
  \multicolumn{1}{c|}{\textbf{}} &
  \multicolumn{1}{c|}{\textbf{}} &
  \multicolumn{1}{c|}{\textbf{}} &
  \multicolumn{1}{c|}{\textbf{}} &
  \multicolumn{1}{c|}{\textbf{}} &
  \multicolumn{1}{c|}{\textbf{}} &
  \multicolumn{1}{c|}{\textbf{}} &
  \multicolumn{1}{c|}{\textbf{}} &
  \multicolumn{1}{c|}{\textbf{}} &
  \multicolumn{1}{c|}{\textbf{}} &
  \multicolumn{1}{c|}{\textbf{}} &
  \multicolumn{1}{c|}{\textbf{}} 
  %\textbf{1} 
  \\
\multicolumn{1}{|l|}{\begin{tabular}[c]{@{}l@{}}\textbf{JNose Test}       {\cite{virginio2019influence}}\end{tabular}} &
  \multicolumn{1}{c|}{\textbf{$\surd$}} &
  \multicolumn{1}{c|}{\textbf{$\surd$}} &
  \multicolumn{1}{c|}{\textbf{$\surd$}} &
  \multicolumn{1}{c|}{\textbf{$\surd$}} &
  \multicolumn{1}{c|}{\textbf{$\surd$}} &
  \multicolumn{1}{c|}{\textbf{}} &
  \multicolumn{1}{c|}{\textbf{$\surd$}} &
  \multicolumn{1}{c|}{\textbf{}} &
  \multicolumn{1}{c|}{\textbf{$\surd$}} &
  \multicolumn{1}{c|}{\textbf{$\surd$}} &
  \multicolumn{1}{c|}{\textbf{$\surd$}} &
  \multicolumn{1}{c|}{\textbf{$\surd$}} &
  \multicolumn{1}{c|}{\textbf{}} &
  \multicolumn{1}{c|}{\textbf{$\surd$}} &
  \multicolumn{1}{c|}{\textbf{}} &
  \multicolumn{1}{c|}{\textbf{}} &
  \multicolumn{1}{c|}{\textbf{}} &
  \multicolumn{1}{c|}{\textbf{}} &
  \multicolumn{1}{c|}{\textbf{$\surd$}} &
  \multicolumn{1}{c|}{\textbf{$\surd$}} &
  \multicolumn{1}{c|}{\textbf{$\surd$}} &
  \multicolumn{1}{c|}{\textbf{}} &
  \multicolumn{1}{c|}{\textbf{$\surd$}} &
  \multicolumn{1}{c|}{\textbf{$\surd$}} &
  \multicolumn{1}{c|}{\textbf{$\surd$}} &
  \multicolumn{1}{c|}{\textbf{}} &
  \multicolumn{1}{c|}{\textbf{$\surd$}} &
  \multicolumn{1}{c|}{\textbf{$\surd$}} &
  \multicolumn{1}{c|}{\textbf{}} &
  \multicolumn{1}{c|}{\textbf{}} &
  \multicolumn{1}{c|}{\textbf{}} &
  \multicolumn{1}{c|}{\textbf{$\surd$}} &
  \multicolumn{1}{c|}{\textbf{}} &
  \multicolumn{1}{c|}{\textbf{$\surd$}} 
  %\textbf{21} 
  \\
\rowcolor{gray!30}
\multicolumn{1}{|l|}{\begin{tabular}[c]{@{}l@{}} \textbf{\textsc{PraDeT}}       {\cite{gambi2018practical} }\end{tabular}} &
  \multicolumn{1}{c|}{\textbf{}} &
  \multicolumn{1}{c|}{\textbf{}} &
  \multicolumn{1}{c|}{\textbf{}} &
  \multicolumn{1}{c|}{\textbf{}} &
  \multicolumn{1}{c|}{\textbf{}} &
  \multicolumn{1}{c|}{\textbf{}} &
  \multicolumn{1}{c|}{\textbf{$\surd$}} &
  \multicolumn{1}{c|}{\textbf{}} &
  \multicolumn{1}{c|}{\textbf{}} &
  \multicolumn{1}{c|}{\textbf{}} &
  \multicolumn{1}{c|}{\textbf{}} &
  \multicolumn{1}{c|}{\textbf{}} &
  \multicolumn{1}{c|}{\textbf{}} &
  \multicolumn{1}{c|}{\textbf{}} &
  \multicolumn{1}{c|}{\textbf{}} &
  \multicolumn{1}{c|}{\textbf{}} &
  \multicolumn{1}{c|}{\textbf{}} &
  \multicolumn{1}{c|}{\textbf{}} &
  \multicolumn{1}{c|}{\textbf{}} &
  \multicolumn{1}{c|}{\textbf{}} &
  \multicolumn{1}{c|}{\textbf{}} &
  \multicolumn{1}{c|}{\textbf{}} &
  \multicolumn{1}{c|}{\textbf{}} &
  \multicolumn{1}{c|}{\textbf{}} &
  \multicolumn{1}{c|}{\textbf{}} &
  \multicolumn{1}{c|}{\textbf{}} &
  \multicolumn{1}{c|}{\textbf{}} &
  \multicolumn{1}{c|}{\textbf{}} &
  \multicolumn{1}{c|}{\textbf{}} &
  \multicolumn{1}{c|}{\textbf{}} &
  \multicolumn{1}{c|}{\textbf{}} &
  \multicolumn{1}{c|}{\textbf{}} &
  \multicolumn{1}{c|}{\textbf{}} &
  \multicolumn{1}{c|}{\textbf{}} 
  %\textbf{1} 
  \\
\multicolumn{1}{|l|}{\begin{tabular}[c]{@{}l@{}}\textbf{RAIDE}       {\cite{Santana2020SBES}}\end{tabular}} &
  \multicolumn{1}{c|}{\textbf{}} &
  \multicolumn{1}{c|}{\textbf{$\surd$}} &
  \multicolumn{1}{c|}{\textbf{}} &
  \multicolumn{1}{c|}{\textbf{}} &
  \multicolumn{1}{c|}{\textbf{$\surd$}} &
  \multicolumn{1}{c|}{\textbf{}} &
  \multicolumn{1}{c|}{\textbf{}} &
  \multicolumn{1}{c|}{\textbf{}} &
  \multicolumn{1}{c|}{\textbf{}} &
  \multicolumn{1}{c|}{\textbf{}} &
  \multicolumn{1}{c|}{\textbf{}} &
  \multicolumn{1}{c|}{\textbf{}} &
  \multicolumn{1}{c|}{\textbf{}} &
  \multicolumn{1}{c|}{\textbf{}} &
  \multicolumn{1}{c|}{\textbf{}} &
  \multicolumn{1}{c|}{\textbf{}} &
  \multicolumn{1}{c|}{\textbf{}} &
  \multicolumn{1}{c|}{\textbf{}} &
  \multicolumn{1}{c|}{\textbf{}} &
  \multicolumn{1}{c|}{\textbf{}} &
  \multicolumn{1}{c|}{\textbf{}} &
  \multicolumn{1}{c|}{\textbf{}} &
  \multicolumn{1}{c|}{\textbf{}} &
  \multicolumn{1}{c|}{\textbf{}} &
  \multicolumn{1}{c|}{\textbf{}} &
  \multicolumn{1}{c|}{\textbf{}} &
  \multicolumn{1}{c|}{\textbf{}} &
  \multicolumn{1}{c|}{\textbf{}} &
  \multicolumn{1}{c|}{\textbf{}} &
  \multicolumn{1}{c|}{\textbf{}} &
  \multicolumn{1}{c|}{\textbf{}} &
  \multicolumn{1}{c|}{\textbf{}} &
  \multicolumn{1}{c|}{\textbf{}} &
  \multicolumn{1}{c|}{\textbf{}} 
  %\textbf{2} 
  \\
\rowcolor{gray!30}
\multicolumn{1}{|l|}{\begin{tabular}[c]{@{}l@{}}\textbf{RTj} \cite{Martinez2020ICSE} \end{tabular}} &
  \multicolumn{1}{c|}{} &
  \multicolumn{1}{c|}{} &
  \multicolumn{1}{c|}{} &
  \multicolumn{1}{c|}{} &
  \multicolumn{1}{c|}{} &
  \multicolumn{1}{c|}{} &
  \multicolumn{1}{c|}{} &
  \multicolumn{1}{c|}{} &
  \multicolumn{1}{c|}{} &
  \multicolumn{1}{c|}{} &
  \multicolumn{1}{c|}{} &
  \multicolumn{1}{c|}{} &
  \multicolumn{1}{c|}{} &
  \multicolumn{1}{c|}{} &
  \multicolumn{1}{c|}{} &
  \multicolumn{1}{c|}{} &
  \multicolumn{1}{c|}{} &
  \multicolumn{1}{c|}{} &
  \multicolumn{1}{c|}{} &
  \multicolumn{1}{c|}{} &
  \multicolumn{1}{c|}{} &
  \multicolumn{1}{c|}{} &
  \multicolumn{1}{c|}{} &
  \multicolumn{1}{c|}{} &
  \multicolumn{1}{c|}{} &
  \multicolumn{1}{c|}{$\surd$} &
  \multicolumn{1}{c|}{} &
  \multicolumn{1}{c|}{} &
  \multicolumn{1}{c|}{} &
  \multicolumn{1}{c|}{} &
  \multicolumn{1}{c|}{} &
  \multicolumn{1}{c|}{} &
  \multicolumn{1}{c|}{} &
  \multicolumn{1}{c|}{} 
  %\textbf{1} 
  \\
\multicolumn{1}{|l|}{\begin{tabular}[c]{@{}l@{}}\textbf{\textsc{SoCRATES}}      {\cite{de2019socrates}}\end{tabular}} &
  \multicolumn{1}{c|}{\textbf{}} &
  \multicolumn{1}{c|}{\textbf{$\surd$}} &
  \multicolumn{1}{c|}{\textbf{}} &
  \multicolumn{1}{c|}{\textbf{}} &
  \multicolumn{1}{c|}{\textbf{}} &
  \multicolumn{1}{c|}{\textbf{}} &
  \multicolumn{1}{c|}{\textbf{}} &
  \multicolumn{1}{c|}{\textbf{}} &
  \multicolumn{1}{c|}{\textbf{}} &
  \multicolumn{1}{c|}{\textbf{}} &
  \multicolumn{1}{c|}{\textbf{}} &
  \multicolumn{1}{c|}{\textbf{$\surd$}} &
  \multicolumn{1}{c|}{\textbf{}} &
  \multicolumn{1}{c|}{\textbf{$\surd$}} &
  \multicolumn{1}{c|}{\textbf{}} &
  \multicolumn{1}{c|}{\textbf{}} &
  \multicolumn{1}{c|}{\textbf{}} &
  \multicolumn{1}{c|}{\textbf{}} &
  \multicolumn{1}{c|}{\textbf{$\surd$}} &
  \multicolumn{1}{c|}{\textbf{$\surd$}} &
  \multicolumn{1}{c|}{\textbf{}} &
  \multicolumn{1}{c|}{\textbf{}} &
  \multicolumn{1}{c|}{\textbf{}} &
  \multicolumn{1}{c|}{\textbf{}} &
  \multicolumn{1}{c|}{\textbf{}} &
  \multicolumn{1}{c|}{\textbf{}} &
  \multicolumn{1}{c|}{\textbf{$\surd$}} &
  \multicolumn{1}{c|}{\textbf{}} &
  \multicolumn{1}{c|}{\textbf{}} &
  \multicolumn{1}{c|}{\textbf{}} &
  \multicolumn{1}{c|}{\textbf{}} &
  \multicolumn{1}{c|}{\textbf{}} &
  \multicolumn{1}{c|}{\textbf{}} &
  \multicolumn{1}{c|}{\textbf{}} 
  %\textbf{6} 
  \\
\rowcolor{gray!30}
\multicolumn{1}{|l|}{\begin{tabular}[c]{@{}l@{}}\textbf{\textsc{Taste}}     {\cite{palomba2018automatic}}\end{tabular}} &
  \multicolumn{1}{c|}{\textbf{}} &
  \multicolumn{1}{c|}{\textbf{}} &
  \multicolumn{1}{c|}{\textbf{}} &
  \multicolumn{1}{c|}{\textbf{}} &
  \multicolumn{1}{c|}{\textbf{}} &
  \multicolumn{1}{c|}{\textbf{}} &
  \multicolumn{1}{c|}{\textbf{}} &
  \multicolumn{1}{c|}{\textbf{}} &
  \multicolumn{1}{c|}{\textbf{}} &
  \multicolumn{1}{c|}{\textbf{}} &
  \multicolumn{1}{c|}{\textbf{}} &
  \multicolumn{1}{c|}{\textbf{$\surd$}} &
  \multicolumn{1}{c|}{\textbf{}} &
  \multicolumn{1}{c|}{\textbf{$\surd$}} &
  \multicolumn{1}{c|}{\textbf{}} &
  \multicolumn{1}{c|}{\textbf{}} &
  \multicolumn{1}{c|}{\textbf{}} &
  \multicolumn{1}{c|}{\textbf{$\surd$}} &
  \multicolumn{1}{c|}{\textbf{}} &
  \multicolumn{1}{c|}{\textbf{}} &
  \multicolumn{1}{c|}{\textbf{}} &
  \multicolumn{1}{c|}{\textbf{}} &
  \multicolumn{1}{c|}{\textbf{}} &
  \multicolumn{1}{c|}{\textbf{}} &
  \multicolumn{1}{c|}{\textbf{}} &
  \multicolumn{1}{c|}{\textbf{}} &
  \multicolumn{1}{c|}{\textbf{}} &
  \multicolumn{1}{c|}{\textbf{}} &
  \multicolumn{1}{c|}{\textbf{}} &
  \multicolumn{1}{c|}{\textbf{}} &
  \multicolumn{1}{c|}{\textbf{}} &
  \multicolumn{1}{c|}{\textbf{}} &
  \multicolumn{1}{c|}{\textbf{}} &
  \multicolumn{1}{c|}{\textbf{}} 
  %\textbf{3}
  \\
\multicolumn{1}{|l|}{\begin{tabular}[c]{@{}l@{}}\textbf{TeCReVis}       {\cite{koochakzadeh2010tecrevis} }\end{tabular}} &
  \multicolumn{1}{c|}{\textbf{}} &
  \multicolumn{1}{c|}{\textbf{}} &
  \multicolumn{1}{c|}{\textbf{}} &
  \multicolumn{1}{c|}{\textbf{}} &
  \multicolumn{1}{c|}{\textbf{}} &
  \multicolumn{1}{c|}{\textbf{}} &
  \multicolumn{1}{c|}{\textbf{}} &
  \multicolumn{1}{c|}{\textbf{}} &
  \multicolumn{1}{c|}{\textbf{}} &
  \multicolumn{1}{c|}{\textbf{}} &
  \multicolumn{1}{c|}{\textbf{}} &
  \multicolumn{1}{c|}{\textbf{}} &
  \multicolumn{1}{c|}{\textbf{}} &
  \multicolumn{1}{c|}{\textbf{}} &
  \multicolumn{1}{c|}{\textbf{}} &
  \multicolumn{1}{c|}{\textbf{}} &
  \multicolumn{1}{c|}{\textbf{}} &
  \multicolumn{1}{c|}{\textbf{}} &
  \multicolumn{1}{c|}{\textbf{}} &
  \multicolumn{1}{c|}{\textbf{}} &
  \multicolumn{1}{c|}{\textbf{}} &
  \multicolumn{1}{c|}{\textbf{}} &
  \multicolumn{1}{c|}{\textbf{}} &
  \multicolumn{1}{c|}{\textbf{}} &
  \multicolumn{1}{c|}{\textbf{}} &
  \multicolumn{1}{c|}{\textbf{}} &
  \multicolumn{1}{c|}{\textbf{}} &
  \multicolumn{1}{c|}{\textbf{}} &
  \multicolumn{1}{c|}{\textbf{}} &
  \multicolumn{1}{c|}{\textbf{$\surd$}} &
  \multicolumn{1}{c|}{\textbf{}} &
  \multicolumn{1}{c|}{\textbf{}} &
  \multicolumn{1}{c|}{\textbf{}} &
  \multicolumn{1}{c|}{\textbf{}} 
  %\textbf{1} 
  \\
\rowcolor{gray!30}
\multicolumn{1}{|l|}{\begin{tabular}[c]{@{}l@{}}\textbf{TEDD}       {\cite{biagiola2019web}}\end{tabular}} &
  \multicolumn{1}{c|}{\textbf{}} &
  \multicolumn{1}{c|}{\textbf{}} &
  \multicolumn{1}{c|}{\textbf{}} &
  \multicolumn{1}{c|}{\textbf{}} &
  \multicolumn{1}{c|}{\textbf{}} &
  \multicolumn{1}{c|}{\textbf{}} &
  \multicolumn{1}{c|}{\textbf{$\surd$}} &
  \multicolumn{1}{c|}{\textbf{}} &
  \multicolumn{1}{c|}{\textbf{}} &
  \multicolumn{1}{c|}{\textbf{}} &
  \multicolumn{1}{c|}{\textbf{}} &
  \multicolumn{1}{c|}{\textbf{}} &
  \multicolumn{1}{c|}{\textbf{}} &
  \multicolumn{1}{c|}{\textbf{}} &
  \multicolumn{1}{c|}{\textbf{}} &
  \multicolumn{1}{c|}{\textbf{}} &
  \multicolumn{1}{c|}{\textbf{}} &
  \multicolumn{1}{c|}{\textbf{}} &
  \multicolumn{1}{c|}{\textbf{}} &
  \multicolumn{1}{c|}{\textbf{}} &
  \multicolumn{1}{c|}{\textbf{}} &
  \multicolumn{1}{c|}{\textbf{}} &
  \multicolumn{1}{c|}{\textbf{}} &
  \multicolumn{1}{c|}{\textbf{}} &
  \multicolumn{1}{c|}{\textbf{}} &
  \multicolumn{1}{c|}{\textbf{}} &
  \multicolumn{1}{c|}{\textbf{}} &
  \multicolumn{1}{c|}{\textbf{}} &
  \multicolumn{1}{c|}{\textbf{}} &
  \multicolumn{1}{c|}{\textbf{}} &
  \multicolumn{1}{c|}{\textbf{}} &
  \multicolumn{1}{c|}{\textbf{}} &
  \multicolumn{1}{c|}{\textbf{}} &
  \multicolumn{1}{c|}{\textbf{}} 
  %\textbf{1} 
  \\
\multicolumn{1}{|l|}{\begin{tabular}[c]{@{}l@{}}\textbf{TeReDetect}       {\cite{koochakzadeh2010tester} }\end{tabular}} &
  \multicolumn{1}{c|}{\textbf{}} &
  \multicolumn{1}{c|}{\textbf{}} &
  \multicolumn{1}{c|}{\textbf{}} &
  \multicolumn{1}{c|}{\textbf{}} &
  \multicolumn{1}{c|}{\textbf{}} &
  \multicolumn{1}{c|}{\textbf{}} &
  \multicolumn{1}{c|}{\textbf{}} &
  \multicolumn{1}{c|}{\textbf{}} &
  \multicolumn{1}{c|}{\textbf{}} &
  \multicolumn{1}{c|}{\textbf{}} &
  \multicolumn{1}{c|}{\textbf{}} &
  \multicolumn{1}{c|}{\textbf{}} &
  \multicolumn{1}{c|}{\textbf{}} &
  \multicolumn{1}{c|}{\textbf{}} &
  \multicolumn{1}{c|}{\textbf{}} &
  \multicolumn{1}{c|}{\textbf{}} &
  \multicolumn{1}{c|}{\textbf{}} &
  \multicolumn{1}{c|}{\textbf{}} &
  \multicolumn{1}{c|}{\textbf{}} &
  \multicolumn{1}{c|}{\textbf{}} &
  \multicolumn{1}{c|}{\textbf{}} &
  \multicolumn{1}{c|}{\textbf{}} &
  \multicolumn{1}{c|}{\textbf{}} &
  \multicolumn{1}{c|}{\textbf{}} &
  \multicolumn{1}{c|}{\textbf{}} &
  \multicolumn{1}{c|}{\textbf{}} &
  \multicolumn{1}{c|}{\textbf{}} &
  \multicolumn{1}{c|}{\textbf{}} &
  \multicolumn{1}{c|}{\textbf{}} &
  \multicolumn{1}{c|}{\textbf{$\surd$}} &
  \multicolumn{1}{c|}{\textbf{}} &
  \multicolumn{1}{c|}{\textbf{}} &
  \multicolumn{1}{c|}{\textbf{}} &
  \multicolumn{1}{c|}{\textbf{}} 
  %\textbf{1}
  \\
\rowcolor{gray!30}
\multicolumn{1}{|l|}{\begin{tabular}[c]{@{}l@{}}\textbf{TestEvoHound}       {\cite{greiler2013strategies}}\end{tabular}} &
  \multicolumn{1}{c|}{\textbf{}} &
  \multicolumn{1}{c|}{\textbf{}} &
  \multicolumn{1}{c|}{\textbf{}} &
  \multicolumn{1}{c|}{\textbf{}} &
  \multicolumn{1}{c|}{\textbf{}} &
  \multicolumn{1}{c|}{\textbf{}} &
  \multicolumn{1}{c|}{\textbf{}} &
  \multicolumn{1}{c|}{\textbf{$\surd$}} &
  \multicolumn{1}{c|}{\textbf{}} &
  \multicolumn{1}{c|}{\textbf{}} &
  \multicolumn{1}{c|}{\textbf{}} &
  \multicolumn{1}{c|}{\textbf{}} &
  \multicolumn{1}{c|}{\textbf{}} &
  \multicolumn{1}{c|}{\textbf{$\surd$}} &
  \multicolumn{1}{c|}{\textbf{}} &
  \multicolumn{1}{c|}{\textbf{}} &
  \multicolumn{1}{c|}{\textbf{}} &
  \multicolumn{1}{c|}{\textbf{$\surd$}} &
  \multicolumn{1}{c|}{\textbf{}} &
  \multicolumn{1}{c|}{\textbf{}} &
  \multicolumn{1}{c|}{\textbf{}} &
  \multicolumn{1}{c|}{\textbf{$\surd$}} &
  \multicolumn{1}{c|}{\textbf{}} &
  \multicolumn{1}{c|}{\textbf{}} &
  \multicolumn{1}{c|}{\textbf{}} &
  \multicolumn{1}{c|}{\textbf{}} &
  \multicolumn{1}{c|}{\textbf{}} &
  \multicolumn{1}{c|}{\textbf{}} &
  \multicolumn{1}{c|}{\textbf{$\surd$}} &
  \multicolumn{1}{c|}{\textbf{}} &
  \multicolumn{1}{c|}{\textbf{}} &
  \multicolumn{1}{c|}{\textbf{}} &
  \multicolumn{1}{c|}{\textbf{$\surd$}} &
  \multicolumn{1}{c|}{\textbf{}} 
  %\textbf{6}
  \\
\multicolumn{1}{|l|}{\begin{tabular}[c]{@{}l@{}}\textbf{TestHound}       {\cite{greiler2013automated}}\end{tabular}} &
  \multicolumn{1}{c|}{\textbf{}} &
  \multicolumn{1}{c|}{\textbf{}} &
  \multicolumn{1}{c|}{\textbf{}} &
  \multicolumn{1}{c|}{\textbf{}} &
  \multicolumn{1}{c|}{\textbf{}} &
  \multicolumn{1}{c|}{\textbf{}} &
  \multicolumn{1}{c|}{\textbf{}} &
  \multicolumn{1}{c|}{\textbf{$\surd$}} &
  \multicolumn{1}{c|}{\textbf{}} &
  \multicolumn{1}{c|}{\textbf{}} &
  \multicolumn{1}{c|}{\textbf{}} &
  \multicolumn{1}{c|}{\textbf{}} &
  \multicolumn{1}{c|}{\textbf{}} &
  \multicolumn{1}{c|}{\textbf{$\surd$}} &
  \multicolumn{1}{c|}{\textbf{}} &
  \multicolumn{1}{c|}{\textbf{}} &
  \multicolumn{1}{c|}{\textbf{}} &
  \multicolumn{1}{c|}{\textbf{$\surd$}} &
  \multicolumn{1}{c|}{\textbf{}} &
  \multicolumn{1}{c|}{\textbf{}} &
  \multicolumn{1}{c|}{\textbf{}} &
  \multicolumn{1}{c|}{\textbf{$\surd$}} &
  \multicolumn{1}{c|}{\textbf{}} &
  \multicolumn{1}{c|}{\textbf{}} &
  \multicolumn{1}{c|}{\textbf{}} &
  \multicolumn{1}{c|}{\textbf{}} &
  \multicolumn{1}{c|}{\textbf{}} &
  \multicolumn{1}{c|}{\textbf{}} &
  \multicolumn{1}{c|}{\textbf{$\surd$}} &
  \multicolumn{1}{c|}{\textbf{}} &
  \multicolumn{1}{c|}{\textbf{}} &
  \multicolumn{1}{c|}{\textbf{}} &
  \multicolumn{1}{c|}{\textbf{$\surd$}} &
  \multicolumn{1}{c|}{\textbf{}} 
  %\textbf{6}
  \\
\rowcolor{gray!30}
\multicolumn{1}{|l|}{\begin{tabular}[c]{@{}l@{}}\textbf{TestQ}       {\cite{breugelmans2008testq}}\end{tabular}} &
  \multicolumn{1}{c|}{\textbf{$\surd$}} &
  \multicolumn{1}{c|}{\textbf{$\surd$}} &
  \multicolumn{1}{c|}{\textbf{}} &
  \multicolumn{1}{c|}{\textbf{}} &
  \multicolumn{1}{c|}{\textbf{}} &
  \multicolumn{1}{c|}{\textbf{$\surd$}} &
  \multicolumn{1}{c|}{\textbf{}} &
  \multicolumn{1}{c|}{\textbf{}} &
  \multicolumn{1}{c|}{\textbf{}} &
  \multicolumn{1}{c|}{\textbf{}} &
  \multicolumn{1}{c|}{\textbf{$\surd$}} &
  \multicolumn{1}{c|}{\textbf{$\surd$}} &
  \multicolumn{1}{c|}{\textbf{$\surd$}} &
  \multicolumn{1}{c|}{\textbf{$\surd$}} &
  \multicolumn{1}{c|}{\textbf{}} &
  \multicolumn{1}{c|}{\textbf{$\surd$}} &
  \multicolumn{1}{c|}{\textbf{$\surd$}} &
  \multicolumn{1}{c|}{\textbf{}} &
  \multicolumn{1}{c|}{\textbf{}} &
  \multicolumn{1}{c|}{\textbf{$\surd$}} &
  \multicolumn{1}{c|}{\textbf{}} &
  \multicolumn{1}{c|}{\textbf{}} &
  \multicolumn{1}{c|}{\textbf{}} &
  \multicolumn{1}{c|}{\textbf{}} &
  \multicolumn{1}{c|}{\textbf{}} &
  \multicolumn{1}{c|}{\textbf{}} &
  \multicolumn{1}{c|}{\textbf{$\surd$}} &
  \multicolumn{1}{c|}{\textbf{}} &
  \multicolumn{1}{c|}{\textbf{}} &
  \multicolumn{1}{c|}{\textbf{}} &
  \multicolumn{1}{c|}{\textbf{}} &
  \multicolumn{1}{c|}{\textbf{}} &
  \multicolumn{1}{c|}{\textbf{}} &
  \multicolumn{1}{c|}{\textbf{$\surd$}} 
  %\textbf{12}
  \\
\multicolumn{1}{|l|}{\begin{tabular}[c]{@{}l@{}}\textbf{\textsc{tsDetect}}       {\cite{peruma2020FSE}}\end{tabular}} &
  \multicolumn{1}{c|}{\textbf{}} &
  \multicolumn{1}{c|}{\textbf{$\surd$}} &
  \multicolumn{1}{c|}{\textbf{$\surd$}} &
  \multicolumn{1}{c|}{\textbf{$\surd$}} &
  \multicolumn{1}{c|}{\textbf{$\surd$}} &
  \multicolumn{1}{c|}{\textbf{}} &
  \multicolumn{1}{c|}{\textbf{}} &
  \multicolumn{1}{c|}{\textbf{}} &
  \multicolumn{1}{c|}{\textbf{$\surd$}} &
  \multicolumn{1}{c|}{\textbf{$\surd$}} &
  \multicolumn{1}{c|}{\textbf{$\surd$}} &
  \multicolumn{1}{c|}{\textbf{$\surd$}} &
  \multicolumn{1}{c|}{\textbf{}} &
  \multicolumn{1}{c|}{\textbf{$\surd$}} &
  \multicolumn{1}{c|}{\textbf{$\surd$}} &
  \multicolumn{1}{c|}{\textbf{}} &
  \multicolumn{1}{c|}{\textbf{}} &
  \multicolumn{1}{c|}{\textbf{}} &
  \multicolumn{1}{c|}{\textbf{$\surd$}} &
  \multicolumn{1}{c|}{\textbf{$\surd$}} &
  \multicolumn{1}{c|}{\textbf{$\surd$}} &
  \multicolumn{1}{c|}{\textbf{}} &
  \multicolumn{1}{c|}{\textbf{$\surd$}} &
  \multicolumn{1}{c|}{\textbf{$\surd$}} &
  \multicolumn{1}{c|}{\textbf{$\surd$}} &
  \multicolumn{1}{c|}{\textbf{}} &
  \multicolumn{1}{c|}{\textbf{$\surd$}} &
  \multicolumn{1}{c|}{\textbf{$\surd$}} &
  \multicolumn{1}{c|}{\textbf{}} &
  \multicolumn{1}{c|}{\textbf{}} &
  \multicolumn{1}{c|}{\textbf{}} &
  \multicolumn{1}{c|}{\textbf{$\surd$}} &
  \multicolumn{1}{c|}{\textbf{}} &
  \multicolumn{1}{c|}{\textbf{}} 
  %\textbf{19}
  \\
\rowcolor{gray!30}
\multicolumn{1}{|l|}{\begin{tabular}[c]{@{}l@{}}\textbf{Unnamed}       {\cite{bavota2012empirical}}\end{tabular}} &
  \multicolumn{1}{c|}{\textbf{}} &
  \multicolumn{1}{c|}{\textbf{$\surd$}} &
  \multicolumn{1}{c|}{\textbf{}} &
  \multicolumn{1}{c|}{\textbf{}} &
  \multicolumn{1}{c|}{\textbf{}} &
  \multicolumn{1}{c|}{\textbf{$\surd$}} &
  \multicolumn{1}{c|}{\textbf{}} &
  \multicolumn{1}{c|}{\textbf{}} &
  \multicolumn{1}{c|}{\textbf{}} &
  \multicolumn{1}{c|}{\textbf{}} &
  \multicolumn{1}{c|}{\textbf{}} &
  \multicolumn{1}{c|}{\textbf{$\surd$}} &
  \multicolumn{1}{c|}{\textbf{$\surd$}} &
  \multicolumn{1}{c|}{\textbf{$\surd$}} &
  \multicolumn{1}{c|}{\textbf{}} &
  \multicolumn{1}{c|}{\textbf{}} &
  \multicolumn{1}{c|}{\textbf{$\surd$}} &
  \multicolumn{1}{c|}{\textbf{}} &
  \multicolumn{1}{c|}{\textbf{$\surd$}} &
  \multicolumn{1}{c|}{\textbf{$\surd$}} &
  \multicolumn{1}{c|}{\textbf{}} &
  \multicolumn{1}{c|}{\textbf{}} &
  \multicolumn{1}{c|}{\textbf{}} &
  \multicolumn{1}{c|}{\textbf{}} &
  \multicolumn{1}{c|}{\textbf{}} &
  \multicolumn{1}{c|}{\textbf{}} &
  \multicolumn{1}{c|}{\textbf{$\surd$}} &
  \multicolumn{1}{c|}{\textbf{}} &
  \multicolumn{1}{c|}{\textbf{}} &
  \multicolumn{1}{c|}{\textbf{}} &
  \multicolumn{1}{c|}{\textbf{$\surd$}} &
  \multicolumn{1}{c|}{\textbf{}} &
  \multicolumn{1}{c|}{\textbf{}} &
  \multicolumn{1}{c|}{\textbf{}} 
  %\textbf{10}
  \\
\multicolumn{1}{|l|}{\textit{\textbf{Total}}} &
  \multicolumn{1}{r|}{\textbf{2}} &
  \multicolumn{1}{r|}{\textbf{6}} &
  \multicolumn{1}{r|}{\textbf{2}} &
  \multicolumn{1}{r|}{\textbf{2}} &
  \multicolumn{1}{r|}{\textbf{3}} &
  \multicolumn{1}{r|}{\textbf{2}} &
  \multicolumn{1}{r|}{\textbf{5}} &
  \multicolumn{1}{r|}{\textbf{2}} &
  \multicolumn{1}{r|}{\textbf{2}} &
  \multicolumn{1}{r|}{\textbf{2}} &
  \multicolumn{1}{r|}{\textbf{3}} &
  \multicolumn{1}{r|}{\textbf{7}} &
  \multicolumn{1}{r|}{\textbf{2}} &
  \multicolumn{1}{r|}{\textbf{9}} &
  \multicolumn{1}{r|}{\textbf{1}} &
  \multicolumn{1}{r|}{\textbf{1}} &
  \multicolumn{1}{r|}{\textbf{2}} &
  \multicolumn{1}{r|}{\textbf{4}} &
  \multicolumn{1}{r|}{\textbf{4}} &
  \multicolumn{1}{r|}{\textbf{5}} &
  \multicolumn{1}{r|}{\textbf{2}} &
  \multicolumn{1}{r|}{\textbf{2}} &
  \multicolumn{1}{r|}{\textbf{2}} &
  \multicolumn{1}{r|}{\textbf{2}} &
  \multicolumn{1}{r|}{\textbf{2}} &
  \multicolumn{1}{c|}{\textbf{2}} &
  \multicolumn{1}{r|}{\textbf{5}} &
  \multicolumn{1}{r|}{\textbf{2}} &
  \multicolumn{1}{r|}{\textbf{2}} &
  \multicolumn{1}{r|}{\textbf{2}} &
  \multicolumn{1}{r|}{\textbf{1}} &
  \multicolumn{1}{r|}{\textbf{2}} &
  \multicolumn{1}{r|}{\textbf{2}} &
  \multicolumn{1}{r|}{\textbf{2}} 
  %\multicolumn{1}{c|}{--}
  \\\hline
\end{tabular}
}
%\vspace{-0.3cm}
\end{table*}

Next, we examine the types of test smells detected by the identified tools in our set. For completeness, we provide, in Table \ref{tab:Test_Definitions}, brief definitions for each unique smell type detected by the identified tools. We also provide their references for more details. When analyzing the definitions of these smell types, we observe that there are smells that are associated with more than one name, but with a similar description of its symptoms. For example, \textit{Assertionless}, \textit{Assertionless Test}, and \textit{Unkown Test} define the absence of an expected assert in the test method. Similarly, \textit{Duplicated Code} and \textit{Test Code Duplication} define the same issue of the existence of code clones. 

Looking at our list of tools, TestLint detects the highest number of smell types (26). JNose Test is the second highest with 21 detected smell types, followed by \textsc{tsDetect} (19 smell types). Furthermore, from Figure \ref{fig:Literature_TimeLine}, it is common to see various tools detecting the same smell types. %we observe instances where more than one tool detects the same smell type. 
Per analogy to code smells, while there is an agreement on the meanings of smells, there is no consensus on identifying them. Therefore, it is evident to see various tools containing different detection strategies for similar smell types. %how to 
Hence, in this analysis, we look at the overlap of detected smell types by the tools in our dataset. The smells, detected by the tools \textit{TestLint}, \textit{OraclePolish}, \textit{TRex}, and \textit{\textsc{PolDet}} are unique to the respective tool. The remaining 16 tools share the detection of some overlapping smells. %that overlap.

In Table \ref{tab:TestSmellTypes_v3}, we identify the overlapping of smells across 16 tools. For each tool, we indicate if the tool detects a specific smell by the \textbf{$\surd$} symbol. From this table, we observe that the three most common smell types are \textit{General Fixture}, \textit{Eager Test}, and \textit{Assertion Roulette}, which are respectively detected by 9, 7, and 6 tools. %It is not surprising to see \textit{General Fixture} being widely popular for detection. 

\subsubsection{\textbf{Supported Programming Languages}}
\noindent

\begin{table*}
\centering
\caption{Distribution of Test Smells Per Programming Languages.}
\vspace{-0.3cm}
%\resizebox{\columnwidth}{!}{%
\begin{adjustbox}{width=1.0\textwidth,center}
% \begin{threeparttable}
\begin{tabular} {|c|llll|c|c|}\hline

\rowcolor{gray!60}
\textbf{Programming Language} & \multicolumn{4}{c|}{\textbf{Supported Test Smell Types}}  & \textbf{Literature Usage} \\ \hline

	                & (01) Assertion Roulette (AR)         & (11) Eager Test (ET)                         & (21) Magic Number Test (MNT)  & (31) Test Maverick (TM)  &   \cite{baker2006trex, Zeiss2006TTCN, Werner2007TTCN, Helmut2007TRex, Neukirchen2008TTCN} \\ 
                    & (02) Assertionless (AL)              & (12) Empty Test (EmT)                        & (22) Mystery Guest (MG)       & (32) Test Pollution (TP) &  \cite{koochakzadeh2010tecrevis, bavota2012empirical, greiler2013automated, greiler2013strategies, zhang2014empirically} \\ 
                 	& (03) Brittle Assertion (BA)          & (13) Exception Handling (EH)                 & (23) Obscure In-line Setup Smell (OISS)  & (33) Test Redundancy (TR)  & \cite{bavota2015test, gyori2015reliable, bell2015efficient, palomba2016diffusion, tahir2016empirical} \\ 
                    & (04) Conditional Test Logic (CTL)    & (14) For Testers Only (FTO)                  & (24) Redundant Assertion (RA)     & (34) Test Run War(TRW)  &  \cite{Spadini2018ICSME, gambi2018practical, virginio2019influence, peruma2019CASCON, qusef2019exploratory} \\ 
\textbf{Java}       & (05) Constructor Initialization (CI) & (15) General Fixture (GF)                    & (25) Redundant Print (RP)  &  (35) TTCN-3 Smells (TTCN) &  \cite{ palomba2018automatic, grano2019scented, biagiola2019web, schvarcbacher2019investigating, Grano2019TSE} \\ 
                    & (06) Dead Field (DF)                 & (16) Ignored Test (IgT)                      & (26) Resource Optimism (RO)    &  (36) Unknown Test (UT) &  \cite{Spadini2020MSR, peruma2020FSE, lambiase2020just, virginio2020empirical, virginio2020jnose}  \\ 
                    & (07) Default Test (DT)               & (17) Indented Test (InT)                     & (27) Rotten Green Tests (RT)  & (37) Unused Input (UI)  &  \cite{Panichella2020ICSME, Soares2020SAST, Kim2020ICSE, Pecorelli2020AVI, Biagiola2020ICST} \\ 
                    & (08) Dependent Test (DepT)           & (18) Indirect Test (IT)                      & (28) Sensitive Equality (SE)  & (38) Vague Header Setup(VHS)  &  \cite{Peruma2020IWoR, Martinez2020ICSE, Fraser2020ICSTW, tufano2016empirical, huo2014improving} \\ 
                    & (09) Duplicate Assert (DA)           & (19) Lack of Cohesion of Test Method (LCM)   & (29) Sleepy Test (ST)  &  (39) Verbose Test (VT) &   \cite{koochakzadeh2010tester, Santana2020SBES}\\ 
                    & (10) Duplicated Code (DC)            & (20) Lazy Test (LT)                          & (30) (31) Test Code Duplication (TCD)         &   &   \\ \hline

%%%%%%%%%%%%%%%%%%%%%%%%%%%%%%%%%%%%%%%%%%%%%%%%%%%%%%%%%%%%%%%%%%%%%%%%%%%%%%%%%%%%%%%%%%%%%%%%%%%%%%%%%%%%%%%%%%%%%%%%%%%%%%%%%%%%%%%%%%%%%%%%%%%%%%%%%%%%%%%%%%%  
\rowcolor{gray!30}                      
\textbf{Scala}     & (01) Assertion Roulette (AR)            & (03) Exception Handling (EH)  & (05) Mystery Guest (MG)       &   & \cite{de2019socrates,de2019assessing} \\ 
\rowcolor{gray!30}
                         & (02) Eager Test (ET)              & (04) General Fixture (GF)     & (06) Sensitive Equality (SE)  &    & \\ \hline
%%%%%%%%%%%%%%%%%%%%%%%%%%%%%%%%%%%%%%%%%%%%%%%%%%%%%%%%%%%%%%%%%%%%%%%%%%%%%%%%%%%%%%%%%%%%%%%%%%%%%%%%%%%%%%%%%%%%%%%%%%%%%%%%%%%%%%%%%%%%%%%%%%%%%%%%%%%%%%%%%%%                         
                         & (01) Abnormal UTF-Use (AUU)                 & (08) Empty Shared-Fixture (ESF)            & (15) Under-the-carpet failing Assertion (UCFA)           & (22) Test-Class Name (TCN)  &  \\  
                         & (02) Anonymous Test (AT)                    & (09) Empty Test-MethodCategory (ETMC)                           & (16) Overcommented Test (OCT)       & (23) Test-MethodCategory Name (TMC) &   \\ 
                         & (03) Assertionless Test (AL)                & (10) Guarded Test (GT) & (17) Overreferencing (OF)       & (24) Transcripting Test (TT) &  \\ 
                         & (04) Comments Only Test (COT)               & (11) Likely ineffective Object-Comparison (LIOC)                             & (18) Proper Organization (PO)  & (25) Unclassied MethodCategory (UMC)  &   \\ 
\textbf{SmallTalk}       & (05) Control Logic (ConL)                   & (13) Long Test (LoT)                &  (19) Returning Assertion (RA)      & (26) Under-the-carpet Assertion (UCA) &  \cite{reichhart2007rule, Delplanque2019ICSE} \\ 
                         & (06) Early Returning Test (ERT)             & (12) Max Instance Variables (MIV)                        &  (20) Rotten Green Tests falls (RT)         & (27) Unused Shared-Fixture Variables (USFV) &  \\ 
                         & (07) Empty MethodCategory (EMC)             & (13) Mixed Selectors (MS)   & (21) Teardown Only Test (TOT)  & (28) Unusual Test Order (UTO) &  \\  \hline
% %%%%%%%%%%%%%%%%%%%%%%%%%%%%%%%%%%%%%%%%%%%%%%%%%%%%%%%%%%%%%%%%%%%%%%%%%%%%%%%%%%%%%%%%%%%%%%%%%%%%%%%%%%%%%%%%%%%%%%%%%%%%%%%%%%%%%%%%%%%%%%%%%%%%%%%%%%%%%%%%%%%                         
\rowcolor{gray!30}                 & (01) Assertion Roulette (AR)                    & (04) Eager Test (ET)         &  (07) General Fixture (GF)  & (10) Mystery Guest (MG)   &  \\ 
\rowcolor{gray!30} \textbf{C++}    & (02) Assertionless Test (ALT)  & (05) Empty Test (EmT)        &  (08) Indented Test (InT)    & (11) Sensitive Equality (SE) &  \cite{breugelmans2008testq}\\ 
 \rowcolor{gray!30}                & (03) Duplicated Code (DC)                       & (06) For Testers Only (FTO)  &  (09) Indirect Test (IT)    & (12) Verbose Test (VT) &  \\ \hline

\end{tabular}
% \begin{tablenotes}
            % \item $\star\,$ These smells derived from Test fixture smells based on \cite{greiler2013automated}.
        % \end{tablenotes}
% \end{threeparttable}
\end{adjustbox}
% {\raggedright \textbf{Note:} Test Fixture (TF), Test Maverick smell, Dead Field smell, Obscure In-line Setup Smell and Vague Header Setup. \par}
        
%}

\label{tab:ProjectBased}
\vspace{-0.2cm}
\end{table*}

% \cite{huo2014improving,schvarcbacher2019investigating,bavota2015test,van2006characterizing}\cite{bavota2012empirical,palomba2017does,van2007detection,grano2019scented,peruma2019distribution}\cite{virginio2019influence,palomba2016diffusion,greiler2013strategies,greiler2013automated,qusef2019exploratory}\cite{spadini2018relation,tahir2016empirical,tufano2016empirical,palomba2018automatic} \anthony{Wajdi -- why do you have four citation groups?} \\ 

Next, we look at the programming languages supported by the various smell types we identify in this study. From Table \ref{tab:ProjectBased}, we observe that the smell types support four programming languages, specifically Java, Scala, Smalltalk, and C++. From this set, Java is the most popular programming language for test smell support, supporting 39 smell types, followed by Smalltalk (28 smell types). In Table \ref{tab:ProjectBased}, we also list the publications (tool and tool adoption) in our dataset that analyze these smell types. From this, we observe that a subset of the Java-supported test smells also support Scala unit test code. Although the developed XUnit guidelines can be applied to various languages \cite{meszaros2007xunit}, including dynamically typed ones such as JavaScript and Python, we did not locate tools that analyzes test suites written in these languages, which represents a noticeable limitation in terms of supporting the high quality of test suites. %, for languages that have taken the lead in terms of number of projects on GitHub \footnote{last visited March 2021: https://madnight.github.io/githut/\#/pull_requests/2020/4}.
%now represent the majority of  , we also observe that a majority of studies on test smells are based on Java systems. 

%In this study, we have performed a comprehensive investigation of the test smells detection tools along with the types of test smells these tools have detected. In Table~\ref{tab:ProjectBased}, we have distributed the test smells in three categories i.e. Java-Based, Scala-Based and SmallTalk-Based. This distribution has been done according to the programming languages used. According to the results of our systematic review, most of the work has been done on Java-Based projects for detecting the test smells i.e. 35 different test smells. Subsequently, researchers have given importance to SmallTalk-Based projects as well, 27 types of test smells are detected in these projects. Scala-Based projects have gain least importance by researchers for the process of detecting test smells. As per our analysis, more work is required on test smells detection using different programming languages so that validity of proposed tool can be done in a broader way.

\begin{tcolorbox}[top=2mm, bottom=2mm, left=2mm, right=2mm]
\textit{Summary}. 
While there has been a steady release of test smell detection tools over the years, there has been an uptick in both tool development and adoption recently, specifically in 2019 and 2020. In terms of detected types, the majority of these tools have an overlap in the types of detected smells, with \textit{General Fixture} being the most commonly detected smell type. While most of the tools detect test smells occurring in Java test suites, there is a lack of support for other popular languages, such as JavaScript and Python. %other programming languages. Finally, a minority of tools provide coverage for a broad set of smell types.
\end{tcolorbox}

%%%%%%%%%%%%%%%%%%%%%%%%%%%%%%%%%%%%%%%%%%%%%%%%%%%%%%%%%%%%%%%%%%%%%%%%%%%%%%%%%%%%%%%%%%%%%%%%%%%%%%%%%%%%%%%%%%%%
%\subsection{Main Features of Test Smell Detection Tools}
%\textbf{RQ2:} \textsl{What are the main features of test smell tools?} \anthony{"Main Features" is too generic. Need to exclusively mention the features at the start of the RQ}
%\wajdi{It described here as the summary of the RQ in introduction, but we started the RQ with small introduction about the xUint and the goal of the question, when we start discussed from the second paragraph here.}

\subsection{\textbf{\RQB}}
\label{SubSection:RQB}
This RQ comprises of two parts that examine the common characteristics of test smell detection tool. The first part examines specific high-level features of such tools, while the second part looks at the types of smell detection techniques implemented by the tools.

\subsubsection{\textbf{Common Characteristics}}
\noindent

\begin{table*}
\centering
\caption{Characteristics of test smell detection tools.}
\vspace{-0.3cm}
\label{tab:Tools_Table_v3}

\resizebox{\textwidth}{!}{%
\begin{threeparttable}

\begin{tabular}{|lllllllllll|}

\rowcolor{gray!60}
\multicolumn{1}{|c}{\cellcolor{gray!60}} &
  \multicolumn{2}{c}{\cellcolor{gray!60}\textbf{Programming Language}} &
  \multicolumn{1}{c}{\cellcolor{gray!60}} &
  \multicolumn{1}{c}{\cellcolor{gray!60}} &
  \multicolumn{1}{c}{\cellcolor{gray!60}} &
  \multicolumn{1}{c}{\cellcolor{gray!60}} &
  \multicolumn{1}{c}{\cellcolor{gray!60}} &
  \multicolumn{1}{c}{\cellcolor{gray!60}} &
  \multicolumn{1}{c|}{\cellcolor{gray!60}} \\
\rowcolor{gray!60}
\multicolumn{1}{|c}{\multirow{-2}{*}{\cellcolor{gray!60}\textbf{Tool}}} &
  \multicolumn{1}{c}{\cellcolor{gray!60}\textbf{Implemented}} &
  \multicolumn{1}{c}{\cellcolor{gray!60}\textbf{Analyzed}} &
  \multicolumn{1}{c}{\multirow{-2}{*}{\cellcolor{gray!60}\textbf{\begin{tabular}[c]{@{}c@{}}Supported \\      Test Framework\end{tabular}}}} &
  \multicolumn{1}{c}{\multirow{-2}{*}{\cellcolor{gray!60}\textbf{Correctness}}} &
  \multicolumn{1}{c}{\multirow{-2}{*}{\cellcolor{gray!60}\textbf{\begin{tabular}[c]{@{}c@{}}Detection \\ Technique\end{tabular}}}} &
  \multicolumn{1}{c}{\multirow{-2}{*}{\cellcolor{gray!60}\textbf{Interface}}} &
  \multicolumn{1}{c}{\multirow{-2}{*}{\cellcolor{gray!60}\textbf{\begin{tabular}[c]{@{}c@{}}Usage \\Guide \end{tabular}}}} &
  \multicolumn{1}{c}{\multirow{-2}{*}{\cellcolor{gray!60}\textbf{\begin{tabular}[c]{@{}c@{}}Adoption in\\Studies\end{tabular}}}} &
    \multicolumn{1}{c}{\multirow{-2}{*}{\cellcolor{gray!60}\textbf{\begin{tabular}[c]{@{}c@{}}Tool\\Website\end{tabular}}}}  \\ \hline

\textbf{DARTS} \textsuperscript{\textdaggerdbl} \cite{lambiase2020just} &
  Java &
  Java &
  JUnit &
  \multicolumn{1}{l}{\begin{tabular}[c]{@{}l@{}}F-Measure: 62\%-76\%\end{tabular}} &
  Information Retrieval &
  IntelliJ plugin &
  Yes &
  -- &
  \cite{Download:DARTS2020} \\ 
%%%%%%%%%%%%%%%%%%%%%%%%%%%%%%%
\rowcolor{gray!30}
\textbf{DrTest} \cite{Delplanque2019ICSE} &
  Smalltalk &
  Pharo \textsuperscript{$\nabla$}&
  SUnit &
  UNK &
  \multicolumn{1}{l}{\begin{tabular}[c]{@{}l@{}}Rule\\Dynamic Tainting\end{tabular}} &
  Pharo plugin &
  Yes &
  -- &
  \cite{Download:DrTest} \\ 
%%%%%%%%%%%%%%%%%%%%%%%%%%%%%%% 
\textbf{DTDetector} \textsuperscript{$\star\,\diamond$} \cite{zhang2014empirically} &
  Java &
  Java &
  JUnit &
  UNK &
  Dynamic Tainting &
  Command-line &
  Yes &
  -- &
  \cite{Download:DTDetector} \\ 
%%%%%%%%%%%%%%%%%%%%%%%%%%%%%%%
\rowcolor{gray!30}
\textbf{ElectricTest} \cite{bell2015efficient} &
  Java &
  Java &
  JUnit &
  UNK &
  Dynamic Tainting  &
  Command-line &
  No &
  -- &
  UNK \\  
%%%%%%%%%%%%%%%%%%%%%%%%%%%%%%%
\textbf{JNose Test} \cite{virginio2020jnose} &
  Java &
  Java &
  JUnit &
  UNK &
  Rule &
  Local web application &
  Yes &
  \cite{virginio2019influence, virginio2020empirical} &
  \cite{Download:JNose2019}  \\
%%%%%%%%%%%%%%%%%%%%%%%%%%%%%%%
\rowcolor{gray!30}
\textbf{OraclePolish} \textsuperscript{$\star$} \cite{huo2014improving} &
  Java &
  Java &
  JUnit &
  UNK &
  Dynamic Tainting &
  Command-line &
  Yes &
  -- &
  \cite{Download:OraclePolish2014} \\
%%%%%%%%%%%%%%%%%%%%%%%%%%%%%%%
\textbf{\textsc{PolDet}} \cite{gyori2015reliable} &
  Java &
  Java &
  JUnit &
  UNK &
  Dynamic Tainting &
  UNK &
  No &
  -- &
  UNK \\ 
%%%%%%%%%%%%%%%%%%%%%%%%%%%%%%%
\rowcolor{gray!30}
\textbf{\textsc{PraDeT}} \cite{gambi2018practical} &
  Java &
  Java &
  JUnit &
  UNK &
  Dynamic Tainting &
  Command-line &
  Yes &
  -- &
  \cite{Download:Pradet2018} \\  
%%%%%%%%%%%%%%%%%%%%%%%%%%%%%%%
\textbf{RAIDE} \textsuperscript{\textdaggerdbl} \cite{Santana2020SBES} &
  Java &
  Java &
  JUnit &
  UNK &
  Rule &
  Eclipse plugin &
  Yes &
  -- &
  \cite{Download:RAIDE2020} \\ 
%%%%%%%%%%%%%%%%%%%%%%%%%%%%%%%
\rowcolor{gray!30}
\textbf{RTj} \textsuperscript{\textdaggerdbl} \cite{Martinez2020ICSE} &
  Java &
  Java &
  JUnit &
  UNK &
  \multicolumn{1}{l}{\begin{tabular}[c]{@{}l@{}}Rule\\Dynamic Tainting\end{tabular}} &
  Command-line &
  Yes &
  -- &
  \cite{Download:RTj} \\ 
%%%%%%%%%%%%%%%%%%%%%%%%%%%%%%%
\textbf{\textsc{SoCRATES}} \cite{de2019socrates} &
  Scala &
  Scala &
  ScalaTest &
  \multicolumn{1}{l}{\begin{tabular}[c]{@{}l@{}}Precision: 98.94\%\\Recall: 89.59\%\end{tabular}} &
  Rule &
  IntelliJ plugin &
  Yes &
  \cite{de2019assessing}  &
  \cite{Download:SOCRATES2019} \\
%%%%%%%%%%%%%%%%%%%%%%%%%%%%%%%
\rowcolor{gray!30}
\textbf{\textsc{Taste}} \cite{palomba2018automatic} &
  UNK &
  Java &
  JUnit &
  \multicolumn{1}{l}{\begin{tabular}[c]{@{}l@{}}Precision: 57\%-75\%\\Recall: 60\%-80\%\end{tabular}} &
  Information Retrieval &
  UNK &
  No &
  \cite{Pecorelli2020AVI} &
  UNK \\
%%%%%%%%%%%%%%%%%%%%%%%%%%%%%%%
\textbf{TeCReVis} \textsuperscript{$\star$} \cite{koochakzadeh2010tecrevis} &
  Java &
  Java &
  JUnit &
  UNK &
  \multicolumn{1}{l}{\begin{tabular}[c]{@{}l@{}}Metrics\\Dynamic Tainting\end{tabular}} &
  Eclipse plugin \textsuperscript{\textdagger} &
  Yes &
  -- &
  \cite{Download:TeReDetect_TeCReVis} \\ 
%%%%%%%%%%%%%%%%%%%%%%%%%%%%%%%
\rowcolor{gray!30}
\textbf{TEDD} \cite{biagiola2019web} &
  Java &
  Java &
  JUnit &
  \multicolumn{1}{l}{\begin{tabular}[c]{@{}l@{}}Precision: 80\%\\Recall: 94\%\end{tabular}} &
  Information Retrieval &
  Command-line &
  Yes &
  \cite{Biagiola2020ICST} &
  \cite{Download:TEDD} \\ 
%%%%%%%%%%%%%%%%%%%%%%%%%%%%%%%
\textbf{TeReDetect} \textsuperscript{$\star$} \cite{koochakzadeh2010tester} &
  Java &
  Java &
  JUnit &
  UNK &
  \multicolumn{1}{l}{\begin{tabular}[c]{@{}l@{}}Metrics\\Dynamic Tainting\end{tabular}} &
  Eclipse plugin \textsuperscript{\textdagger} &
  Yes &
  -- &
  \cite{Download:TeReDetect_TeCReVis} \\ 
%%%%%%%%%%%%%%%%%%%%%%%%%%%%%%%
\rowcolor{gray!30}
\textbf{TestEvoHound} \cite{greiler2013strategies} &
  Java &
  Java &
  JUnit, TestNG &
  UNK &
  Metrics &
  UNK &
  No &
  -- &
  UNK  \\
%%%%%%%%%%%%%%%%%%%%%%%%%%%%%%%
\textbf{TestHound} \textsuperscript{\textdaggerdbl$\,\star$} \cite{greiler2013automated} &
  Java &
  Java &
  JUnit, TestNG &
  UNK &
  Metrics &
  Desktop application &
  No &
  -- &
  \cite{Download:TestHound2013} \\ 
%%%%%%%%%%%%%%%%%%%%%%%%%%%%%%%
\rowcolor{gray!30}
\textbf{TestLint} \textsuperscript{$\star$} \cite{reichhart2007rule} &
  Smalltalk &
  Smalltalk &
  Sunit &
  UNK &
  \multicolumn{1}{l}{\begin{tabular}[c]{@{}l@{}}Rule\\Dynamic Tainting\end{tabular}} &
  UNK &
  Yes &
  -- &
  \cite{Download:TestLint2007} \\ 
%%%%%%%%%%%%%%%%%%%%%%%%%%%%%%%
\textbf{TestQ} \textsuperscript{$\star$} \cite{breugelmans2008testq} &
  Python &
  C++, Java &
  \multicolumn{1}{l}{\begin{tabular}[c]{@{}l@{}}CppUnit, JUnit,\\Qtest\end{tabular}} &
  UNK &
  Metrics &
  Desktop application &
  Yes &
  -- &
  \cite{Download:TestQ2008} \\  
%%%%%%%%%%%%%%%%%%%%%%%%%%%%%%%
\rowcolor{gray!30}
\textbf{TRex} \textsuperscript{\textdaggerdbl$\,$\textsection$\,\star$} \cite{baker2006trex} &
  Java &
  Java &
  TTCN-3 &
  UNK &
  Rule &
  Eclipse plugin &
  Yes &
  \cite{Zeiss2006TTCN,Werner2007TTCN,Neukirchen2008TTCN,Helmut2007TRex} &
  \cite{Download:TRex} \\  
%%%%%%%%%%%%%%%%%%%%%%%%%%%%%%%
\textbf{\textsc{tsDetect}} \cite{peruma2020FSE} &
  Java &
  Java &
  JUnit &
  \multicolumn{1}{l}{\begin{tabular}[c]{@{}l@{}}Precision: 85\%-100\%\\Recall: 90\%-100\%\end{tabular}} &
  Rule &
  Command-line &
  Yes &
    \multicolumn{1}{l}{\begin{tabular}[c]{@{}l@{}}\cite{peruma2019CASCON,schvarcbacher2019investigating,Spadini2020MSR,Kim2020ICSE}\\\cite{Fraser2020ICSTW,Panichella2020ICSME,Soares2020SAST,Peruma2020IWoR}\end{tabular}} &
  \cite{Download:tsDetect2019} \\
%%%%%%%%%%%%%%%%%%%%%%%%%%%%%%%
\rowcolor{gray!30}
\textbf{Unnamed} \cite{bavota2012empirical} &
  UNK &
  Java &
  JUnit &
  \multicolumn{1}{l}{\begin{tabular}[c]{@{}l@{}}Precision: 88\%\\Recall: 100\%\end{tabular}} &
  Rule &
  Command-line &
  No &
    \multicolumn{1}{l}{\begin{tabular}[c]{@{}l@{}}\cite{bavota2015test,tahir2016empirical, palomba2016diffusion,tufano2016empirical}\\\cite{Spadini2018ICSME,qusef2019exploratory, grano2019scented, Grano2019TSE, Panichella2020ICSME}\end{tabular}} &
  UNK \\
%%%%%%%%%%%%%%%%%%%%%%%%%%%%%%%
  \hline
  
\end{tabular}%
\begin{tablenotes}
            \item \textdaggerdbl$\,$ Provides support for refactoring. | 
            \textdagger$\,$ Embedded inside the CodeCover \cite{CodeCover} plugin. | 
            \textsection$\,$ Enhanced version released in \cite{Helmut2007TRex}. | 
            $\star\,$ Included in the catalog of Garousi and K{\"u}{\c{c}}{\"u}k \cite{garousi2018smells}.
             \item $\diamond\,$ Also known as `TestIsolation' in the catalog of Garousi and K{\"u}{\c{c}}{\"u}k \cite{garousi2018smells}. | 
             $\nabla\,$ Pharo is a dialect of Smalltalk.
        \end{tablenotes}
\end{threeparttable}
}
\vspace{-0.3cm}
\end{table*}

While the number and types of test smells detected by the tools are essential in selecting an appropriate tool, there are other features that can be considered to make a more informed decision. Similar to prior literature \cite{fernandes2016review,azadi2019architectural,ain2019systematic}, we review our set of test smell detection tools with respect to the following characteristics:
%While there can be plenty of features that can be considered, this analysis looks at a specific set. Listed below are the features that we considered for the selection: 
\begin{enumerate}[wide, labelwidth=1pt, labelindent=0pt]
    \item \textbf{Programming Language} - This feature comprises of the programming language that the tool is implemented with and the programming language(s) the tool supports.
    \item \textbf{Supported Test Framework} - These frameworks provide an environment for developers to write unit tests. As part of the detection strategy the tool may be depended on the presence of specific framework API's in the test code.
    \item \textbf{Correctness} - Provides insight into how accurately the tool can detect smells. We look for instances where the tool authors provide values for precision and recall or F-measure.
    \item \textbf{Detection Technique} - Strategy the tool utilizes to analyze test code for the presence of smells.
    \item \textbf{Interface} - Indicates how developers interact with the tool.
    \item \textbf{Usages Guide Availability} - Indicates if documentation on how to use the tool is available (either in the tool's publication or website).
    \item \textbf{Adoption in Research Studies} - This provides insight into the popularity of the tool in the research community.
    \item \textbf{Tool Website} - Supplementary documentation about the tool, such as (where available) its source code repository, installation/execution instructions, etc.
\end{enumerate}

Table \ref{tab:Tools_Table_v3} detail our findings for each tool. In case we cannot locate the needed information, we label it as `UNK' in the table.

It is evident from Table \ref{tab:Tools_Table_v3} that the majority of test smell detection tools ($\approx$ 86\%) focus on detecting test smells exclusively for Java-based systems and are mostly focused on identifying any deviation from the guidelines of JUnit testing framework. This further corroborates our prior RQ finding where we show that most test smell types are geared towards Java systems, and thereby most research around test smells focuses on datasets composed of Java systems. Additionally, there are three tools, namely TestQ, TestHound, and TestEvoHound, which support two or more testing frameworks. In terms of correctness, most tools do not publish details around their accuracy. Additionally, the majority of tools do not report on performance speeds and execution times. Our findings show that only six tools published their detection accuracy, in terms of precision, recall or F-measure. From this set, only DARTS, \textsc{Taste}, and \textsc{tsDetect} report values for each smell type it supports; hence the correctness score is reported as a range. From our set of 22 detection tools, only five tools provide refactoring support. TestHound provides textual information on smells correction. While RAIDE provides a semi-automated correction, RTj, DARTS and TRex provide the automated refactoring of their detected smells. However, none of these tools provide details concerning the accuracy of their refactoring capabilities. From detection standpoint, we observe that tools, focusing on test dependency and rotten green test smells, use a dynamic detection strategy, while most static analysis-based smells prefer to utilize a rules-based approach. We discuss in detail the different techniques later on. In terms of adoption in research studies, only seven of the tools have been utilized by the research community to study test smells. 

Next, in terms of how developers run/interact with these tools, there are two categories-- graphical user interface (GUI) and command-line (i.e., non-GUI) tools. Most of the GUI tools are in the form of IDE plugins or web and desktop applications. In terms of tool availability, we searched for a link to the tool website or binaries. In case the link is absent or no longer functional, we contacted the publication's corresponding author. From these 22 publications, we were able to only locate 17 tools. It should be noted that, except for TestLint, the website for these 17 tools points to the tool's source code repository. Further, when examining the project repositories, we observe that \textsc{tsDetect} was the most forked repository (21 forks). Furthermore, an examination of a tool's publications, website, and the README file in the source code repository (where available) yields only 16 tools presenting guidelines on how to setup and/or execute the tool. Finally, from this set, only \textsc{tsDetect} and the tool by Bavota et al. \cite{bavota2012empirical} show a high adoption rate, having been used in at least eight other studies. The majority of the tools are only limited to the studies to which they were first introduced.

\subsubsection{\textbf{Smell Detection Techniques}}
\noindent

%An important part of a smell detection tool is the technique the tool utilizes to detect smells in the source code. Our analysis shows that while most of the tools in our study perform a static analysis of the source code, there are also instances of dynamic analysis. Further, the techniques the tools utilize to detect test smells can be grouped into four categories-- Metrics, Rules/Heuristic, Information Retrieval, and Dynamic Tainting. Our rationale for providing this analysis is to provide future smell detection tool developers with insight into how the current tools detect smells.

Each tool represents an implementation of a smell detection strategy. Each detection strategy reflects an interpretation of how the smell type manifests in the source code, i.e., how the smell symptoms can be identified. Our analysis shows that, while most of the tools in our study rely on static analysis of source code, some tools identify smells through dynamic analysis. These detection strategies can be grouped into four categories, namely Metrics, Rules/Heuristic, Information Retrieval, and Dynamic Tainting. With this clustering of strategies, we aim to familiarize future smell detection tool researchers/developers with smell detection techniques.

\textbf{Metrics.} The use of metrics to profile smells, is one of the early techniques, and popular ones, for all code smells in general. In a nutshell, the smell symptoms are measured through their impact on structural and semantic measurements, where their values go beyond pre-defined threshold. These metrics, and their corresponding threshold values, are combined into a rules-based tree to make a binary decision, of whether the code under analysis suffers from a smell or not. In a typical metrics-based smell detection approach, the source code is parsed and converted into an abstract syntax tree (AST). This AST is then subjected to a metrics-based analysis to identify and capture the test smells. For instance, Van Rompaey et al. \cite{van2007detection}, utilize metrics such as Number of Object Used in setup, Number of Production-Type, Number of Fixture Objects, Number Of Fixture Production Types, and Average Fixture Usage to detect smells such as \textit{General Fixture} and \textit{Eager Test}. The threshold used are chosen by the user or empirically derived from a representative set. %In Table \ref{tab:Metrics_Table}, we provide some examples of how smell types are profiled using metrics. %details around the metrics utilized by tools in detecting specific test smells.  

\textbf{Rules/Heuristic.} Rules or heuristics smell detection augments the metrics-based techniques with patterns that can be found in the source code \cite{sharma2018survey}. The smell is detected when the input matches a pre-defined set of metric thresholds with the existence of some code patterns. For example, the \textit{Assertion Roulette} smell is detected by defining a heuristic that checks whether a test method contains several assertion statements without an explanation message as a parameter for each assertion method.

\textbf{Information Retrieval.} In this technique, the main steps include extracting information/content from the test code and normalizing it. As part of the extraction process, the textual content from each JUnit test class (e.g., source code identifier and source comments) are taken as the necessary features for identifying the test smells. These characteristics are normalized via multiple text pre-processing steps. These steps include stemming, decomposing identifier names, stop word, and programming keyword removal. The normalized text is then weighted using Term Frequency and Inverse Document Frequency. The end-result is applying machine learning algorithms to extract textual features that can discriminate between classes, i.e., smell types. 

\textbf{Dynamic Tainting.} It monitors the source-code while it executes. Dynamic tainting enables the analysis of the actual data from the code based on run-time information. In particular, it works in two steps: (1) run the source-code along with predefined taint value/mark (i.e., user input), and (2) reason which executions are affected by that value/mark \cite{schwartz2010all}. 

Our categorization of the tools, in our study, based on the four smell detection techniques are shown in Table \ref{tab:Tools_Table_v3}. The Rule/Heuristic-based technique is a frequently adopted detection technique as the definition for most smells are based on well-defined rules \cite{van2001refactoring,meszaros2007xunit}. The metric-based mechanism is less frequently utilized due to: (1) not all known metrics proposed by \cite{van2006characterizing,van2007detection,greiler2013automated,breugelmans2008testq} have the ability to detect all test smells, since they can go beyond traditional design anomalies, and (2) the reliance on determining an appropriate set of thresholds is considered to be a significant challenge \cite{sharma2018survey}. Looking at tools that utilize a dynamic-based detection technique, we observe that most test dependency and rotten green test detection tools utilize this technique. Finally, DARTS, \textsc{Taste}, and TEDD utilize Information Retrieval techniques. However, as these tools rely on source code feature extraction, the lack of such information could affect the detection accuracy \cite{palomba2018automatic}.

\begin{tcolorbox}[top=2mm, bottom=2mm, left=2mm, right=2mm]
\textit{Summary}. 
JUnit is the most popular testing framework supported by test smell detection tools, with most static analysis based tools opting to utilize a rules-based detection strategy to identify smells. Even though most of the tools publish their source code, information about the tool's accuracy is seldom available. Our analysis only shows that only six tools publish their scores related to correctness. Finally, only a very few of these tools have a high adoption rate in other research studies. 
\end{tcolorbox}
%\input{Tables/Metrics_Table}

%%%%%%%%%%%%%%%%%%%%%%%%%%%%%%%%%%%%%%%%%%%%%%%%%%%%%%%%%%%%%%%%%%%%%%%%%%%%%%%%%%%%%%%%%%%%%%%%%%%%%%%%%%%%%%%%%%%%

% \section{An experimental study}
% \subsection{Case Study}

%%%%%%%%%%%%%%%%%%%%%%%%%%%%%%%%%%%%%%%%%%%%%%%%%%%%%%%%%%%%%%%%%%%%%%%%%%%%%%%%%%%%%%%%%%%%%%%%%%%%%%%%%%%%%%%%%%%%

% \begin{figure}[t]
%   \centering
%   \includegraphics[width=0.5\textwidth]{Images/2020-10-10_00-29-28.png}
%   \caption{Distribution .}
% \label{fig:Accuracy}
% \end{figure}

% %%%%%%%%%%%%%%%%%%%%%%%%%%%%%%%%%%%%%%%%%%%%%%%%%%%%

% %%%%%%%%%%%%%%%%%%%%%%%%%%%%%%%%%%%%%%%%%%%%%%%%%%%%

% \begin{figure}[t]
%   \centering
%   \includegraphics[width=0.5\textwidth]{Images/index.png}
%   \caption{Distribution .}
% \label{fig:Accuracy}
% \end{figure}

% %%%%%%%%%%%%%%%%%%%%%%%%%%%%%%%%%%%%%%%%%%%%%%%%%%%%

% %%%%%%%%%%%%%%%%%%%%%%%%%%%%%%%%%%%%%%%%%%%%%%%%%%%%

% \begin{figure}[t]
%   \centering
%   \includegraphics[width=0.5\textwidth]{Images/newplot-4.png}
%   \caption{Distribution .}
% \label{fig:Accuracy}
% \end{figure}

% %%%%%%%%%%%%%%%%%%%%%%%%%%%%%%%%%%%%%%%%%%%%%%%%%%%%

% \input{Tables/VenueTable}
% \input{Tables/AuthorsTable}

\section{Discussion}
\label{Discussions}

As a systematic mapping study, our findings provide a high-level understanding of the current state of test smell detection tools. %and serve as a platform for future studies in this area. 
In brief, as seen from our RQ findings, the research community has produced many test smell detection tools that support the detection of various test smell types. These tools, in turn, have been utilized in studies on test smells to understand how they influence software development. However, our findings also demonstrate areas of concern and expansion in this field. In this section, through a series of takeaways, we discuss how our findings can support the research and developer community in selecting the right tool as well as provide future directions for implementing and maintaining future test smell tools.  

\noindent\textbf{Takeaway 1: Standardization of smell names and definitions.} From RQ$_1$, we observe an overlap of the smell types detected by the tools. For instance, \textit{Assertion Roulette} is detected by six JUnit supported tools. However, the implementation of the detection rules may vary between these tools. Additionally, there can be instances where some test smell types with the same/similar definitions are known by different names. This fragmentation of smell definitions is not unique to test smells;  Sobrinho et al. \cite{Sobrinho2021TSE} experience this in code smells. This phenomenon provides the opportunity for future research in this area to compare and contrast such smell types. The agreement on smells definitions, does not necessarily induce similar interpretations. Since there is no consensus on how to measure smells, each smell type can be identified using different detection strategies, and the choice of the strategy becomes part of the developer's preferences. %Additionally, there is no oracle set that helps adopters with reviewing and comparing tools detection accuracy. Such benchmarks exist in other software engineering areas, such as defect prediction, regression testing, and vulnerabilities. So, there is a need for designing such oracle in the context of test smells. %implemented by the tools. This \christian{Maybe we can give some kind of opinion about how future research should resolve this} % MWM: Done

\noindent\textbf{Takeaway 2: Improve support for non-Java programming languages and testing frameworks.}
While our findings from RQ$_1$ show the existence of multiple test smell detection tools, our RQ$_2$ findings show that most of these tools are limited to supporting Java systems that utilize the JUnit testing framework, thereby narrowing test smell research to Java systems. Hence, restricting research to a single environment/language will not accurately reflect reality \cite{Silva2020TSE}. While it can be argued that as most test smells are based on xUnit guidelines \cite{meszaros2007xunit} research findings on Java systems can carry over to other similar languages (e.g., C\#), actual practitioners of non-Java systems gain no benefit without a tool to use in their development workflow. Furthermore, recent trends have shown a rise in the popularity of dynamically typed programming languages (e.g., Python and JavaScript) \cite{PYPL, TIOBE} giving more urgency for the research community to construct tools that support non-traditional research languages.

\noindent\textbf{Takeaway 3: Do not reinvent the wheel.}
Researchers/practitioners need to evaluate if implementing another detection tool is necessary for their specific needs or if modifying an existing tool would suffice. Reusing existing tools will not only save effort, but also develops more mature and robust frameworks. For instance, Spadini et al. \cite{schvarcbacher2019investigating} integrate \textsc{tsDetect} into a code quality monitoring system, while the tool implemented by Tufano et al. \cite{tufano2016empirical}, HistoryMiner, utilizes the tool of Bavota et al. \cite{bavota2012empirical} to detect test smells in the lifetime of a project. Additionally, some of the tools in our dataset are based on other tools in the dataset. For instance, JNose Test and RAID are built on top of \textsc{tsDetect} and DARTS is based on \textsc{Taste}. It is important that, when introducing new tools, tool maintainers should design their tools to be ready for customization. For instance, Spadini et al. \cite{Spadini2020MSR} customize \textsc{tsDetect} by introducing thresholds to meet their research objective. This can help reduce the release of near-duplicate tools. It also further strengthens the case for the importance of public availability of a tool's source code. Having access to the code enables the improvement of the tool's quality. It also facilitates extensions in improving current detection strategies or introducing the detection of new smell types, which, in the long run, improves the tool's usefulness.
%\christian{I think this point would be better served if we can make a suggestion for what the community -should- do instead of making more test smell tools. Is it our opinion that these tools need something that they currently don't have? We mentioned more programming languages already. Given the tests that they already cover, maybe we could suggest that test smell tools begin expanding into other types of smell? Like linguistic anti-patterns or something that isn't covered by the set above. If we can find a weakness in their coverage, we should point it out here. }\anthony{updated text. pls. review}\mohamed{I made my pass and I like it now.}

\noindent\textbf{Takeaway 4: Improve transparency on the quality of tools.}
As reported in RQ$_2$, only a few tools report on the correctness of the tool. Furthermore, clarity around bias mitigation is not completely addressed for the tools that do report correctness scores. While our objective here is not to discredit the validity of the current set of test smell detection tools, we only highlight inconsistencies that might lead to research studies obtaining varying results based on the tool in use. As stated by Panichella et al. \cite{Panichella2020ICSME}, there is a need for a community-maintained gold-set/standard of smelly test files to validate the current and future smell detection tools. We understand that the process involved in the creation of a community-curated gold-set might be time-consuming. Hence, in the meantime, we recommend that the peer-review process be adjusted, such that providing metrics for precision and recall at the smell type level (instead of an overall tool accuracy score) along with the evaluation dataset is made mandatory.

%\christian{What kind of metrics and mitigation do we recommend based on what we found out?}\anthony{updated}\mohamed{It makes sense to me now}

\noindent\textbf{Takeaway 5: Expand from just detecting test smells to interactive refactoring.}

Granted that the primary purpose of these tools is the detection of test smells, developers will also immensely benefit from suggested refactoring templates for each smell type. While there have been initial efforts to include such functionality in detection tools, it does not generalize to the current popular detected smells, and more research is needed to elaborate on their ability to appropriately change the test files without introducing any regression. %especially around the quality of the performed refactorings. 

\section{Related Work} %\anthony{\textbf{Attn: Ali/Mohamed:} Have we missed any related studies? }
\label{sec:RelatedWork}
In this section, we examine studies that perform literature reviews in the area of test smells and show how we differentiate our work.
%%In this section, we review studies similar to ours and how we differentiate our work. We split the related work into three parts-- (1) studies exclusively focused on testing and test smells, (2) generalized studies on smells in code (which may include test smells), and (3) studies providing a catalog of detection-based tools.

% Manual process in detecting and correcting smells can make broad test suites difficult in testing and professionals need supporting tools especially in this field. A few SLR studies have presented some information about tools to identify test smell types \cite{garousi2018smells, sharma2018survey, shafique2015systematic}. However, none of these studies have discussed and explained how different the tools are in detecting test smell types and the techniques used in the existing tools, especially when there are more than twenty test smell types to be detected. To the best of our knowledge, this study is the first to present a systematic literature review and perform an evaluation of test smells detection tools. Several studies \cite{fernandes2016review, de2015evaluation, ain2019systematic, ardito2020tool, raulamo2017choosing, shafique2015systematic} were identified, which could be considered in relation to our study.

%\subsection{Software Testing \& Test Smells}
Garousi and Mäntylä \cite{garousi2016systematic} conduct a tertiary study in the field of software testing. Through an analysis of 101 secondary studies, the authors identify popular testing methods and phases in secondary studies. Furthermore, the authors also identify areas in software testing underrepresented in secondary studies, such as performance, mobile among others. However, while the authors provide information around testing-reflated tools, they do not include test smell detection tools in their analysis. In a survey on test smells, Garousi and K{\"u}{\c{c}}{\"u}k \cite{garousi2018smells,garousi2019smells} analyze a total of 166 articles from industry and academia. In their study, the authors present a catalog of test smells including details around prevention, detection, and  correction. As part of their study, the authors also provide a listing of test smell detection tools. However, this listing is limited to a brief description of the tool and its associated download URL.

In a study on the software smells, Sharma et al. \cite{sharma2018survey} analyze 445 primary studies from academia. Included in this study are test smells. However, as this is a generalized study on smells, the authors do not perform a deep-dive on test smells. In this study, the authors provide an extensive catalog of the various types of software smells and provide a classification for these smells. Even though the authors provide a catalog of smell-related tools, test smell detection tools are not present in the catalog. %A large-scale systematic literature on bad smells conducted by Sobrinho et al. \cite{Sobrinho2021TSE} reports that studies of bad smells are popular among the research community. %While the smell duplicate code is frequently studied, the smells long method and large class are also frequently studied. Other findings include fragmentation of smell definitions and the non-availability of tool source code. 
As part of their set of bad smells, the authors also include test smells.

While these studies either provide an in-depth look into the research around test smells or include test smells as part of a generalized study on code smells, the information presented around test smell detection tools is limited. The singularity focus of detection tools in our study enables us to expand the list of available peer-reviewed test smell detection tools and present more details around such tools, including comparing tool features.

\section{Threats To Validity}
\label{sec:ThreatsValidity}
Even though we limit our literature search to six digital libraries, these libraries provide us with publications from a diverse set of scientific conferences, workshops, and journals. Further, these libraries are utilized in similar studies. Due to our strict inclusion/exclusion criteria, there exists a possibility specific tools are not included in this study. However, our criteria ensures that we only present peer-reviewed publications. In other words, grey literature is not in scope for this study. Furthermore, even though our search keywords reflect our study's specificness (i.e., focus on test smell detection tools), we capture the tools that are part of the catalog of Garousi and K{\"u}{\c{c}}{\"u}k \cite{garousi2018smells}. Additionally, to identify the words that should appear in the search string, we perform a pilot test. An essential part of our study is reviewing publications to determine the publications needed for our study and extracting information related to our tool feature analysis. To mitigate reviewer bias, each publication was evaluated by three authors. Our review process included a discussion phase to resolve selection conflicts. Furthermore, any publication that did not explicitly mention using a tool to detect test smells was excluded from our dataset. 
\section{Conclusion and Future Work}
\label{sec:Conclusion}

This study identifies 22 test smell detection tools made available by the research community through a comprehensive search of peer-reviewed scientific publications. As part of our analysis, we identify the smell types detected by these tools and highlight the smell types that overlap. Additional comparisons between these tools show that most of these tools support the JUnit framework, and while the source code is made available, details around the tool's detection correctness are not always made public. We envision our findings act as a one-stop source for test smell detection tools and empower researchers/practitioners in selecting the appropriate tool for their requirements. Future work in this area includes a hands-on evaluation of each tool to determine the extent to which tools detect common smell types and create a benchmark for such smell types.
%In this study, we have presented latest test smell detection tools, test smells types and the approaches used for developing these tools. Because of different test smell detection techniques have been used by researchers, we have categorized the selected studies into three groups. Subsequently, we have comprehensively analyzed the 12 test smell detection tools by considering various features, such as the type of test smell, total number of test smells detected by each tool, detection technique, test framework and availability of each tool. Thus, test smell detection tools, their important features and the approaches are publicized under single research which is rarely available to the best of our knowledge. This research will definitely facilitate researchers, practitioners and developers to select appropriate test smell detection tools and techniques as per their requirements. Moreover, it has been analyzed that developed tools have mostly focused on detecting Java based test smells only. Hence, work can be done on developing test smell detection tools based on multiple programming languages. In future, machine learning and artificial intelligence based techniques can be explored for detecting test smells as they possess the ability of making intelligent decisions on the basis of training data.   

% \begin{acks}
% \end{acks}

\bibliographystyle{ACM-Reference-Format}
\bibliography{base}


\begin{thebibliography}{91}

%%% the Web address.

\ifx \showCODEN    \undefined \def \showCODEN     #1{\unskip}     \fi
\ifx \showDOI      \undefined \def \showDOI       #1{#1}\fi
\ifx \showISBNx    \undefined \def \showISBNx     #1{\unskip}     \fi
\ifx \showISBNxiii \undefined \def \showISBNxiii  #1{\unskip}     \fi
\ifx \showISSN     \undefined \def \showISSN      #1{\unskip}     \fi
\ifx \showLCCN     \undefined \def \showLCCN      #1{\unskip}     \fi
\ifx \shownote     \undefined \def \shownote      #1{#1}          \fi
\ifx \showarticletitle \undefined \def \showarticletitle #1{#1}   \fi
\ifx \showURL      \undefined \def \showURL       {\relax}        \fi
% invisible to TeX
\providecommand\bibfield[2]{#2}
\providecommand\bibinfo[2]{#2}
\providecommand\natexlab[1]{#1}
\providecommand\showeprint[2][]{arXiv:#2}

\bibitem[\protect\citeauthoryear{??}{Pro}{[n.d.]}]%
        {ProjectWebiste}
 \bibinfo{year}{[n.d.]}\natexlab{}.
\newblock \bibinfo{howpublished}{\url{https://doi.org/10.5281/zenodo.4726288}}.
\newblock


\bibitem[\protect\citeauthoryear{??}{Cod}{[n.d.]}]%
        {CodeCover}
 \bibinfo{year}{[n.d.]}\natexlab{}.
\newblock \bibinfo{title}{CodeCover}.
\newblock \bibinfo{howpublished}{\url{http://codecover.org/}}.
\newblock
\newblock
\shownote{(Accessed: 03/02/2021).}


\bibitem[\protect\citeauthoryear{??}{Dow}{[n.d.]a}]%
        {Download:DARTS2020}
 \bibinfo{year}{[n.d.]}\natexlab{a}.
\newblock \bibinfo{title}{DARTS}.
\newblock
  \bibinfo{howpublished}{\url{https://github.com/StefanoLambiase/DARTS}}.
\newblock
\newblock
\shownote{(Accessed: 03/02/2021).}


\bibitem[\protect\citeauthoryear{??}{Dow}{[n.d.]b}]%
        {Download:DrTest}
 \bibinfo{year}{[n.d.]}\natexlab{b}.
\newblock \bibinfo{title}{DrTest}.
\newblock
  \bibinfo{howpublished}{\url{https://github.com/juliendelplanque/DrTests}}.
\newblock
\newblock
\shownote{(Accessed: 03/02/2021).}


\bibitem[\protect\citeauthoryear{??}{Dow}{[n.d.]c}]%
        {Download:DTDetector}
 \bibinfo{year}{[n.d.]}\natexlab{c}.
\newblock \bibinfo{title}{DTDetector}.
\newblock \bibinfo{howpublished}{\url{https://github.com/winglam/dtdetector}}.
\newblock
\newblock
\shownote{(Accessed: 03/02/2021).}


\bibitem[\protect\citeauthoryear{??}{Dow}{[n.d.]d}]%
        {Download:JNose2019}
 \bibinfo{year}{[n.d.]}\natexlab{d}.
\newblock \bibinfo{title}{JNose}.
\newblock \bibinfo{howpublished}{\url{https://github.com/arieslab/jnose}}.
\newblock
\newblock
\shownote{(Accessed: 03/02/2021).}


\bibitem[\protect\citeauthoryear{??}{Dow}{[n.d.]e}]%
        {Download:OraclePolish2014}
 \bibinfo{year}{[n.d.]}\natexlab{e}.
\newblock \bibinfo{title}{OraclePolish}.
\newblock \bibinfo{howpublished}{\url{https://bitbucket.org/udse/}}.
\newblock
\newblock
\shownote{(Accessed: 03/02/2021).}


\bibitem[\protect\citeauthoryear{??}{PYP}{[n.d.]}]%
        {PYPL}
 \bibinfo{year}{[n.d.]}\natexlab{}.
\newblock \bibinfo{title}{Popularity of Programming Language index}.
\newblock \bibinfo{howpublished}{\url{http://pypl.github.io/PYPL.html}}.
\newblock
\newblock
\shownote{(Accessed: 02/25/2021).}


\bibitem[\protect\citeauthoryear{??}{TIO}{[n.d.]}]%
        {TIOBE}
 \bibinfo{year}{[n.d.]}\natexlab{}.
\newblock \bibinfo{title}{popularity of programming languages index}.
\newblock \bibinfo{howpublished}{\url{https://www.tiobe.com/tiobe-index/}}.
\newblock
\newblock
\shownote{(Accessed: 02/25/2021).}


\bibitem[\protect\citeauthoryear{??}{Dow}{[n.d.]f}]%
        {Download:Pradet2018}
 \bibinfo{year}{[n.d.]}\natexlab{f}.
\newblock \bibinfo{title}{PRADET}.
\newblock
  \bibinfo{howpublished}{\url{https://github.com/gmu-swe/pradet-replication}}.
\newblock
\newblock
\shownote{(Accessed: 03/02/2021).}


\bibitem[\protect\citeauthoryear{??}{Dow}{[n.d.]g}]%
        {Download:RAIDE2020}
 \bibinfo{year}{[n.d.]}\natexlab{g}.
\newblock \bibinfo{title}{RAIDE}.
\newblock \bibinfo{howpublished}{\url{https://raideplugin.github.io/RAIDE/}}.
\newblock
\newblock
\shownote{(Accessed: 03/02/2021).}


\bibitem[\protect\citeauthoryear{??}{Dow}{[n.d.]h}]%
        {Download:RTj}
 \bibinfo{year}{[n.d.]}\natexlab{h}.
\newblock \bibinfo{title}{RTj}.
\newblock \bibinfo{howpublished}{\url{https://github.com/UPHF/RTj}}.
\newblock
\newblock
\shownote{(Accessed: 03/02/2021).}


\bibitem[\protect\citeauthoryear{??}{Dow}{[n.d.]i}]%
        {Download:SOCRATES2019}
 \bibinfo{year}{[n.d.]}\natexlab{i}.
\newblock \bibinfo{title}{SOCRATES}.
\newblock \bibinfo{howpublished}{\url{https://github.com/jonas-db/socrates}}.
\newblock
\newblock
\shownote{(Accessed: 03/02/2021).}


\bibitem[\protect\citeauthoryear{??}{Dow}{[n.d.]j}]%
        {Download:TEDD}
 \bibinfo{year}{[n.d.]}\natexlab{j}.
\newblock \bibinfo{title}{TEDD}.
\newblock
  \bibinfo{howpublished}{\url{https://github.com/matteobiagiola/FSE19-submission-material-TEDD}}.
\newblock
\newblock
\shownote{(Accessed: 03/02/2021).}


\bibitem[\protect\citeauthoryear{??}{Dow}{[n.d.]k}]%
        {Download:TeReDetect_TeCReVis}
 \bibinfo{year}{[n.d.]}\natexlab{k}.
\newblock \bibinfo{title}{TeReDetect \& TeCReVis}.
\newblock
  \bibinfo{howpublished}{\url{https://sourceforge.net/projects/codecover/}}.
\newblock
\newblock
\shownote{(Accessed: 03/02/2021).}


\bibitem[\protect\citeauthoryear{??}{Dow}{[n.d.]l}]%
        {Download:TestHound2013}
 \bibinfo{year}{[n.d.]}\natexlab{l}.
\newblock \bibinfo{title}{TestHound}.
\newblock
  \bibinfo{howpublished}{\url{https://github.com/SERG-Delft/TestHound}}.
\newblock
\newblock
\shownote{(Accessed: 03/02/2021).}


\bibitem[\protect\citeauthoryear{??}{Dow}{[n.d.]m}]%
        {Download:TestLint2007}
 \bibinfo{year}{[n.d.]}\natexlab{m}.
\newblock \bibinfo{title}{TestLint}.
\newblock
  \bibinfo{howpublished}{\url{http://scg.unibe.ch/wiki/alumni/stefanreichhart/testsmells}}.
\newblock
\newblock
\shownote{(Accessed: 03/02/2021).}


\bibitem[\protect\citeauthoryear{??}{Dow}{[n.d.]n}]%
        {Download:TestQ2008}
 \bibinfo{year}{[n.d.]}\natexlab{n}.
\newblock \bibinfo{title}{TestQ}.
\newblock
  \bibinfo{howpublished}{\url{https://code.google.com/archive/p/tsmells/}}.
\newblock
\newblock
\shownote{(Accessed: 03/02/2021).}


\bibitem[\protect\citeauthoryear{??}{Dow}{[n.d.]o}]%
        {Download:TRex}
 \bibinfo{year}{[n.d.]}\natexlab{o}.
\newblock \bibinfo{title}{TRex}.
\newblock
  \bibinfo{howpublished}{\url{https://www.trex.informatik.uni-goettingen.de/trac}}.
\newblock
\newblock
\shownote{(Accessed: 03/02/2021).}


\bibitem[\protect\citeauthoryear{??}{Dow}{[n.d.]p}]%
        {Download:tsDetect2019}
 \bibinfo{year}{[n.d.]}\natexlab{p}.
\newblock \bibinfo{title}{tsDetect}.
\newblock \bibinfo{howpublished}{\url{https://testsmells.github.io/}}.
\newblock
\newblock
\shownote{(Accessed: 03/02/2021).}


\bibitem[\protect\citeauthoryear{Ain, Butt, Anwar, Azam, and Maqbool}{Ain
  et~al\mbox{.}}{2019}]%
        {ain2019systematic}
\bibfield{author}{\bibinfo{person}{Qurat~Ul Ain}, \bibinfo{person}{Wasi~Haider
  Butt}, \bibinfo{person}{Muhammad~Waseem Anwar}, \bibinfo{person}{Farooque
  Azam}, {and} \bibinfo{person}{Bilal Maqbool}.}
  \bibinfo{year}{2019}\natexlab{}.
\newblock \showarticletitle{A systematic review on code clone detection}.
\newblock \bibinfo{journal}{\emph{IEEE Access}}  \bibinfo{volume}{7}
  (\bibinfo{year}{2019}), \bibinfo{pages}{86121--86144}.
\newblock


\bibitem[\protect\citeauthoryear{Akbar and Kak}{Akbar and Kak}{2020}]%
        {akbar2020large}
\bibfield{author}{\bibinfo{person}{Shayan~A Akbar} {and}
  \bibinfo{person}{Avinash~C Kak}.} \bibinfo{year}{2020}\natexlab{}.
\newblock \showarticletitle{A Large-Scale Comparative Evaluation of IR-Based
  Tools for Bug Localization}. In \bibinfo{booktitle}{\emph{Proceedings of the
  17th International Conference on Mining Software Repositories}}.
  \bibinfo{pages}{21--31}.
\newblock


\bibitem[\protect\citeauthoryear{Avgeriou, Taibi, Ampatzoglou, Fontana, Besker,
  Chatzigeorgiou, Lenarduzzi, Martini, Moschou, Pigazzini,
  et~al\mbox{.}}{Avgeriou et~al\mbox{.}}{2020}]%
        {avgeriou2020overview}
\bibfield{author}{\bibinfo{person}{Paris~C Avgeriou}, \bibinfo{person}{Davide
  Taibi}, \bibinfo{person}{Apostolos Ampatzoglou},
  \bibinfo{person}{Francesca~Arcelli Fontana}, \bibinfo{person}{Terese Besker},
  \bibinfo{person}{Alexandros Chatzigeorgiou}, \bibinfo{person}{Valentina
  Lenarduzzi}, \bibinfo{person}{Antonio Martini}, \bibinfo{person}{Nasia
  Moschou}, \bibinfo{person}{Ilaria Pigazzini}, {et~al\mbox{.}}}
  \bibinfo{year}{2020}\natexlab{}.
\newblock \showarticletitle{An overview and comparison of technical debt
  measurement tools}.
\newblock \bibinfo{journal}{\emph{IEEE Software}} (\bibinfo{year}{2020}).
\newblock


\bibitem[\protect\citeauthoryear{Azadi, Fontana, and Taibi}{Azadi
  et~al\mbox{.}}{2019}]%
        {azadi2019architectural}
\bibfield{author}{\bibinfo{person}{Umberto Azadi},
  \bibinfo{person}{Francesca~Arcelli Fontana}, {and} \bibinfo{person}{Davide
  Taibi}.} \bibinfo{year}{2019}\natexlab{}.
\newblock \showarticletitle{Architectural smells detected by tools: a catalogue
  proposal}. In \bibinfo{booktitle}{\emph{2019 IEEE/ACM International
  Conference on Technical Debt (TechDebt)}}. IEEE, \bibinfo{pages}{88--97}.
\newblock


\bibitem[\protect\citeauthoryear{Baker, Evans, Grabowski, Neukirchen, and
  Zeiss}{Baker et~al\mbox{.}}{2006}]%
        {baker2006trex}
\bibfield{author}{\bibinfo{person}{Paul Baker}, \bibinfo{person}{Dominic
  Evans}, \bibinfo{person}{Jens Grabowski}, \bibinfo{person}{Helmut
  Neukirchen}, {and} \bibinfo{person}{Benjamin Zeiss}.}
  \bibinfo{year}{2006}\natexlab{}.
\newblock \showarticletitle{TRex-the refactoring and metrics tool for TTCN-3
  test specifications}. In \bibinfo{booktitle}{\emph{Testing: Academic \&
  Industrial Conference-Practice And Research Techniques (TAIC PART'06)}}.
  IEEE, \bibinfo{pages}{90--94}.
\newblock


\bibitem[\protect\citeauthoryear{Barn, Barat, and Clark}{Barn
  et~al\mbox{.}}{2017}]%
        {Barn2017ISEC}
\bibfield{author}{\bibinfo{person}{Balbir Barn}, \bibinfo{person}{Souvik
  Barat}, {and} \bibinfo{person}{Tony Clark}.} \bibinfo{year}{2017}\natexlab{}.
\newblock \showarticletitle{Conducting Systematic Literature Reviews and
  Systematic Mapping Studies}. In \bibinfo{booktitle}{\emph{Proceedings of the
  10th Innovations in Software Engineering Conference}} (Jaipur, India)
  \emph{(\bibinfo{series}{ISEC '17})}. \bibinfo{publisher}{Association for
  Computing Machinery}, \bibinfo{address}{New York, NY, USA},
  \bibinfo{pages}{212–213}.
\newblock
\showISBNx{9781450348560}


\bibitem[\protect\citeauthoryear{Bavota, Qusef, Oliveto, De~Lucia, and
  Binkley}{Bavota et~al\mbox{.}}{2012}]%
        {bavota2012empirical}
\bibfield{author}{\bibinfo{person}{Gabriele Bavota}, \bibinfo{person}{Abdallah
  Qusef}, \bibinfo{person}{Rocco Oliveto}, \bibinfo{person}{Andrea De~Lucia},
  {and} \bibinfo{person}{David Binkley}.} \bibinfo{year}{2012}\natexlab{}.
\newblock \showarticletitle{An empirical analysis of the distribution of unit
  test smells and their impact on software maintenance}. In
  \bibinfo{booktitle}{\emph{2012 28th IEEE International Conference on Software
  Maintenance (ICSM)}}. IEEE, \bibinfo{pages}{56--65}.
\newblock


\bibitem[\protect\citeauthoryear{Bavota, Qusef, Oliveto, De~Lucia, and
  Binkley}{Bavota et~al\mbox{.}}{2015}]%
        {bavota2015test}
\bibfield{author}{\bibinfo{person}{Gabriele Bavota}, \bibinfo{person}{Abdallah
  Qusef}, \bibinfo{person}{Rocco Oliveto}, \bibinfo{person}{Andrea De~Lucia},
  {and} \bibinfo{person}{Dave Binkley}.} \bibinfo{year}{2015}\natexlab{}.
\newblock \showarticletitle{Are test smells really harmful? An empirical
  study}.
\newblock \bibinfo{journal}{\emph{Empirical Software Engineering}}
  \bibinfo{volume}{20}, \bibinfo{number}{4} (\bibinfo{year}{2015}),
  \bibinfo{pages}{1052--1094}.
\newblock


\bibitem[\protect\citeauthoryear{Bell, Kaiser, Melski, and Dattatreya}{Bell
  et~al\mbox{.}}{2015}]%
        {bell2015efficient}
\bibfield{author}{\bibinfo{person}{Jonathan Bell}, \bibinfo{person}{Gail
  Kaiser}, \bibinfo{person}{Eric Melski}, {and} \bibinfo{person}{Mohan
  Dattatreya}.} \bibinfo{year}{2015}\natexlab{}.
\newblock \showarticletitle{Efficient dependency detection for safe Java test
  acceleration}. In \bibinfo{booktitle}{\emph{Proceedings of the 2015 10th
  Joint Meeting on Foundations of Software Engineering}}.
  \bibinfo{pages}{770--781}.
\newblock


\bibitem[\protect\citeauthoryear{Biagiola, Stocco, Mesbah, Ricca, and
  Tonella}{Biagiola et~al\mbox{.}}{2019}]%
        {biagiola2019web}
\bibfield{author}{\bibinfo{person}{Matteo Biagiola}, \bibinfo{person}{Andrea
  Stocco}, \bibinfo{person}{Ali Mesbah}, \bibinfo{person}{Filippo Ricca}, {and}
  \bibinfo{person}{Paolo Tonella}.} \bibinfo{year}{2019}\natexlab{}.
\newblock \showarticletitle{Web test dependency detection}. In
  \bibinfo{booktitle}{\emph{Proceedings of the 2019 27th ACM Joint Meeting on
  European Software Engineering Conference and Symposium on the Foundations of
  Software Engineering}}. \bibinfo{pages}{154--164}.
\newblock


\bibitem[\protect\citeauthoryear{{Biagiola}, {Stocco}, {Ricca}, and
  {Tonella}}{{Biagiola} et~al\mbox{.}}{2020}]%
        {Biagiola2020ICST}
\bibfield{author}{\bibinfo{person}{M. {Biagiola}}, \bibinfo{person}{A.
  {Stocco}}, \bibinfo{person}{F. {Ricca}}, {and} \bibinfo{person}{P.
  {Tonella}}.} \bibinfo{year}{2020}\natexlab{}.
\newblock \showarticletitle{Dependency-Aware Web Test Generation}. In
  \bibinfo{booktitle}{\emph{2020 IEEE 13th International Conference on Software
  Testing, Validation and Verification (ICST)}}. \bibinfo{pages}{175--185}.
\newblock


\bibitem[\protect\citeauthoryear{Breugelmans and Van~Rompaey}{Breugelmans and
  Van~Rompaey}{2008}]%
        {breugelmans2008testq}
\bibfield{author}{\bibinfo{person}{Manuel Breugelmans} {and}
  \bibinfo{person}{Bart Van~Rompaey}.} \bibinfo{year}{2008}\natexlab{}.
\newblock \showarticletitle{TestQ: Exploring Structural and Maintenance
  Characteristics of Unit Test Suites}. In \bibinfo{booktitle}{\emph{WASDeTT-1:
  1st International Workshop on Advanced Software Development Tools and
  Techniques}}.
\newblock


\bibitem[\protect\citeauthoryear{d.~P.~{Sobrinho}, {De Lucia}, and
  d.~A.~{Maia}}{d.~P.~{Sobrinho} et~al\mbox{.}}{2021}]%
        {Sobrinho2021TSE}
\bibfield{author}{\bibinfo{person}{E.~V. d. P.~{Sobrinho}}, \bibinfo{person}{A.
  {De Lucia}}, {and} \bibinfo{person}{M. d. A.~{Maia}}.}
  \bibinfo{year}{2021}\natexlab{}.
\newblock \showarticletitle{A Systematic Literature Review on Bad Smells–5
  W's: Which, When, What, Who, Where}.
\newblock \bibinfo{journal}{\emph{IEEE Transactions on Software Engineering}}
  \bibinfo{volume}{47}, \bibinfo{number}{1} (\bibinfo{year}{2021}),
  \bibinfo{pages}{17--66}.
\newblock


\bibitem[\protect\citeauthoryear{De~Bleser, Di~Nucci, and De~Roover}{De~Bleser
  et~al\mbox{.}}{2019a}]%
        {de2019assessing}
\bibfield{author}{\bibinfo{person}{Jonas De~Bleser}, \bibinfo{person}{Dario
  Di~Nucci}, {and} \bibinfo{person}{Coen De~Roover}.}
  \bibinfo{year}{2019}\natexlab{a}.
\newblock \showarticletitle{Assessing diffusion and perception of test smells
  in scala projects}. In \bibinfo{booktitle}{\emph{2019 IEEE/ACM 16th
  International Conference on Mining Software Repositories (MSR)}}. IEEE,
  \bibinfo{pages}{457--467}.
\newblock


\bibitem[\protect\citeauthoryear{De~Bleser, Di~Nucci, and De~Roover}{De~Bleser
  et~al\mbox{.}}{2019b}]%
        {de2019socrates}
\bibfield{author}{\bibinfo{person}{Jonas De~Bleser}, \bibinfo{person}{Dario
  Di~Nucci}, {and} \bibinfo{person}{Coen De~Roover}.}
  \bibinfo{year}{2019}\natexlab{b}.
\newblock \showarticletitle{SoCRATES: Scala radar for test smells}. In
  \bibinfo{booktitle}{\emph{Proceedings of the Tenth ACM SIGPLAN Symposium on
  Scala}}. \bibinfo{pages}{22--26}.
\newblock


\bibitem[\protect\citeauthoryear{{Delplanque}, {Ducasse}, {Polito}, {Black},
  and {Etien}}{{Delplanque} et~al\mbox{.}}{2019}]%
        {Delplanque2019ICSE}
\bibfield{author}{\bibinfo{person}{J. {Delplanque}}, \bibinfo{person}{S.
  {Ducasse}}, \bibinfo{person}{G. {Polito}}, \bibinfo{person}{A.~P. {Black}},
  {and} \bibinfo{person}{A. {Etien}}.} \bibinfo{year}{2019}\natexlab{}.
\newblock \showarticletitle{Rotten Green Tests}. In
  \bibinfo{booktitle}{\emph{2019 IEEE/ACM 41st International Conference on
  Software Engineering (ICSE)}}. \bibinfo{pages}{500--511}.
\newblock
\urldef\tempurl%
\url{https://doi.org/10.1109/ICSE.2019.00062}
\showDOI{\tempurl}


\bibitem[\protect\citeauthoryear{Fernandes, Oliveira, Vale, Paiva, and
  Figueiredo}{Fernandes et~al\mbox{.}}{2016}]%
        {fernandes2016review}
\bibfield{author}{\bibinfo{person}{Eduardo Fernandes},
  \bibinfo{person}{Johnatan Oliveira}, \bibinfo{person}{Gustavo Vale},
  \bibinfo{person}{Thanis Paiva}, {and} \bibinfo{person}{Eduardo Figueiredo}.}
  \bibinfo{year}{2016}\natexlab{}.
\newblock \showarticletitle{A review-based comparative study of bad smell
  detection tools}. In \bibinfo{booktitle}{\emph{Proceedings of the 20th
  International Conference on Evaluation and Assessment in Software
  Engineering}}. \bibinfo{pages}{1--12}.
\newblock


\bibitem[\protect\citeauthoryear{{Fraser}, {Gambi}, and {Rojas}}{{Fraser}
  et~al\mbox{.}}{2020}]%
        {Fraser2020ICSTW}
\bibfield{author}{\bibinfo{person}{G. {Fraser}}, \bibinfo{person}{A. {Gambi}},
  {and} \bibinfo{person}{J.~M. {Rojas}}.} \bibinfo{year}{2020}\natexlab{}.
\newblock \showarticletitle{Teaching Software Testing with the Code Defenders
  Testing Game: Experiences and Improvements}. In
  \bibinfo{booktitle}{\emph{2020 IEEE International Conference on Software
  Testing, Verification and Validation Workshops (ICSTW)}}.
  \bibinfo{pages}{461--464}.
\newblock
\urldef\tempurl%
\url{https://doi.org/10.1109/ICSTW50294.2020.00082}
\showDOI{\tempurl}


\bibitem[\protect\citeauthoryear{Gambi, Bell, and Zeller}{Gambi
  et~al\mbox{.}}{2018}]%
        {gambi2018practical}
\bibfield{author}{\bibinfo{person}{Alessio Gambi}, \bibinfo{person}{Jonathan
  Bell}, {and} \bibinfo{person}{Andreas Zeller}.}
  \bibinfo{year}{2018}\natexlab{}.
\newblock \showarticletitle{Practical test dependency detection}. In
  \bibinfo{booktitle}{\emph{2018 IEEE 11th International Conference on Software
  Testing, Verification and Validation (ICST)}}. IEEE, \bibinfo{pages}{1--11}.
\newblock


\bibitem[\protect\citeauthoryear{{Garousi}, { K{\"u}{\c{c}}{\"u}k}, and
  {Felderer}}{{Garousi} et~al\mbox{.}}{2019}]%
        {garousi2019smells}
\bibfield{author}{\bibinfo{person}{V. {Garousi}}, \bibinfo{person}{B. {
  K{\"u}{\c{c}}{\"u}k}}, {and} \bibinfo{person}{M. {Felderer}}.}
  \bibinfo{year}{2019}\natexlab{}.
\newblock \showarticletitle{What We Know About Smells in Software Test Code}.
\newblock \bibinfo{journal}{\emph{IEEE Software}} \bibinfo{volume}{36},
  \bibinfo{number}{3} (\bibinfo{year}{2019}), \bibinfo{pages}{61--73}.
\newblock


\bibitem[\protect\citeauthoryear{Garousi and K{\"u}{\c{c}}{\"u}k}{Garousi and
  K{\"u}{\c{c}}{\"u}k}{2018}]%
        {garousi2018smells}
\bibfield{author}{\bibinfo{person}{Vahid Garousi} {and}
  \bibinfo{person}{Bar{\i}{\c{s}} K{\"u}{\c{c}}{\"u}k}.}
  \bibinfo{year}{2018}\natexlab{}.
\newblock \showarticletitle{Smells in software test code: A survey of knowledge
  in industry and academia}.
\newblock \bibinfo{journal}{\emph{Journal of systems and software}}
  (\bibinfo{year}{2018}).
\newblock


\bibitem[\protect\citeauthoryear{Garousi and M{\"a}ntyl{\"a}}{Garousi and
  M{\"a}ntyl{\"a}}{2016}]%
        {garousi2016systematic}
\bibfield{author}{\bibinfo{person}{Vahid Garousi} {and} \bibinfo{person}{Mika~V
  M{\"a}ntyl{\"a}}.} \bibinfo{year}{2016}\natexlab{}.
\newblock \showarticletitle{A systematic literature review of literature
  reviews in software testing}.
\newblock \bibinfo{journal}{\emph{Information and Software Technology}}
  (\bibinfo{year}{2016}), \bibinfo{pages}{195--216}.
\newblock


\bibitem[\protect\citeauthoryear{Grano, Palomba, Di~Nucci, De~Lucia, and
  Gall}{Grano et~al\mbox{.}}{2019}]%
        {grano2019scented}
\bibfield{author}{\bibinfo{person}{Giovanni Grano}, \bibinfo{person}{Fabio
  Palomba}, \bibinfo{person}{Dario Di~Nucci}, \bibinfo{person}{Andrea
  De~Lucia}, {and} \bibinfo{person}{Harald~C Gall}.}
  \bibinfo{year}{2019}\natexlab{}.
\newblock \showarticletitle{Scented since the beginning: On the diffuseness of
  test smells in automatically generated test code}.
\newblock \bibinfo{journal}{\emph{Journal of Systems and Software}}
  \bibinfo{volume}{156} (\bibinfo{year}{2019}).
\newblock


\bibitem[\protect\citeauthoryear{{Grano}, {Palomba}, and {Gall}}{{Grano}
  et~al\mbox{.}}{2019}]%
        {Grano2019TSE}
\bibfield{author}{\bibinfo{person}{G. {Grano}}, \bibinfo{person}{F. {Palomba}},
  {and} \bibinfo{person}{H.~C. {Gall}}.} \bibinfo{year}{2019}\natexlab{}.
\newblock \showarticletitle{Lightweight Assessment of Test-Case Effectiveness
  using Source-Code-Quality Indicators}.
\newblock \bibinfo{journal}{\emph{IEEE Transactions on Software Engineering}}
  (\bibinfo{year}{2019}), \bibinfo{pages}{1--1}.
\newblock
\urldef\tempurl%
\url{https://doi.org/10.1109/TSE.2019.2903057}
\showDOI{\tempurl}


\bibitem[\protect\citeauthoryear{Greiler, Van~Deursen, and Storey}{Greiler
  et~al\mbox{.}}{2013a}]%
        {greiler2013automated}
\bibfield{author}{\bibinfo{person}{Michaela Greiler}, \bibinfo{person}{Arie
  Van~Deursen}, {and} \bibinfo{person}{Margaret-Anne Storey}.}
  \bibinfo{year}{2013}\natexlab{a}.
\newblock \showarticletitle{Automated detection of test fixture strategies and
  smells}. In \bibinfo{booktitle}{\emph{2013 IEEE Sixth International
  Conference on Software Testing, Verification and Validation}}. IEEE,
  \bibinfo{pages}{322--331}.
\newblock


\bibitem[\protect\citeauthoryear{Greiler, Zaidman, Van~Deursen, and
  Storey}{Greiler et~al\mbox{.}}{2013b}]%
        {greiler2013strategies}
\bibfield{author}{\bibinfo{person}{Michaela Greiler}, \bibinfo{person}{Andy
  Zaidman}, \bibinfo{person}{Arie Van~Deursen}, {and}
  \bibinfo{person}{Margaret-Anne Storey}.} \bibinfo{year}{2013}\natexlab{b}.
\newblock \showarticletitle{Strategies for avoiding text fixture smells during
  software evolution}. In \bibinfo{booktitle}{\emph{2013 10th Working
  Conference on Mining Software Repositories (MSR)}}. IEEE,
  \bibinfo{pages}{387--396}.
\newblock


\bibitem[\protect\citeauthoryear{Gyori, Shi, Hariri, and Marinov}{Gyori
  et~al\mbox{.}}{2015}]%
        {gyori2015reliable}
\bibfield{author}{\bibinfo{person}{Alex Gyori}, \bibinfo{person}{August Shi},
  \bibinfo{person}{Farah Hariri}, {and} \bibinfo{person}{Darko Marinov}.}
  \bibinfo{year}{2015}\natexlab{}.
\newblock \showarticletitle{Reliable testing: Detecting state-polluting tests
  to prevent test dependency}. In \bibinfo{booktitle}{\emph{Proceedings of the
  2015 International Symposium on Software Testing and Analysis}}.
  \bibinfo{pages}{223--233}.
\newblock


\bibitem[\protect\citeauthoryear{Huo and Clause}{Huo and Clause}{2014}]%
        {huo2014improving}
\bibfield{author}{\bibinfo{person}{Chen Huo} {and} \bibinfo{person}{James
  Clause}.} \bibinfo{year}{2014}\natexlab{}.
\newblock \showarticletitle{Improving oracle quality by detecting brittle
  assertions and unused inputs in tests}. In
  \bibinfo{booktitle}{\emph{Proceedings of the 22nd ACM SIGSOFT International
  Symposium on Foundations of Software Engineering}}.
  \bibinfo{pages}{621--631}.
\newblock


\bibitem[\protect\citeauthoryear{{Kim}}{{Kim}}{2020}]%
        {Kim2020ICSE}
\bibfield{author}{\bibinfo{person}{D.~J. {Kim}}.}
  \bibinfo{year}{2020}\natexlab{}.
\newblock \showarticletitle{An Empirical Study on the Evolution of Test Smell}.
  In \bibinfo{booktitle}{\emph{2020 IEEE/ACM 42nd International Conference on
  Software Engineering: Companion Proceedings (ICSE-Companion)}}.
  \bibinfo{pages}{149--151}.
\newblock


\bibitem[\protect\citeauthoryear{Koochakzadeh and Garousi}{Koochakzadeh and
  Garousi}{2010a}]%
        {koochakzadeh2010tecrevis}
\bibfield{author}{\bibinfo{person}{Negar Koochakzadeh} {and}
  \bibinfo{person}{Vahid Garousi}.} \bibinfo{year}{2010}\natexlab{a}.
\newblock \showarticletitle{Tecrevis: a tool for test coverage and test
  redundancy visualization}. In \bibinfo{booktitle}{\emph{International
  Academic and Industrial Conference on Practice and Research Techniques}}.
  Springer, \bibinfo{pages}{129--136}.
\newblock


\bibitem[\protect\citeauthoryear{Koochakzadeh and Garousi}{Koochakzadeh and
  Garousi}{2010b}]%
        {koochakzadeh2010tester}
\bibfield{author}{\bibinfo{person}{Negar Koochakzadeh} {and}
  \bibinfo{person}{Vahid Garousi}.} \bibinfo{year}{2010}\natexlab{b}.
\newblock \showarticletitle{A tester-assisted methodology for test redundancy
  detection}.
\newblock \bibinfo{journal}{\emph{Advances in Software Engineering}}
  \bibinfo{volume}{2010} (\bibinfo{year}{2010}).
\newblock


\bibitem[\protect\citeauthoryear{Lacerda, Petrillo, Pimenta, and
  Guéhéneuc}{Lacerda et~al\mbox{.}}{2020}]%
        {Lacerda2020JSS}
\bibfield{author}{\bibinfo{person}{Guilherme Lacerda}, \bibinfo{person}{Fabio
  Petrillo}, \bibinfo{person}{Marcelo Pimenta}, {and}
  \bibinfo{person}{Yann~Gaël Guéhéneuc}.} \bibinfo{year}{2020}\natexlab{}.
\newblock \showarticletitle{Code smells and refactoring: A tertiary systematic
  review of challenges and observations}.
\newblock \bibinfo{journal}{\emph{Journal of Systems and Software}}
  \bibinfo{volume}{167} (\bibinfo{year}{2020}), \bibinfo{pages}{110610}.
\newblock
\showISSN{0164-1212}
\urldef\tempurl%
\url{https://www.sciencedirect.com/science/article/pii/S0164121220300881}
\showURL{%
\tempurl}


\bibitem[\protect\citeauthoryear{Lambiase, Cupito, Pecorelli, De~Lucia, and
  Palomba}{Lambiase et~al\mbox{.}}{2020}]%
        {lambiase2020just}
\bibfield{author}{\bibinfo{person}{Stefano Lambiase}, \bibinfo{person}{Andrea
  Cupito}, \bibinfo{person}{Fabiano Pecorelli}, \bibinfo{person}{Andrea
  De~Lucia}, {and} \bibinfo{person}{Fabio Palomba}.}
  \bibinfo{year}{2020}\natexlab{}.
\newblock \showarticletitle{Just-In-Time Test Smell Detection and Refactoring:
  The DARTS Project}. In \bibinfo{booktitle}{\emph{Proceedings of the 28th
  International Conference on Program Comprehension}}.
  \bibinfo{pages}{441--445}.
\newblock


\bibitem[\protect\citeauthoryear{Luo, Hariri, Eloussi, and Marinov}{Luo
  et~al\mbox{.}}{2014}]%
        {Luo2014FSE}
\bibfield{author}{\bibinfo{person}{Qingzhou Luo}, \bibinfo{person}{Farah
  Hariri}, \bibinfo{person}{Lamyaa Eloussi}, {and} \bibinfo{person}{Darko
  Marinov}.} \bibinfo{year}{2014}\natexlab{}.
\newblock \showarticletitle{An Empirical Analysis of Flaky Tests}. In
  \bibinfo{booktitle}{\emph{Proceedings of the 22nd ACM SIGSOFT International
  Symposium on Foundations of Software Engineering}} (Hong Kong, China)
  \emph{(\bibinfo{series}{FSE 2014})}. \bibinfo{publisher}{Association for
  Computing Machinery}, \bibinfo{address}{New York, NY, USA}.
\newblock
\showISBNx{9781450330565}


\bibitem[\protect\citeauthoryear{Martinez, Etien, Ducasse, and
  Fuhrman}{Martinez et~al\mbox{.}}{2020}]%
        {Martinez2020ICSE}
\bibfield{author}{\bibinfo{person}{Matias Martinez}, \bibinfo{person}{Anne
  Etien}, \bibinfo{person}{St\'{e}phane Ducasse}, {and}
  \bibinfo{person}{Christopher Fuhrman}.} \bibinfo{year}{2020}\natexlab{}.
\newblock \showarticletitle{RTj: A Java Framework for Detecting and Refactoring
  Rotten Green Test Cases}. In \bibinfo{booktitle}{\emph{Proceedings of the
  ACM/IEEE 42nd International Conference on Software Engineering: Companion
  Proceedings}} (Seoul, South Korea) \emph{(\bibinfo{series}{ICSE '20})}.
  \bibinfo{publisher}{Association for Computing Machinery},
  \bibinfo{address}{New York, NY, USA}, \bibinfo{pages}{69–72}.
\newblock
\showISBNx{9781450371223}


\bibitem[\protect\citeauthoryear{Meszaros}{Meszaros}{2007}]%
        {meszaros2007xunit}
\bibfield{author}{\bibinfo{person}{Gerard Meszaros}.}
  \bibinfo{year}{2007}\natexlab{}.
\newblock \bibinfo{booktitle}{\emph{xUnit test patterns: Refactoring test
  code}}.
\newblock \bibinfo{publisher}{Pearson Education}.
\newblock


\bibitem[\protect\citeauthoryear{Neukirchen and Bisanz}{Neukirchen and
  Bisanz}{2007}]%
        {Helmut2007TRex}
\bibfield{author}{\bibinfo{person}{Helmut Neukirchen} {and}
  \bibinfo{person}{Martin Bisanz}.} \bibinfo{year}{2007}\natexlab{}.
\newblock \showarticletitle{Utilising Code Smells to Detect Quality Problems in
  TTCN-3 Test Suites}. In \bibinfo{booktitle}{\emph{Testing of Software and
  Communicating Systems}}, \bibfield{editor}{\bibinfo{person}{Alexandre
  Petrenko}, \bibinfo{person}{Margus Veanes}, \bibinfo{person}{Jan Tretmans},
  {and} \bibinfo{person}{Wolfgang Grieskamp}} (Eds.).
  \bibinfo{publisher}{Springer Berlin Heidelberg}, \bibinfo{address}{Berlin,
  Heidelberg}, \bibinfo{pages}{228--243}.
\newblock
\showISBNx{978-3-540-73066-8}


\bibitem[\protect\citeauthoryear{Neukirchen, Zeiss, and Grabowski}{Neukirchen
  et~al\mbox{.}}{2008}]%
        {Neukirchen2008TTCN}
\bibfield{author}{\bibinfo{person}{Helmut Neukirchen},
  \bibinfo{person}{Benjamin Zeiss}, {and} \bibinfo{person}{Jens Grabowski}.}
  \bibinfo{year}{2008}\natexlab{}.
\newblock \showarticletitle{An approach to quality engineering of TTCN-3 test
  specifications}.
\newblock \bibinfo{journal}{\emph{International Journal on Software Tools for
  Technology Transfer}} \bibinfo{number}{4} (\bibinfo{year}{2008}).
\newblock


\bibitem[\protect\citeauthoryear{Palomba, Di~Nucci, Panichella, Oliveto, and
  De~Lucia}{Palomba et~al\mbox{.}}{2016}]%
        {palomba2016diffusion}
\bibfield{author}{\bibinfo{person}{Fabio Palomba}, \bibinfo{person}{Dario
  Di~Nucci}, \bibinfo{person}{Annibale Panichella}, \bibinfo{person}{Rocco
  Oliveto}, {and} \bibinfo{person}{Andrea De~Lucia}.}
  \bibinfo{year}{2016}\natexlab{}.
\newblock \showarticletitle{On the diffusion of test smells in automatically
  generated test code: An empirical study}. In \bibinfo{booktitle}{\emph{2016
  IEEE/ACM 9th International Workshop on Search-Based Software Testing
  (SBST)}}. IEEE, \bibinfo{pages}{5--14}.
\newblock


\bibitem[\protect\citeauthoryear{Palomba, Zaidman, and De~Lucia}{Palomba
  et~al\mbox{.}}{2018}]%
        {palomba2018automatic}
\bibfield{author}{\bibinfo{person}{Fabio Palomba}, \bibinfo{person}{Andy
  Zaidman}, {and} \bibinfo{person}{Andrea De~Lucia}.}
  \bibinfo{year}{2018}\natexlab{}.
\newblock \showarticletitle{Automatic test smell detection using information
  retrieval techniques}. In \bibinfo{booktitle}{\emph{2018 IEEE International
  Conference on Software Maintenance and Evolution (ICSME)}}. IEEE,
  \bibinfo{pages}{311--322}.
\newblock


\bibitem[\protect\citeauthoryear{{Panichella}, {Panichella}, {Fraser},
  {Sawant}, and {Hellendoorn}}{{Panichella} et~al\mbox{.}}{2020}]%
        {Panichella2020ICSME}
\bibfield{author}{\bibinfo{person}{A. {Panichella}}, \bibinfo{person}{S.
  {Panichella}}, \bibinfo{person}{G. {Fraser}}, \bibinfo{person}{A.~A.
  {Sawant}}, {and} \bibinfo{person}{V.~J. {Hellendoorn}}.}
  \bibinfo{year}{2020}\natexlab{}.
\newblock \showarticletitle{Revisiting Test Smells in Automatically Generated
  Tests: Limitations, Pitfalls, and Opportunities}. In
  \bibinfo{booktitle}{\emph{2020 IEEE International Conference on Software
  Maintenance and Evolution (ICSME)}}. \bibinfo{pages}{523--533}.
\newblock
\urldef\tempurl%
\url{https://doi.org/10.1109/ICSME46990.2020.00056}
\showDOI{\tempurl}


\bibitem[\protect\citeauthoryear{Parkkila, Ikonen, and Porras}{Parkkila
  et~al\mbox{.}}{2015}]%
        {Parkkila2015CSR}
\bibfield{author}{\bibinfo{person}{Janne Parkkila}, \bibinfo{person}{Jouni
  Ikonen}, {and} \bibinfo{person}{Jari Porras}.}
  \bibinfo{year}{2015}\natexlab{}.
\newblock \showarticletitle{Where is the research on connecting game
  worlds?—A systematic mapping study}.
\newblock \bibinfo{journal}{\emph{Computer Science Review}}
  \bibinfo{volume}{18} (\bibinfo{year}{2015}), \bibinfo{pages}{46--58}.
\newblock
\showISSN{1574-0137}
\urldef\tempurl%
\url{https://doi.org/10.1016/j.cosrev.2015.10.002}
\showDOI{\tempurl}


\bibitem[\protect\citeauthoryear{Pecorelli, Di~Lillo, Palomba, and
  De~Lucia}{Pecorelli et~al\mbox{.}}{2020}]%
        {Pecorelli2020AVI}
\bibfield{author}{\bibinfo{person}{Fabiano Pecorelli},
  \bibinfo{person}{Gianluca Di~Lillo}, \bibinfo{person}{Fabio Palomba}, {and}
  \bibinfo{person}{Andrea De~Lucia}.} \bibinfo{year}{2020}\natexlab{}.
\newblock \showarticletitle{VITRuM: A Plug-In for the Visualization of
  Test-Related Metrics}. In \bibinfo{booktitle}{\emph{Proceedings of the
  International Conference on Advanced Visual Interfaces}} (Salerno, Italy)
  \emph{(\bibinfo{series}{AVI '20})}. \bibinfo{publisher}{Association for
  Computing Machinery}, \bibinfo{address}{New York, NY, USA}, Article
  \bibinfo{articleno}{101}.
\newblock
\showISBNx{9781450375351}


\bibitem[\protect\citeauthoryear{Peruma, Almalki, Newman, Mkaouer, Ouni, and
  Palomba}{Peruma et~al\mbox{.}}{2019}]%
        {peruma2019CASCON}
\bibfield{author}{\bibinfo{person}{Anthony Peruma}, \bibinfo{person}{Khalid
  Almalki}, \bibinfo{person}{Christian~D. Newman},
  \bibinfo{person}{Mohamed~Wiem Mkaouer}, \bibinfo{person}{Ali Ouni}, {and}
  \bibinfo{person}{Fabio Palomba}.} \bibinfo{year}{2019}\natexlab{}.
\newblock \showarticletitle{On the Distribution of Test Smells in Open Source
  Android Applications: An Exploratory Study}. In
  \bibinfo{booktitle}{\emph{Proceedings of the 29th Annual International
  Conference on Computer Science and Software Engineering}} (Toronto, Ontario,
  Canada) \emph{(\bibinfo{series}{CASCON ’19})}. \bibinfo{publisher}{IBM
  Corp.}, \bibinfo{address}{USA}, \bibinfo{pages}{193–202}.
\newblock


\bibitem[\protect\citeauthoryear{Peruma, Almalki, Newman, Mkaouer, Ouni, and
  Palomba}{Peruma et~al\mbox{.}}{2020a}]%
        {peruma2020FSE}
\bibfield{author}{\bibinfo{person}{Anthony Peruma}, \bibinfo{person}{Khalid
  Almalki}, \bibinfo{person}{Christian~D. Newman},
  \bibinfo{person}{Mohamed~Wiem Mkaouer}, \bibinfo{person}{Ali Ouni}, {and}
  \bibinfo{person}{Fabio Palomba}.} \bibinfo{year}{2020}\natexlab{a}.
\newblock \showarticletitle{TsDetect: An Open Source Test Smells Detection
  Tool}. In \bibinfo{booktitle}{\emph{Proceedings of the 28th ACM Joint Meeting
  on European Software Engineering Conference and Symposium on the Foundations
  of Software Engineering}} (Virtual Event, USA)
  \emph{(\bibinfo{series}{ESEC/FSE 2020})}. \bibinfo{publisher}{Association for
  Computing Machinery}, \bibinfo{address}{New York, NY, USA}, 5.
\newblock
\showISBNx{9781450370431}
\urldef\tempurl%
\url{https://doi.org/10.1145/3368089.3417921}
\showDOI{\tempurl}


\bibitem[\protect\citeauthoryear{Peruma, Newman, Mkaouer, Ouni, and
  Palomba}{Peruma et~al\mbox{.}}{2020b}]%
        {Peruma2020IWoR}
\bibfield{author}{\bibinfo{person}{Anthony Peruma},
  \bibinfo{person}{Christian~D. Newman}, \bibinfo{person}{Mohamed~Wiem
  Mkaouer}, \bibinfo{person}{Ali Ouni}, {and} \bibinfo{person}{Fabio Palomba}.}
  \bibinfo{year}{2020}\natexlab{b}.
\newblock \showarticletitle{An Exploratory Study on the Refactoring of Unit
  Test Files in Android Applications}. In \bibinfo{booktitle}{\emph{Proceedings
  of the IEEE/ACM 42nd International Conference on Software Engineering
  Workshops}} (Seoul, Republic of Korea) \emph{(\bibinfo{series}{ICSEW'20})}.
  \bibinfo{publisher}{Association for Computing Machinery},
  \bibinfo{address}{New York, NY, USA}, \bibinfo{pages}{350–357}.
\newblock
\showISBNx{9781450379632}
\urldef\tempurl%
\url{https://doi.org/10.1145/3387940.3392189}
\showDOI{\tempurl}


\bibitem[\protect\citeauthoryear{Petersen, Feldt, Mujtaba, and
  Mattsson}{Petersen et~al\mbox{.}}{2008}]%
        {Petersen2008EASE}
\bibfield{author}{\bibinfo{person}{Kai Petersen}, \bibinfo{person}{Robert
  Feldt}, \bibinfo{person}{Shahid Mujtaba}, {and} \bibinfo{person}{Michael
  Mattsson}.} \bibinfo{year}{2008}\natexlab{}.
\newblock \showarticletitle{Systematic Mapping Studies in Software
  Engineering}. In \bibinfo{booktitle}{\emph{Proceedings of the 12th
  International Conference on Evaluation and Assessment in Software
  Engineering}} (Italy) \emph{(\bibinfo{series}{EASE'08})}.
  \bibinfo{publisher}{BCS Learning \& Development Ltd.},
  \bibinfo{address}{Swindon, GBR}, \bibinfo{pages}{68–77}.
\newblock


\bibitem[\protect\citeauthoryear{Pressman and Bruce R.~Maxim}{Pressman and
  Bruce R.~Maxim}{2014}]%
        {pressman2014software}
\bibfield{author}{\bibinfo{person}{R.S. Pressman} {and} \bibinfo{person}{D.
  Bruce R.~Maxim}.} \bibinfo{year}{2014}\natexlab{}.
\newblock \bibinfo{booktitle}{\emph{Software Engineering: A Practitioner's
  Approach}}.
\newblock \bibinfo{publisher}{McGraw-Hill Education}.
\newblock
\showISBNx{9780078022128}
\showLCCN{2013035493}


\bibitem[\protect\citeauthoryear{Qusef, Elish, and Binkley}{Qusef
  et~al\mbox{.}}{2019}]%
        {qusef2019exploratory}
\bibfield{author}{\bibinfo{person}{Abdallah Qusef}, \bibinfo{person}{Mahmoud~O
  Elish}, {and} \bibinfo{person}{David Binkley}.}
  \bibinfo{year}{2019}\natexlab{}.
\newblock \showarticletitle{An Exploratory Study of the Relationship Between
  Software Test Smells and Fault-Proneness}.
\newblock \bibinfo{journal}{\emph{IEEE Access}}  \bibinfo{volume}{7}
  (\bibinfo{year}{2019}), \bibinfo{pages}{139526--139536}.
\newblock


\bibitem[\protect\citeauthoryear{Reichhart, G{\^\i}rba, and Ducasse}{Reichhart
  et~al\mbox{.}}{2007}]%
        {reichhart2007rule}
\bibfield{author}{\bibinfo{person}{Stefan Reichhart}, \bibinfo{person}{Tudor
  G{\^\i}rba}, {and} \bibinfo{person}{St{\'e}phane Ducasse}.}
  \bibinfo{year}{2007}\natexlab{}.
\newblock \showarticletitle{Rule-based Assessment of Test Quality.}
\newblock \bibinfo{journal}{\emph{Journal of Object Technology}}
  \bibinfo{volume}{6}, \bibinfo{number}{9} (\bibinfo{year}{2007}),
  \bibinfo{pages}{231--251}.
\newblock


\bibitem[\protect\citeauthoryear{Santana, Martins, Rocha, Virg\'{\i}nio, Cruz,
  Costa, and Machado}{Santana et~al\mbox{.}}{2020}]%
        {Santana2020SBES}
\bibfield{author}{\bibinfo{person}{Railana Santana}, \bibinfo{person}{Luana
  Martins}, \bibinfo{person}{Larissa Rocha}, \bibinfo{person}{T\'{a}ssio
  Virg\'{\i}nio}, \bibinfo{person}{Adriana Cruz}, \bibinfo{person}{Heitor
  Costa}, {and} \bibinfo{person}{Ivan Machado}.}
  \bibinfo{year}{2020}\natexlab{}.
\newblock \showarticletitle{RAIDE: A Tool for Assertion Roulette and Duplicate
  Assert Identification and Refactoring}. In
  \bibinfo{booktitle}{\emph{Proceedings of the 34th Brazilian Symposium on
  Software Engineering}} (Natal, Brazil) \emph{(\bibinfo{series}{SBES '20})}.
  \bibinfo{publisher}{Association for Computing Machinery},
  \bibinfo{address}{New York, NY, USA}, \bibinfo{pages}{374–379}.
\newblock
\showISBNx{9781450387538}


\bibitem[\protect\citeauthoryear{Schvarcbacher, Spadini, Bruntink, and
  Oprescu}{Schvarcbacher et~al\mbox{.}}{2019}]%
        {schvarcbacher2019investigating}
\bibfield{author}{\bibinfo{person}{Martin Schvarcbacher},
  \bibinfo{person}{Davide Spadini}, \bibinfo{person}{Magiel Bruntink}, {and}
  \bibinfo{person}{Ana Oprescu}.} \bibinfo{year}{2019}\natexlab{}.
\newblock \showarticletitle{Investigating developer perception on test smells
  using better code hub-Work in progress}. In \bibinfo{booktitle}{\emph{2019
  Seminar Series on Advanced Techniques and Tools for Software Evolution,
  SATTOSE}}.
\newblock


\bibitem[\protect\citeauthoryear{Schwartz, Avgerinos, and Brumley}{Schwartz
  et~al\mbox{.}}{2010}]%
        {schwartz2010all}
\bibfield{author}{\bibinfo{person}{Edward~J Schwartz},
  \bibinfo{person}{Thanassis Avgerinos}, {and} \bibinfo{person}{David
  Brumley}.} \bibinfo{year}{2010}\natexlab{}.
\newblock \showarticletitle{All you ever wanted to know about dynamic taint
  analysis and forward symbolic execution (but might have been afraid to ask)}.
  In \bibinfo{booktitle}{\emph{2010 IEEE symposium on Security and privacy}}.
  IEEE, \bibinfo{pages}{317--331}.
\newblock


\bibitem[\protect\citeauthoryear{Sharma and Spinellis}{Sharma and
  Spinellis}{2018}]%
        {sharma2018survey}
\bibfield{author}{\bibinfo{person}{Tushar Sharma} {and}
  \bibinfo{person}{Diomidis Spinellis}.} \bibinfo{year}{2018}\natexlab{}.
\newblock \showarticletitle{A survey on software smells}.
\newblock \bibinfo{journal}{\emph{Journal of Systems and Software}}
  \bibinfo{volume}{138} (\bibinfo{year}{2018}), \bibinfo{pages}{158--173}.
\newblock


\bibitem[\protect\citeauthoryear{{Siegmund}, {Siegmund}, and {Apel}}{{Siegmund}
  et~al\mbox{.}}{2015}]%
        {Siegmund2015ICSE}
\bibfield{author}{\bibinfo{person}{J. {Siegmund}}, \bibinfo{person}{N.
  {Siegmund}}, {and} \bibinfo{person}{S. {Apel}}.}
  \bibinfo{year}{2015}\natexlab{}.
\newblock \showarticletitle{Views on Internal and External Validity in
  Empirical Software Engineering}. In \bibinfo{booktitle}{\emph{2015 IEEE/ACM
  37th IEEE International Conference on Software Engineering}}.
\newblock
\urldef\tempurl%
\url{https://doi.org/10.1109/ICSE.2015.24}
\showDOI{\tempurl}


\bibitem[\protect\citeauthoryear{{Silva}, {Silva}, {De Souza Santos}, {Terra},
  and {Valente}}{{Silva} et~al\mbox{.}}{2020}]%
        {Silva2020TSE}
\bibfield{author}{\bibinfo{person}{D. {Silva}}, \bibinfo{person}{J. {Silva}},
  \bibinfo{person}{G.~J. {De Souza Santos}}, \bibinfo{person}{R. {Terra}},
  {and} \bibinfo{person}{M.~T.~O. {Valente}}.} \bibinfo{year}{2020}\natexlab{}.
\newblock \showarticletitle{RefDiff 2.0: A Multi-language Refactoring Detection
  Tool}.
\newblock \bibinfo{journal}{\emph{IEEE Transactions on Software Engineering}}
  (\bibinfo{year}{2020}), \bibinfo{pages}{1--1}.
\newblock
\urldef\tempurl%
\url{https://doi.org/10.1109/TSE.2020.2968072}
\showDOI{\tempurl}


\bibitem[\protect\citeauthoryear{Soares, Ribeiro, Amaral, Gheyi, Fernandes,
  Garcia, Fonseca, and Santos}{Soares et~al\mbox{.}}{2020}]%
        {Soares2020SAST}
\bibfield{author}{\bibinfo{person}{Elvys Soares}, \bibinfo{person}{M\'{a}rcio
  Ribeiro}, \bibinfo{person}{Guilherme Amaral}, \bibinfo{person}{Rohit Gheyi},
  \bibinfo{person}{Leo Fernandes}, \bibinfo{person}{Alessandro Garcia},
  \bibinfo{person}{Baldoino Fonseca}, {and} \bibinfo{person}{Andr\'{e}
  Santos}.} \bibinfo{year}{2020}\natexlab{}.
\newblock \showarticletitle{Refactoring Test Smells: A Perspective from
  Open-Source Developers}. In \bibinfo{booktitle}{\emph{Proceedings of the 5th
  Brazilian Symposium on Systematic and Automated Software Testing}} (Natal,
  Brazil) \emph{(\bibinfo{series}{SAST 20})}. \bibinfo{publisher}{Association
  for Computing Machinery}, \bibinfo{address}{New York, NY, USA},
  \bibinfo{pages}{50–59}.
\newblock
\showISBNx{9781450387552}


\bibitem[\protect\citeauthoryear{{Spadini}, {Palomba}, {Zaidman}, {Bruntink},
  and {Bacchelli}}{{Spadini} et~al\mbox{.}}{2018}]%
        {Spadini2018ICSME}
\bibfield{author}{\bibinfo{person}{D. {Spadini}}, \bibinfo{person}{F.
  {Palomba}}, \bibinfo{person}{A. {Zaidman}}, \bibinfo{person}{M. {Bruntink}},
  {and} \bibinfo{person}{A. {Bacchelli}}.} \bibinfo{year}{2018}\natexlab{}.
\newblock \showarticletitle{On the Relation of Test Smells to Software Code
  Quality}. In \bibinfo{booktitle}{\emph{2018 IEEE International Conference on
  Software Maintenance and Evolution (ICSME)}}. \bibinfo{pages}{1--12}.
\newblock
\showISSN{1063-6773}


\bibitem[\protect\citeauthoryear{Spadini, Schvarcbacher, Oprescu, Bruntink, and
  Bacchelli}{Spadini et~al\mbox{.}}{2020}]%
        {Spadini2020MSR}
\bibfield{author}{\bibinfo{person}{Davide Spadini}, \bibinfo{person}{Martin
  Schvarcbacher}, \bibinfo{person}{Ana-Maria Oprescu}, \bibinfo{person}{Magiel
  Bruntink}, {and} \bibinfo{person}{Alberto Bacchelli}.}
  \bibinfo{year}{2020}\natexlab{}.
\newblock \showarticletitle{Investigating Severity Thresholds for Test Smells}.
  In \bibinfo{booktitle}{\emph{Proceedings of the 17th International Conference
  on Mining Software Repositories}} (Seoul, Republic of Korea)
  \emph{(\bibinfo{series}{MSR '20})}. \bibinfo{publisher}{Association for
  Computing Machinery}, \bibinfo{address}{New York, NY, USA},
  \bibinfo{pages}{311–321}.
\newblock
\showISBNx{9781450375177}
\urldef\tempurl%
\url{https://doi.org/10.1145/3379597.3387453}
\showDOI{\tempurl}


\bibitem[\protect\citeauthoryear{Tahir, Counsell, and MacDonell}{Tahir
  et~al\mbox{.}}{2016}]%
        {tahir2016empirical}
\bibfield{author}{\bibinfo{person}{Amjed Tahir}, \bibinfo{person}{Steve
  Counsell}, {and} \bibinfo{person}{Stephen~G MacDonell}.}
  \bibinfo{year}{2016}\natexlab{}.
\newblock \showarticletitle{An empirical study into the relationship between
  class features and test smells}. In \bibinfo{booktitle}{\emph{2016 23rd
  Asia-Pacific Software Engineering Conference (APSEC)}}. IEEE,
  \bibinfo{pages}{137--144}.
\newblock


\bibitem[\protect\citeauthoryear{Tufano, Palomba, Bavota, Di~Penta, Oliveto,
  De~Lucia, and Poshyvanyk}{Tufano et~al\mbox{.}}{2016}]%
        {tufano2016empirical}
\bibfield{author}{\bibinfo{person}{Michele Tufano}, \bibinfo{person}{Fabio
  Palomba}, \bibinfo{person}{Gabriele Bavota}, \bibinfo{person}{Massimiliano
  Di~Penta}, \bibinfo{person}{Rocco Oliveto}, \bibinfo{person}{Andrea
  De~Lucia}, {and} \bibinfo{person}{Denys Poshyvanyk}.}
  \bibinfo{year}{2016}\natexlab{}.
\newblock \showarticletitle{An empirical investigation into the nature of test
  smells}. In \bibinfo{booktitle}{\emph{Proceedings of the 31st IEEE/ACM
  International Conference on Automated Software Engineering}}.
  \bibinfo{pages}{4--15}.
\newblock


\bibitem[\protect\citeauthoryear{Van~Deursen, Moonen, Van Den~Bergh, and
  Kok}{Van~Deursen et~al\mbox{.}}{2001}]%
        {van2001refactoring}
\bibfield{author}{\bibinfo{person}{Arie Van~Deursen}, \bibinfo{person}{Leon
  Moonen}, \bibinfo{person}{Alex Van Den~Bergh}, {and} \bibinfo{person}{Gerard
  Kok}.} \bibinfo{year}{2001}\natexlab{}.
\newblock \showarticletitle{Refactoring test code}. In
  \bibinfo{booktitle}{\emph{Proceedings of the 2nd international conference on
  extreme programming and flexible processes in software engineering
  (XP2001)}}. \bibinfo{pages}{92--95}.
\newblock


\bibitem[\protect\citeauthoryear{Van~Rompaey, Du~Bois, and Demeyer}{Van~Rompaey
  et~al\mbox{.}}{2006}]%
        {van2006characterizing}
\bibfield{author}{\bibinfo{person}{Bart Van~Rompaey}, \bibinfo{person}{Bart
  Du~Bois}, {and} \bibinfo{person}{Serge Demeyer}.}
  \bibinfo{year}{2006}\natexlab{}.
\newblock \showarticletitle{Characterizing the relative significance of a test
  smell}. In \bibinfo{booktitle}{\emph{2006 22nd IEEE International Conference
  on Software Maintenance}}. IEEE, \bibinfo{pages}{391--400}.
\newblock


\bibitem[\protect\citeauthoryear{Van~Rompaey, Du~Bois, Demeyer, and
  Rieger}{Van~Rompaey et~al\mbox{.}}{2007}]%
        {van2007detection}
\bibfield{author}{\bibinfo{person}{Bart Van~Rompaey}, \bibinfo{person}{Bart
  Du~Bois}, \bibinfo{person}{Serge Demeyer}, {and} \bibinfo{person}{Matthias
  Rieger}.} \bibinfo{year}{2007}\natexlab{}.
\newblock \showarticletitle{On the detection of test smells: A metrics-based
  approach for general fixture and eager test}.
\newblock \bibinfo{journal}{\emph{IEEE Transactions on Software Engineering}}
  \bibinfo{volume}{33}, \bibinfo{number}{12} (\bibinfo{year}{2007}),
  \bibinfo{pages}{800--817}.
\newblock


\bibitem[\protect\citeauthoryear{Virg{\'\i}nio, Martins, Rocha, Santana, Cruz,
  Costa, and Machado}{Virg{\'\i}nio et~al\mbox{.}}{2020a}]%
        {virginio2020jnose}
\bibfield{author}{\bibinfo{person}{T{\'a}ssio Virg{\'\i}nio},
  \bibinfo{person}{Luana Martins}, \bibinfo{person}{Larissa Rocha},
  \bibinfo{person}{Railana Santana}, \bibinfo{person}{Adriana Cruz},
  \bibinfo{person}{Heitor Costa}, {and} \bibinfo{person}{Ivan Machado}.}
  \bibinfo{year}{2020}\natexlab{a}.
\newblock \showarticletitle{JNose: Java Test Smell Detector}. In
  \bibinfo{booktitle}{\emph{Proceedings of the 34th Brazilian Symposium on
  Software Engineering}}. \bibinfo{pages}{564--569}.
\newblock


\bibitem[\protect\citeauthoryear{Virg{\'\i}nio, Martins, Soares, Santana,
  Costa, and Machado}{Virg{\'\i}nio et~al\mbox{.}}{2020b}]%
        {virginio2020empirical}
\bibfield{author}{\bibinfo{person}{T{\'a}ssio Virg{\'\i}nio},
  \bibinfo{person}{Luana~Almeida Martins}, \bibinfo{person}{Larissa~Rocha
  Soares}, \bibinfo{person}{Railana Santana}, \bibinfo{person}{Heitor Costa},
  {and} \bibinfo{person}{Ivan Machado}.} \bibinfo{year}{2020}\natexlab{b}.
\newblock \showarticletitle{An empirical study of automatically-generated tests
  from the perspective of test smells}. In
  \bibinfo{booktitle}{\emph{Proceedings of the 34th Brazilian Symposium on
  Software Engineering}}. \bibinfo{pages}{92--96}.
\newblock


\bibitem[\protect\citeauthoryear{Virg{\'\i}nio, Santana, Martins, Soares,
  Costa, and Machado}{Virg{\'\i}nio et~al\mbox{.}}{2019}]%
        {virginio2019influence}
\bibfield{author}{\bibinfo{person}{T{\'a}ssio Virg{\'\i}nio},
  \bibinfo{person}{Railana Santana}, \bibinfo{person}{Luana~Almeida Martins},
  \bibinfo{person}{Larissa~Rocha Soares}, \bibinfo{person}{Heitor Costa}, {and}
  \bibinfo{person}{Ivan Machado}.} \bibinfo{year}{2019}\natexlab{}.
\newblock \showarticletitle{On the influence of Test Smells on Test Coverage}.
  In \bibinfo{booktitle}{\emph{Proceedings of the XXXIII Brazilian Symposium on
  Software Engineering}}. \bibinfo{pages}{467--471}.
\newblock


\bibitem[\protect\citeauthoryear{Werner, Grabowski, Neukirchen, R{\"o}ttger,
  Waack, and Zeiss}{Werner et~al\mbox{.}}{2007}]%
        {Werner2007TTCN}
\bibfield{author}{\bibinfo{person}{Edith Werner}, \bibinfo{person}{Jens
  Grabowski}, \bibinfo{person}{Helmut Neukirchen}, \bibinfo{person}{Nils
  R{\"o}ttger}, \bibinfo{person}{Stephan Waack}, {and}
  \bibinfo{person}{Benjamin Zeiss}.} \bibinfo{year}{2007}\natexlab{}.
\newblock \showarticletitle{TTCN-3 Quality Engineering: Using Learning
  Techniques to Evaluate Metric Sets}. In \bibinfo{booktitle}{\emph{SDL 2007:
  Design for Dependable Systems}}, \bibfield{editor}{\bibinfo{person}{Emmanuel
  Gaudin}, \bibinfo{person}{Elie Najm}, {and} \bibinfo{person}{Rick Reed}}
  (Eds.). \bibinfo{publisher}{Springer Berlin Heidelberg},
  \bibinfo{address}{Berlin, Heidelberg}, \bibinfo{pages}{54--68}.
\newblock
\showISBNx{978-3-540-74984-4}


\bibitem[\protect\citeauthoryear{Wohlin}{Wohlin}{2014}]%
        {10.1145/2601248.2601268}
\bibfield{author}{\bibinfo{person}{Claes Wohlin}.}
  \bibinfo{year}{2014}\natexlab{}.
\newblock \showarticletitle{Guidelines for Snowballing in Systematic Literature
  Studies and a Replication in Software Engineering}. In
  \bibinfo{booktitle}{\emph{Proceedings of the 18th International Conference on
  Evaluation and Assessment in Software Engineering}} (London, England, United
  Kingdom) \emph{(\bibinfo{series}{EASE '14})}. \bibinfo{publisher}{Association
  for Computing Machinery}, \bibinfo{address}{New York, NY, USA}, Article
  \bibinfo{articleno}{38}, \bibinfo{numpages}{10}~pages.
\newblock
\showISBNx{9781450324762}
\urldef\tempurl%
\url{https://doi.org/10.1145/2601248.2601268}
\showDOI{\tempurl}


\bibitem[\protect\citeauthoryear{Zeiss, Neukirchen, Grabowski, Evans, and
  Baker}{Zeiss et~al\mbox{.}}{2006}]%
        {Zeiss2006TTCN}
\bibfield{author}{\bibinfo{person}{Benjamin Zeiss}, \bibinfo{person}{Helmut
  Neukirchen}, \bibinfo{person}{Jens Grabowski}, \bibinfo{person}{Dominic
  Evans}, {and} \bibinfo{person}{Paul Baker}.} \bibinfo{year}{2006}\natexlab{}.
\newblock \showarticletitle{Refactoring and Metrics for TTCN-3 Test Suites}. In
  \bibinfo{booktitle}{\emph{System Analysis and Modeling: Language Profiles}},
  \bibfield{editor}{\bibinfo{person}{Reinhard Gotzhein} {and}
  \bibinfo{person}{Rick Reed}} (Eds.). \bibinfo{publisher}{Springer Berlin
  Heidelberg}, \bibinfo{address}{Berlin, Heidelberg},
  \bibinfo{pages}{148--165}.
\newblock
\showISBNx{978-3-540-68373-5}


\bibitem[\protect\citeauthoryear{Zhang, Jalali, Wuttke, Mu{\c{s}}lu, Lam,
  Ernst, and Notkin}{Zhang et~al\mbox{.}}{2014}]%
        {zhang2014empirically}
\bibfield{author}{\bibinfo{person}{Sai Zhang}, \bibinfo{person}{Darioush
  Jalali}, \bibinfo{person}{Jochen Wuttke}, \bibinfo{person}{K{\i}van{\c{c}}
  Mu{\c{s}}lu}, \bibinfo{person}{Wing Lam}, \bibinfo{person}{Michael~D Ernst},
  {and} \bibinfo{person}{David Notkin}.} \bibinfo{year}{2014}\natexlab{}.
\newblock \showarticletitle{Empirically revisiting the test independence
  assumption}. In \bibinfo{booktitle}{\emph{Proceedings of the 2014
  International Symposium on Software Testing and Analysis}}.
  \bibinfo{pages}{385--396}.
\newblock


\end{thebibliography}

\end{document}